%% file: PhD_Thesis.tex
\documentclass[a4paper]{scrreprt}
\usepackage{amsmath}
\usepackage{graphicx,epsfig}
\usepackage{color}
\usepackage{mathrsfs}
\usepackage{setspace}
\usepackage{fancyhdr}

\newcommand{\be}{\begin{equation}}
\newcommand{\ee}{\end{equation}}
\newcommand{\bd}{\begin{equation*}}
\newcommand{\ed}{\end{equation*}}
\newcommand{\bea}{\begin{eqnarray}}
\newcommand{\eea}{\end{eqnarray}}
\newcommand{\bead}{\begin{eqnarray*}}
\newcommand{\eead}{\end{eqnarray*}}

\newcommand{\gapp}{\mathrel{\raise.3ex\hbox{$>$}\mkern-14mu
              \lower0.6ex\hbox{$\sim$}}}
\newcommand{\lapp}{\mathrel{\raise.3ex\hbox{$<$}\mkern-14mu
              \lower0.6ex\hbox{$\sim$}}}

\begin{document}

\pagenumbering{roman}

\begin{titlepage}
\begin{center}
\vspace*{1in}
{\Large Quantum Mechanical Effects in Gravitational Collapse}
\par
\vspace{1in}
{\large by}
\par
\vspace{1in}
{\large Eric Sean Greenwood}
\par
\vspace{0.5in}
{\large January 5, 2010}
\par
\vfill
A dissertation submitted to the Faculty of the Graduate School of the University at Buffalo, State University of New York in partial fulfillment of the requirements for the degree of Doctor of Philosophy.
\par
\vspace{0.5in}
Department of Physics
\end{center}
\end{titlepage}

\newpage

%\listoftables

\pagestyle{fancy}
\fancyhead[LO]{\slshape}
\fancyfoot[C]{\thepage}

\chapter*{Dedication}
\addcontentsline{toc}{chapter}{Dedication}

This body of work is dedicated to my father, Richard D. Greenwood Sr., and my grandfather, Albert Kneaskern. My grandfather for his faith in me beyond anyone else I have ever known, even before I had matured enough to realize the implication of his faith. For my father who taught me the true nature of hard work and devotion. Together they have embodied everything that I have ever aspired to in life. Yours memories shall forever live in me, may you both rest in peace. 

\newpage

\chapter*{Acknowledgements}
\addcontentsline{toc}{chapter}{Acknowledgements}

I would like to thank my committee members and the faculty and staff in the physics department at the University at Buffalo. 

I would like to thank first and foremost my advisor Dr.~Dejan Stojkovic for taking me under his wing and giving me an opportunity to blossom as both an individual and a physicist. His insight and understanding of the material is something that I myself only hope to achieve as a physicist. I would like to thank Dr.~Ulrich Baur for being my surrogate adviser for a year; his mentoring during that time was invaluable to me for both his insight and his willingness to explore subjects outside his specialization. I would like to thank Dr.~Doreen Wackeroth for her help, kindness, wisdom and tolerance of me in her classes (sorry for being such a trouble maker during your lectures!). I would also like to thank Dr.~Francis Gasparini for his compassion and console during some very difficult times during my duration at the University at Buffalo. His understanding and advise where inspirational to me during these trying times and having faith in me to teach both summer and regular semester courses. 

I would also like to thank my family for their constant support and devotion over the years before and during this body of work. Your support was and is very comforting to me. A very special thank you to my friends that have become an intricate part of my life during the torture which is graduate school. I especially want to thank my friends Tyler Glembo, Andr\'as Sablauer and Kenneth Smith, for without them I would have never made it this far in both my schooling and my life.

\newpage

\tableofcontents

\listoffigures
\addcontentsline{toc}{chapter}{List of Figures}

\newpage

\chapter*{Abstract}
\addcontentsline{toc}{chapter}{Abstract}

In this thesis we investigate quantum mechanical effects to various aspects of gravitational collapse. These quantum mechanical effects are implemented in the context of the Functional Schr\"odinger formalism. The Functional Schr\"odinger formalism allows us to investigate the time-dependent evolutions of the quantum mechanical effects, which is beyond the scope of the usual methods used to investigate the quantum mechanical corrections of gravitational collapse. Utilizing the time-dependent nature of the Functional Schr\"odinger formalism, we study the quantization of a spherically symmetric domain wall from the view point of an asymptotic and infalling observer, in the absence of radiation. To build a more realistic picture, we then study the time-dependent nature of the induced radiation during the collapse using a semi-classical approach. Using the domain wall and the induced radiation, we then study the time-dependent evolution of the entropy of the domain wall. Finally we make some remarks about the possible inclusion of backreaction into the system. 

\newpage

\pagenumbering{arabic}

\chapter{Introduction}
\label{ch:intro}

This thesis is based on work done in a series of nine papers, which have been published in several different journals, see Refs.\cite{Stojkovic,GreenStoj,Greenwood,EG_ent,WangGreenStoj,EG_RNrad,GreenHalHao,GreenHalPolStoj}. The work here is varied and takes several different aspects into account, however, for this thesis we concentrate on only a subset of these papers. The subset of interest here are those papers which include gravitational collapse of a massive shell only. Even though this is only a subset of the possible parameters that a black hole can have, this subset displays most of the interesting features that illustrate the core of our work.  

\newpage

\include{formal}

\newpage

\include{number}

\newpage

\include{model}

\newpage

\include{Classical}

\newpage

\include{quantum}

\newpage

\include{Radiation}

\newpage

\include{Entropy}

\newpage

\include{BR}

\newpage

\chapter{Conclusion}
\label{ch:Conclusion}

In this thesis we have investigated quantum mechanical effects of gravitational collapse by utilizing the time-dependent nature of the Functional Schr\"odinger formalism. As stressed throughout this thesis, the Functional Schr\"odinger formalism allows us to investigate the intermediate regimes that the standard methods cannot. Therefore we can obtain a better understanding of what is happening during the evolution of the collapse, at least in the context of the Functional Schr\"odinger formalism. As we have seen in the previous chapters, the effects in this intermediate regime are robust and give good insight into the process of collapse.

This thesis is not meant to be an exhaustive list of the different types of gravitational collapse. Here we solely concentrated on a massive domain wall, while ignoring all other observable quantities (such as charge and angular momentum). However, one can ``easily" incorporate these observable into the system as well. For example, one can repeat the steps above for the case of a massive-charged domain wall (i.e. Reissner-Nordstr\"om). This has been done for the classical and quantum solutions in Ref.\cite{WangGreenStoj} and for the semi-classical radiation in Ref.\cite{EG_RNrad}. The analysis can also be repeated for different topologies, other than spherically symmetric domain walls, as well as for different asymptotic space-times (such as de Sitter or anti-de Sitter). In Ref.\cite{GreenHalHao} the classical and quantum solutions are studied for a $(3+1)$-dimensional BTZ black string in AdS space. It is well know that a $(3+1)$-dimensional BTZ black string has the topology of a cylinder and is asymptotic to AdS space-time, due to the negative cosmological constant. 

It is also important to note here that the Functional Schr\"odinger formalism is not restricted to gravitational collapse. One could also apply the formalism to expanding systems as well, which are essentially collapsing systems in reverse. In this case, one can investigate an expanding de Sitter or anti-de Sitter universe and consider the radiation and entropy during the evolution of expansion. For subsequent work on the radiation given off during expansion, see ``Time dependent fluctuations and particle production in cosmological de Sitter and anti-de Sitter spaces," by E. Greenwood, D. Dai and D. Stojkovic (submitted for publication in Phys.~Rev.~{\bf D}). In the case of de Sitter expansion, which is represented by the Freedman-Robertson-Walker metric, the horizon is the largest comoving distance which light emitted now can reach the observer at any time in the future. It is expected that that de Sitter space can produce thermal radiation as well (for some counter arguments see Refs.\cite{Sekiwa,Volovik}). In the case of anti-de Sitter expansion, unlike de Sitter expansion, the space-time does not contain an event horizon. Therefore, one would not expect thermal radiation with a constant temperature. However, due to the time-dependent metric, particle production is still expected. Here it is expected that after a short time of expansion, the universe starts recollapsing and ends up forming a black hole, see for example Refs.\cite{Coleman_deLuccia,AbbottColeman}. The Functional Schr\"odinger formalism can be applied to these situations as well to help shed light on these questions.

\newpage

\appendix

%\chapter*{Appendix}
%\addcontentsline{toc}{chapter}{Appendix}

\include{Invariant}

\newpage

\include{Gauss_Codazzi}

\newpage

\include{rho_t}

\newpage

\include{rho_tau}

\end{document}

%% file: formal.tex
\chapter{Formalism}
\label{ch:formal}

The most important part of our research is the formalism used to study the quantum mechanical effects of gravitational collapse. To study these quantum effects of gravitational collapse we will institute the Functional Schr\"odinger equation. In this part of the thesis we will first derive the Functional Schr\"odinger equation, which will be the primary equation used to study the quantum effects of gravitational collapse.

The main purpose of the Functional Sch\"odinger equation is to introduce the ``observer" time into the Wheeler-de Witt equation. It is well known in General Relativity that for different foliations of space-time, different physical observations occur. For example, one can consider the equations of motion for a black hole. Let us consider an object which is falling into a black hole from different points of view. If one chooses the time as observed by an asymptotic observer as the desired foliation of space-time, upon solving the equations of motion of the object, one finds that it takes an infinite amount of time for the object to fall into the black (even if the object is a photon). However, if one choses the time as observed by a freely falling observer (one that is falling into the black hole along a geodesic) as the desired foliation of space-time, upon solving the equations of motion of the object in this case, one finds that it takes a finite amount of time for the object to fall into the black hole. Therefore, one can see that it is important and instructive to consider different foliations of the space-time to learn different aspects of the system of gravitational collapse. The Functional Schr\"odinger equation allows for one to specify the particular foliation of space-time that is of interest and study the system from that view point.

A second purpose of the Functional Schr\"odinger equation is to allow one to investigate the time evolution of the system. Typically this is not done in the study of gravitational collapse. The preferred method of study is to consider an initial static asymptotically flat space-time, let the system evolve (with no knowledge of the evolution), then consider a final different static asymptotically flat space-time. Then by comparing these two different space-times, one can in principle have some understanding of what the evolution was like between these two events. The Functional Schr\"odinger equation will in principle allow us to study the total time evolution of the system, not just the static asymptotically flat regions of space-time.

\section{Functional Schr\"odinger Equation}

In this section we will derive the Functional Schr\"odinger equation.

The Wheeler-de Witt equation for a closed universe is given by, see Ref.\cite{DeWitt},
\be
  H\Psi=0
  \label{WdW}
\ee
where $H$ is the total Hamiltonian and $\Psi$ is the total wavefunction for all the ingredients of the system, including the observer's degrees of freedom denoted by ${\cal{O}}$. Eq.(\ref{WdW}) is a consequence of the idea that there is no ``God" time, or no preferred time, or no super-observer time. Therefore Eq.(\ref{WdW}) is written in a gauge independent fashion, since there is no preferred observer to observe the system.

In general we can write the wavefunction in Eq.(\ref{WdW}) as
\be
  \Psi=\Psi\left(X^{\alpha},g_{\mu\nu},\Phi,{\cal{O}}\right).
  \label{wPsi}
\ee
Here $X^{\alpha}=X^{\alpha}(\zeta^a)$ describes the location of the wall as a function of the internal wall world volume coordinates $\zeta^a$, $g_{\mu\nu}$ is the metric, and $\Phi$ is a scalar field. The Roman indices go over the internal domain wall world volume coordinates and the Greek indices go over space-time coordinates. Note that the wavefunctional in Eq.(\ref{wPsi}) is a functional of the fields but not the space-time coordinates. In general, the total Hamiltonian is a linear combination of the Hamiltonian of the system itself and that of the Hamiltonian of the observer. Therefore we will separate the Hamiltonian into two parts, one for the system and the other for the observer, which can be written as
\be
  H=H_{sys}+H_{obs}.
  \label{commutation}
\ee
Any (weak) interaction terms between the observer and the wall-metric-scalar system are included in $H_{sys}$. The observer is assumed to not significantly affect the evolution of the system and vice versa. In mathematical language this means that we are assuming that the Hamiltonian for the system and the observer commute with each other
\be
  \left[H_{sys},H_{obs}\right]=0.
\ee
The total wavefunction Eq.(\ref{wPsi}) can be written as a sum over eigenstates
\be
  \Psi=\sum_kc_k\Psi^k_{sys}(sys,t)\Psi_{obs}^k({\cal{O}},t)
  \label{totPsi}
\ee
where $k$ labels the eigenstates and $c_k$ are complex coefficients. 

To solve the full Wheeler-de Witt equation is very difficult since it involves all the degrees of freedom, both that of the system and the observer. Here we shall utilize the frequently employed strategy of truncating the field degrees of freedom to a finite subset, hence we will be consider with the minsuperspace version of the Wheeler-de Witt equation. As long as we keep all the relevant degrees of freedom that are of interest, this is a useful truncation. Since we are only considering a subset of the total degrees of freedom, this is now considered an ``open" system. In an ``open" system, one can then define an appropriate ``observer" time in which one chooses to make measurements. Therefore we can write the Schr\"odinger equation for the observation as
\be
  i\frac{\partial\Psi^k_{obs}}{\partial t}\equiv H_{obs}\Psi_{obs}^k.
  \label{ObsSchrod}
\ee
This is convenient, however, we wish to make observations on the system not on the observer. To transform this to observation on the system we will make use Eq.(\ref{WdW}).

To introduce the observer time on observations of the system, we use Eq.(\ref{WdW}), Eq.(\ref{totPsi}) and Eq.(\ref{ObsSchrod}), therefore we can write
\begin{align}
  H\Psi&=\left(H_{sys}+H_{obs}\right)\sum_kc_k\Psi^k_{sys}(sys,t)\Psi_{obs}^k({\cal{O}},t)\nonumber\\
      &=H_{sys}\sum_kc_k\Psi^k_{sys}(sys,t)\Psi_{obs}^k({\cal{O}},t)+H_{obs}\sum_kc_k\Psi^k_{sys}(sys,t)\Psi_{obs}^k({\cal{O}},t)\nonumber\\
      &=H_{sys}\sum_kc_k\Psi^k_{sys}(sys,t)\Psi_{obs}^k({\cal{O}},t)+\sum_kc_k\Psi^k_{sys}(sys,t)H_{obs}\Psi_{obs}^k({\cal{O}},t)\nonumber\\
      &=H_{sys}\sum_kc_k\Psi^k_{sys}(sys,t)\Psi_{obs}^k({\cal{O}},t)+i\sum_kc_k\Psi^k_{sys}(sys,t)\partial_t\Psi_{obs}^k({\cal{O}},t)
      \label{der}
\end{align}
where we made use of Eq.(\ref{commutation}) and in the last line we used Eq.(\ref{ObsSchrod}).

Now consider the integral of the last term in Eq.(\ref{der}), we have
\begin{align}
  \int_{t_i}^{t_f}dt\sum_kc_k\Psi_{sys}^k(sys,t)\partial_t\Psi_{obs}^k({\cal{O}},t)=&\sum_kc_k\Psi_{sys}^k(sys,t)\Psi_{obs}^k({\cal{O}},t)\Big{|}_{t_i}^{t_f}\nonumber\\
         &-\int_{t_i}^{t_f}dt\sum_k\left(\partial_t\Psi_{sys}^k(sys,t)\right)\Psi_{obs}^k({\cal{O}},t)\nonumber\\
         =&\Psi\Big{|}_{t_i}^{t_f}-\int_{t_i}^{t_f}dt\sum_k\left(\partial_t\Psi_{sys}^k(sys,t)\right)\Psi_{obs}^k({\cal{O}},t).
         \label{by_parts}
\end{align}
However, by virtue of the Wheeler-de Witt equation, the total wavefunction $\Psi$ is time-independent. Therefore the first term on the right hand side in Eq.(\ref{by_parts}) is zero. Shrinking the integral we then have,
\be
  \sum_kc_k\Psi_{sys}^k(sys,t)\partial_t\Psi_{obs}^k({\cal{O}},t)=-\sum_kc_k\left(\partial_t\Psi_{sys}^k(sys,t)\right)\Psi_{obs}^k({\cal{O}},t).
  \label{equiv}
\ee
Substituting Eq.(\ref{equiv}) into Eq.(\ref{der}) we can then write,
\be
  H_{sys}\sum_kc_k\Psi_{sys}^k(sys,t)\Psi_{obs}^k({\cal{O}},t)=i\sum_kc_k\left(\partial_t\Psi_{sys}^k(sys,t)\right)\Psi_{obs}^k({\cal{O}},t)
\ee
or interms of one $k$ value, we then arrive at the Functional Schr\"odinger equation
\be
  H_{sys}\Psi_{sys}^k=i\frac{\partial\Psi_{sys}^k}{\partial t}.
\ee
For convenience, from now on we will denote the system wavefunction simply by $\Psi$ and drop the superscript $k$ and the subscript ``sys". Similarly $H$ will now denote $H_{sys}$, and the Schr\"odinger equation reads
\be
  H\Psi=i\frac{\partial\Psi}{\partial t}.
  \label{FSE}
\ee

\section{Discussion}

Here we have derived the Functional Schr\"odinger equation. As discussed above, the purpose of the Functional Schr\"odinger equation is to introduce the ``observer" time into the Wheeler-de Witt equation, Eq.(\ref{WdW}). This will allow us to be able to use the classical Hamiltonian of the system of gravitational collapse, then study the evolution of the system from the view point of any observer of our choosing. This has two benefits: First the formalism allows us to choose the ``observer" we wish to study. As discussed earlier, different ``observers" will observer different phenomena which are of interest. Secondly, the formalism will allow us to evolve the system quantum mechanically over time. One of the benefits of this approach is that we can, in principle, observe thermodynamic properties of the system in a time-dependent fashion, which we will discuss in Chapters \ref{ch:radiation} and \ref{ch:entropy}. As we will discuss, this is something which is beyond the scope of the usual methods used to study the thermodynamic properties of the system.

%% file: number.tex
\chapter{Occupation Number}
\label{ch:number}

Throughout the text we will be interested in the number of particles created during the gravitational collapse of our object, i.e. the radiation. Therefore we will derive the occupation number of the particles created during the time of collapse in this chapter for future convenience. Throughout this text we will be interested in systems with spherical symmetry, since this is the simplest case to consider. Here we note that due to the spherical symmetry, gravitational radiation is excluded from the system, thus the radiation which we will consider will be from the excitation of particles due to the time-dependent nature of the gravitational metric. 

To consider the radiation we will consider a quantum scalar field $\phi$ in the background of the gravitational collapsing object, which is given by
\be
  S=\int d^4x\frac{1}{2}\sqrt{-g}g^{\mu\nu}\partial_{\mu}\phi\partial_{\nu}\phi
  \label{ScalarAct}
\ee
where $S$ is the action of the scalar field. The reason we are considering only a scalar field is that this is the simplest and easiest case, which gives insight into most of the physically significant phenomena. By considering more complicated fields, one arrives at the so-called gray-body factors, see for example Ref.\cite{Birrell}, which are dependent on the type of field used. Here we derive the number of particles induced as a function of observer time ``$t$". Here ``$t$" is used for any foliation of space-time used throughout this body of work, whether the time is that of an asymptotic observer or that of an infalling observer. 

In most cases we arrive at the Hamiltonian of the system from Eq.(\ref{ScalarAct}), which is of the form of a sum of uncoupled simple harmonic oscillator. To simplify the notation, we consider one eigenmode of the simple harmonic oscillator given by
\be
  H=\frac{p^2}{2m}+\frac{m}{2}\omega^2(t)x^2
  \label{SHO}
\ee
where $p$ is the momentum conjugate to $x$ and $x$ is the eigenmode. Using the standard quantization procedure, upon inserting Eq.(\ref{SHO}) into Eq.(\ref{FSE}), we can then write
\be
  \left[-\frac{1}{2m}\frac{\partial^2}{\partial x^2}+\frac{m}{2}\omega^2(t)x^2\right]\psi(x,t)=i\frac{\psi(x,t)}{\partial t}.
  \label{SHOFS}
\ee
Here Eq.(\ref{SHOFS}) can be solved exactly by utilizing the invariant operator method first developed by Lewis and Reisenfeld, see Ref.\cite{Lewis} and Appendix \ref{ch:Invariant}. Using this method, Dantas, Pedrosa and Baseia showed, see Ref.\cite{Pedrosa}, that the exact solution to Eq.(\ref{SHOFS}) at late times is given by
\be
  \psi(x,t)=e^{i\alpha(t)}\left(\frac{m}{\pi\rho^2}\right)^{1/4}\exp\left[i\frac{m}{2}\left(\frac{\rho_t}{\rho}+\frac{i}{\rho^2}\right)x^2\right]
  \label{PedWave}
\ee
where $\rho_t$ denotes the derivative of $\rho(t)$ with respect to $t$, and $\rho$ is given by the real solution of the non-linear auxilarly equation
\be
  \rho_{tt}+\omega^2(t)\rho=\frac{1}{\rho^3}
  \label{gen_rho}
\ee
with intitial conditions
\be
  \rho(t_i)=\frac{1}{\sqrt{\omega_0}}, \hspace{2mm} \rho_t(t_i)=0
  \label{IC}
\ee
where $t_i$ is the initial time. The time-dependent phase $\alpha$ is given by
\be
  \alpha(t)=-\frac{1}{2}\int_{t_i}^t\frac{dt'}{\rho^2(t')}.
\ee

To find the occupation number of the induced radiation, consider an observer with detectors that are designed to register particles of different frequencies for the free scalar field $\Phi$ at earlier times. Such an observer will interpret the wavefunction of a given mode $x$ at late times in terms of simple harmonic oscillator states, $\{\varphi_n\}$, at final frequency $\bar{\omega}$. Here $\bar{\omega}$, is the value of the frequency evaluated at a time $t_f$ as seen by the observer. The number of quanta in eigenmode $x$ can be evaluated by decomposing the wavefunction Eq.(\ref{PedWave}) in terms of the states $\{\varphi_n\}$, and by evaluating the occupation number of that mode. To implement this, we start by writing the wavefunction for a given mode at time $t>t_f$ in terms of the simple harmonic oscillator basis at $t=0$
\be
  \psi(x,t)=\sum_nc_n(t)\varphi_n(x)
  \label{Phi_Exp}
\ee
where
\be
  c_n=\int dx\varphi_n^*(x)\psi(x,t)
  \label{c_n}
\ee
is the overlap, i.e. inner product, between the initial and final state of the wavefunction. The occupation number at eigenfrequency $\bar{\omega}$ by the time $t>t_f$, is given by the expectation value
\be
  N(t,\bar{\omega})=\sum_nn\left|c_n\right|^2.
  \label{occnum}
\ee

To evaluate the sum in Eq.(\ref{occnum}), we use the simple harmonic oscillator basis states but at a frequency $\bar{\omega}$ to keep track of the different $\omega$'s in the calculation. To evaluate the occupation numbers at time $t>t_f$, we need only set $\bar{\omega}=\omega(t_f)$. So the simple harmonic oscillator basis states are written as (see for example Appendix A.4 of Ref.\cite{Sakurai})
\be
  \varphi(b)=\left(\frac{m\bar{\omega}}{\pi}\right)^{1/4}\frac{e^{-m\bar{\omega}b^2/2}}{\sqrt{2^nn!}}{\cal{H}}_n\left(\sqrt{m\bar{\omega}}b\right)
\ee
where ${\cal{H}}_n$ are the Hermite polynomials. Then Eq.(\ref{c_n}) and Eq.(\ref{PedWave}) together gives
\bea
  c_n&=&\left(\frac{1}{\pi^2\bar{\omega}\rho^2}\right)^{1/4}\frac{e^{i\alpha}}{\sqrt{2^nn!}}\int d\xi e^{-P\xi^2/2}{\cal{H}}_n(\xi)\nonumber\\
         &=&\left(\frac{1}{\pi^2\bar{\omega}\rho^2}\right)^{1/4}\frac{e^{i\alpha}}{\sqrt{2^nn!}}I_n
\eea
where
\be
  P=1-\frac{i}{\bar{\omega}}\left(\frac{\rho_{\eta}}{\rho}+\frac{i}{\rho^2}\right).
  \label{P}
\ee

To find $I_n$ consider the corresponding integral over the generating function for the Hermite polynomials
\bea
  J(z)&=&\int d\xi e^{-P\xi^2/2}e^{-z^2+2z\xi}\nonumber\\
         &=&\sqrt{\frac{2\pi}{P}}e^{-z^2(1-2/P)}.
\eea
Since
\be
  e^{-z^2+2z\xi}=\sum_{n=0}^{\infty}\frac{z^n}{n!}{\cal{H}}_n(\xi)
\ee
we can then write
\be
  \int d\xi e^{-P\xi^2/2}{\cal{H}}_n(\xi)=\frac{d^n}{dz^n}J(z)\Big{|}_{z=0}.
\ee
Therefore
\be
  I_n=\sqrt{\frac{2\pi}{P}}\left(1-\frac{2}{P}\right)^{n/2}{\cal{H}}_n(0).
\ee
Since
\be
  {\cal{H}}_n(0)=(-1)^{n/2}\sqrt{2^nn!}\frac{(n-1)!!}{\sqrt{n!}}, \hspace{2mm} n=even
\ee
and ${\cal{H}}_n(0)=0$ for $n=odd$, we find the coefficients $c_n$ for even values of $n$,
\be
  c_n=\frac{(-1)^{n/2}e^{i\alpha}}{(\bar{\omega}\rho^2)^{1/4}}\sqrt{\frac{2}{P}}\left(1-\frac{2}{P}\right)^{n/2}\frac{(n-1)!!}{\sqrt{n!}}.
\ee
For odd $n$, $c_n=0$.

We can now find the number of particles produced during the collapse. Let 
\be
  \chi=\left|1-\frac{2}{P}\right|.
  \label{chi}
\ee
Then using Eq.(\ref{occnum}) we have
\bea
  N(t,\bar{\omega})&=&\sum_{n=even}n\left|c_n\right|^2\nonumber\\
      &=&\frac{2}{\sqrt{\bar{\omega}\rho^2}|P|}\chi\frac{d}{d\chi}\sum_{n=even}\frac{(n-1)!!}{n!!}\chi^n\nonumber\\
      &=&\frac{2}{\sqrt{\bar{\omega}\rho^2}|P|}\chi\frac{d}{d\chi}\frac{1}{\sqrt{1-\chi^2}}\nonumber\\
      &=&\frac{2}{\sqrt{\bar{\omega}\rho^2}|P|}\frac{\chi^2}{(1-\chi^2)^{3/2}}.
      \label{OccNum}
\eea
Now inserting Eq.(\ref{P}) and Eq.(\ref{chi}) leads to 
\be
  N(t,\bar{\omega})=\frac{\bar{\omega}\rho^2}{\sqrt{2}}\left[\left(1-\frac{1}{\bar{\omega}\rho^2}\right)^2+\left(\frac{\rho_t}{\bar{\omega}\rho}\right)^2\right].
\ee

\section{Discussion}

Here we derived the occupation number of the radiation induced during the time of gravitational collapse. The occupation number is measured by an observer with a detector at late times $t>t_f$. As stated earlier, this was done for convenience since we will use this quantity several times during this text.

%% file: model.tex
\chapter{Model}
\label{ch:model}

To study a concrete realization of black hole formation we consider a spherically symmetric Nambu-Goto domain wall (representing a shell of matter) that is collapsing. To include the possibility of (spherically symmetric) radiation, as discussed in the previous chapter (Chapter \ref{ch:number}), we consider a massless scalar field, $\Phi$, that is coupled to the gravitational field but not directly to the domain wall. The action for the system is then given as
\be
  S=\int d^4x\sqrt{-g}\left[-\frac{1}{16\pi G}{\cal{R}}+\frac{1}{2}(\partial_{\mu}\Phi)^2\right]-\sigma\int d^3\zeta\sqrt{-\gamma}+S_{obs}.
  \label{gen_action}
\ee
The first term is the Einstein-Hilbert action for the gravitational field, the second is the scalar field action, the third is the domain wall action in terms of the wall world volume coordinates, $\zeta^a$ ($a=0,1,2$), the wall tension $\sigma$, and the induced world volume metric
\be
  \gamma_{ab}=g_{\mu\nu}\partial_aX^{\mu}\partial_bX^{\nu}.
\ee
As stated in Chapter \ref{ch:formal}, the coordinates $X^{\mu}=X^{\mu}(\zeta^a)$ describe the location of the wall. The term $S_{obs}$ in Eq.(\ref{gen_action}) denotes the action for the observer.

As discussed earlier, a general treatment of full Wheeler-de Witt equation, Eq.(\ref{WdW}), is very difficult. So, we shall use the frequently employed strategy of truncating the field degrees of freedom to a finite set, typically including only the relevant degrees of freedom. In other words, we will consider the minisuperspace version of the Wheeler-de Witt equation. As long as we keep all the relevant degrees of freedom, this is a useful truncation. Since we are considering spherically symmetric domain walls, we will assume spherical symmetry for all the fields. Thus, the wall is described by the radial degree of freedom $R(t)$ only. 

The metric for the wall is then taken to be the solution to Einstein equations for a spherical domain wall. In Ref.\cite{Ipser} the metric, as follows from the spherical symmetry, outside the wall is given by
\be
  ds^2=-\left(1-\frac{R_s}{r}\right)dt^2+\left(1-\frac{R_s}{r}\right)^{-1}dr^2+r^2d\Omega^2, \hspace{2mm} r>R(t)
  \label{Met_out}
\ee
where $R_s=2GM$ is the Schwarzschild radius in terms of the mass $M$ of the wall, and
\be
  d\Omega^2=d\theta^2+\sin^2\theta d\phi^2.
\ee
By Birkhoff's theorem, in the interior of the spherical domain wall the line element of the metric is flat, i.e. Minkowski, which is given by
\be
  ds^2=-dT^2+dr^2+r^2d\Omega^2, \hspace{2mm} r<R(t).
  \label{Met_in}
\ee
Here $T$ is the interior time coordinate, not to be confused with temperature. The interior time coordinate is related to the asymptotic observer time coordinate $t$ via the proper time $\tau$ of the domain wall. By matching the coordinates for the interior and exterior at the wall, in analogy with the Isreal junction condition (see Ref.\cite{Israel}), and assuming that the wall is infinitely thin, we have the relations
\be
  \frac{dT}{d\tau}=\sqrt{1+\left(\frac{dR}{d\tau}\right)^2}
  \label{dTdtau}
\ee
and
\be
  \frac{dt}{d\tau}=\frac{1}{B}\sqrt{B+\left(\frac{dR}{d\tau}\right)^2}
  \label{dtdtau}
\ee
where
\be
  B\equiv1-\frac{R_s}{R}.
  \label{B}
\ee
By taking the ratio of Eq.(\ref{dTdtau}) and Eq.(\ref{dtdtau}), the relationship between the interior time $T$ and the asymptotic time $t$ is given by
\be
  \frac{dT}{dt}=\frac{\sqrt{1+R_{\tau}^2}B}{\sqrt{B+R_{\tau}^2}}=\sqrt{B-\frac{(1-B)}{B}\dot{R}^2}
  \label{dTdt}
\ee
where $R_{\tau}=dR/d\tau$ and $\dot{R}=dR/dt$. 

Since we are restricting the system to fields with spherical symmetry only, we need not include other metric degrees of freedom. Thus, the scalar field can also be truncated to be the spherically symmetric modes
\be
  \Phi=\Phi(r,t).
\ee

In Ref.\cite{Ipser}, Ipser and Sikivie integrated the equations of motion for the spherically symmetric domain wall. They found that the mass is actually a constant of motion and is given by, see Appendix \ref{ch:GC} for a sketch of the method used,
\be
  M=\frac{1}{2}4\pi\sigma R^2\left[\sqrt{1+R_{\tau}^2}+\sqrt{B+R_{\tau}^2}\right]
  \label{mass}
\ee
where it is assumed that max($R$)$>1/4\pi\sigma G$. This assumption is just used to ensure that one does not start off inside of the collapsing spherical domain wall.

By virtue of Eq.(\ref{B}), Eq.(\ref{mass}) is implicit since $R_s=2GM$. Solving for $M$ explicitly in terms of $R_{\tau}$ gives
\be
  M=4\pi\sigma R^2\left[\sqrt{1+R_{\tau}^2}-2\pi\sigma GR\right].
  \label{Mass_tau}
\ee
However, making use of Eq.(\ref{dTdt}) we can solve for $M$ in terms of $R_T=dR/dT$
\be
  M=4\pi\sigma R^2\left[\frac{1}{\sqrt{1-R_T^2}}-2\pi\sigma GR\right].
  \label{Mass_T}
\ee

Before we proceed we wish to discuss the physical relevance of Eq.(\ref{Mass_tau}). First consider the case where $R_{\tau}=0$, i.e. for a static domain wall. The first term in the square bracket is just the total rest mass of the shell. When the shell is moving, i.e. $R_{\tau}\not=0$, the first term in the square bracket takes the kinetic energy of the domain wall into account. The last term in the square bracket is the self-gravity, or the binding energy of the domain wall. Therefore we can identify Eq.(\ref{Mass_tau}) (Eq.(\ref{Mass_T})) as the total energy of the system, hence the Hamiltonian of the system. Thus, we will refer to Eq.(\ref{Mass_tau}) (Eq.(\ref{Mass_T})) as the Hamiltonian.

\section{Discussion}

Here we developed the classical Hamiltonian for a massive spherically symmetric domain wall undergoing gravitational collapse. As stated in the last paragraph, Eq.(\ref{Mass_tau}) is the conserved mass of the system, however, it can be interpreted as the Hamiltonian. For the remainder of the text, we will use Eq.(\ref{Mass_tau}) (Eq.(\ref{Mass_T})) as the Hamiltonian for the system.

%% file: Classical.tex
\chapter{Classical Treatment}
\label{ch:Classical}

In this chapter we wish to study the classical equations of motion of the collapsing spherically symmetric domain wall. To do so, we will consider the cases for two different foliations of space-time. Here we consider the collapse of the spherically symmetric domain wall from the point of view of an asymptotic observer (one who is at rest with respect to the collapse) and from the point of view of an infalling observer (one who is riding along with the shell).  This will be done by considering Eq.(\ref{Mass_T}) and Eq.(\ref{Mass_tau}). 

A naive approach to obtaining the dynamics for the spherical domain wall is to insert Eq.(\ref{Met_out}) and Eq.(\ref{Met_in}), as well as Eq.(\ref{Mass_tau}) into the original action Eq.(\ref{gen_action}). Upon doing so it is known that this approach does not give the correct dynamics for gravitating systems. Therefore we will take an alternative approach for finding the action. We will find the action that does in fact lead to the correct mass conservation law. The form of the action can be deduced from Eq.(\ref{Mass_T}) (Eq.(\ref{Mass_tau})).

\section{Asymptotic Observer}

First we will consider the equations of motion from the view point of the asymptotic observer. The asymptotic observer is any observer stationary with respect to the collapsing domain wall, typically taken to be located at infinity. Here we summarize the work originally done in Ref.\cite{Stojkovic}.

From Eq.(\ref{Mass_T}) we find the effective action for the spherically symmetric domain wall to be
\be
  S_{eff}=-4\pi\sigma\int dTR^2\left[\sqrt{1-R_T^2}-2\pi\sigma GR\right].
  \label{Seff_T}
\ee
Using Eq.(\ref{dTdt}) we can write Eq.(\ref{Seff_T}) in terms of the asymptotic observer time $t$ as
\be
  S_{eff}=-4\pi\sigma\int dtR^2\left[\sqrt{B-\frac{\dot{R}^2}{B}}-2\pi\sigma GR\sqrt{B-\frac{(1-B)}{B}\dot{R}^2}\right].
  \label{Seff_t}
\ee 
From Eq.(\ref{Seff_t}) the effective Lagrangian for the system is
\be
  L_{eff}=-4\pi\sigma R^2\left[\sqrt{B-\frac{\dot{R}^2}{B}}-2\pi\sigma GR\sqrt{B-\frac{(1-B)}{B}\dot{R}^2}\right].
  \label{Leff_t}
\ee

The generalized momentum $\Pi_R$ can be derived from Eq.(\ref{Leff_t}) in the usual manner, this is given by
\be
  \Pi_R=\frac{4\pi\sigma R^2\dot{R}}{\sqrt{B}}\left[\frac{1}{\sqrt{B^2-\dot{R}^2}}-\frac{2\pi\sigma GR(1-B)}{\sqrt{B^2-(1-B)\dot{R}^2}}\right].
  \label{Pi_R}
\ee
Therefore from the Lagrangian, Eq.(\ref{Leff_t}), and the generalized momentum, Eq.(\ref{Pi_R}), the  Hamiltonian can then be written as
\be
  H=4\pi\sigma B^{3/2}R^2\left[\frac{1}{\sqrt{B^2-\dot{R}^2}}-\frac{2\pi\sigma GR(1-B)}{\sqrt{B^2-(1-B)\dot{R}^2}}\right].
  \label{Ham_Rdot}
\ee

For later convenience, we wish to find the Hamiltonian as a function of $(R,\Pi_R)$. To do so, we need to eliminate $\dot{R}$ in favor of $\Pi_R$ using Eq.(\ref{Pi_R}). This can be done, in principle, but is very messy (the solution will involve solutions of a quartic polynomial). However, we will be interested in what is happening as the shell approaches the horizon, i.e. when $R$ is close to $R_s$ and hence $B\rightarrow0$, since this is the most interesting region of study. In this limit one can see that the denominators of the two terms in Eq.(\ref{Pi_R}) are equal. So we can rewrite Eq.(\ref{Pi_R}) as
\be
  \Pi_R\approx\frac{4\pi\mu R^2\dot{R}}{\sqrt{B}\sqrt{B^2-\dot{R}^2}}
  \label{P_R Hor}
\ee
where
\be
  \mu\equiv\sigma(1-2\pi\sigma GR_s).
\ee
In the region $R\sim R_s$ the Hamiltonian, Eq.(\ref{Ham_Rdot}), is then approximately given by
\bea
  H&\approx&\frac{4\pi\mu B^{3/2}R^2}{\sqrt{B^2-\dot{R}^2}}\label{Ham Hor}\\
     &=&\sqrt{(B\Pi_R)^2+B(4\pi\mu R^2)^2}.
     \label{Ham Hor Pi_R}
\eea
Here we note that the Hamiltonian, written in the form of Eq.(\ref{Ham Hor Pi_R}), has the form of the energy of a relativistic particle with a position dependent mass.

Since the mass is a constant of motion, the Hamiltonian is a conserved quantity, so from Eq.(\ref{Ham Hor}) we can write
\be
  h=\frac{B^{3/2}R^2}{\sqrt{B^2-\dot{R}^2}}
  \label{h}
\ee
where $h\equiv H/4\pi\mu$ is a constant. 

Solving Eq.(\ref{h}) for $\dot{R}$ we obtain
\be
  \dot{R}=\pm B\sqrt{1-\frac{BR^4}{h^2}}.
  \label{dotRgen}
\ee
In the region $R\sim R_s$, this takes the form
\be
  \dot{R}\approx\pm B\left(1-\frac{1}{2}\frac{BR^4}{h^2}\right).
\ee

Since in the region $R\sim R_s$, $B\rightarrow0$, the dynamics for the collapsing spherically symmetric domain wall in this region can be obtained by solving the expression
\be
  \dot{R}=\pm B.
  \label{Rdot}
\ee
To leading order in $R-R_s$, the solution is
\be
  R(t)\approx R_s+(R_0-R_s)e^{\pm t/R_s}
  \label{R(t)}
\ee
where $R_0$ is the radius of the shell at $t=0$. Since we are interested in the collapsing shell, we take the negative sign in the exponential term, Eq.(\ref{R(t)}). 

Here we wish to make some comments on Eq.(\ref{R(t)}). By virtue of the negative sign in the exponential, this then implies that, from the classical point of view, the asymptotic observer never sees the formation of the horizon of the black hole, since Eq.(\ref{R(t)}) equals $R_s$ only as $t\rightarrow\infty$. This is in agreement with the fact that it takes an infinite amount of time for a photon to reach the horizon of a pre-existing black hole, as seen by an asymptotic observer (see for example Ref.\cite{Wheeler}). Therefore, Eq.(\ref{R(t)}) makes sense from this point of view. 

\begin{figure}[htbp]
\includegraphics{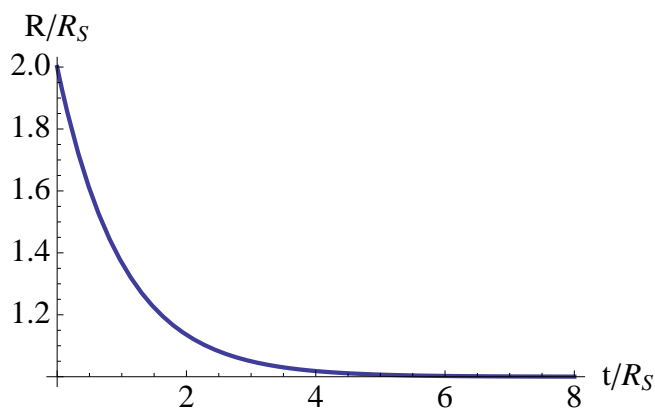}
\caption{Here we plot the the solution in Eq.(\ref{R(t)}).}
\label{Rt}
\end{figure}

In Figure \ref{Rt} we plot the position of the domain wall for the asymptotic observer. Here we see the asymptotic behavior of the time dependence of the position of the domain wall. As shown in Eq.(\ref{R(t)}), the domain wall asymptotes to the Schwarzschild radius, taking an infinite amount of time for the domain wall to reach the Schwarzschild radius.

\begin{figure}[htbp]
\includegraphics{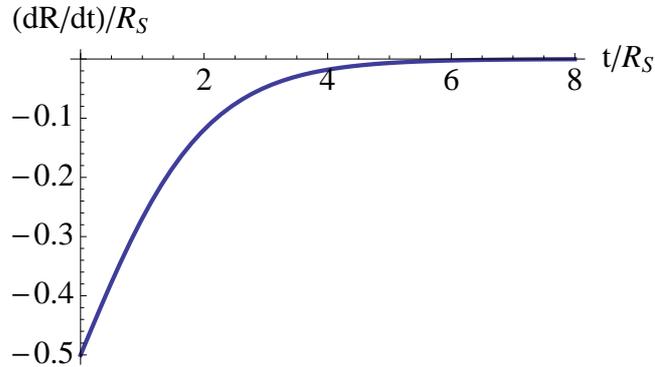}
\caption{Here we plot the the solution in Eq.(\ref{Rdot}).}
\label{R_dot}
\end{figure}

Figure \ref{R_dot} shows the corresponding velocity of the domain wall as seen by the asymptotic observer. Here we see that the velocity of the domain wall asymptotes to zero as the domain wall collapses toward the Schwarzschild radius, as given in Eq.(\ref{Rdot}).

\section{Infalling Observer}

Now we turn our attention to the infalling observer case, where the conserved mass is given by Eq.(\ref{Mass_tau}). Here we point out the misnomer in the name infalling. The infalling observe here is not to be confused with the traditional view point of an infalling observer, one who is traveling along a geodesic, or a freely falling observer. The observer in this case is infalling from the fact that the observer is attached to the domain wall and is infalling with the wall. Therefore, eventhough the observer is in a locally Minkowski reference frame, the overall reference frame is still Schwarzschild, since at any point in time the observer is in a Schwarzschild reference frame. Here we summarize the work originally done in Ref.\cite{GreenStoj}.

The effective action consistent with Eq.(\ref{Mass_tau}) is 
\be
  S_{eff}=-4\pi\sigma\int d\tau R^2\left[\sqrt{1+R_{\tau}^2}-R_{\tau}\sinh^{-1}(R_{\tau})-2\pi\sigma GR\right].
  \label{Seff_tau}
\ee
Therefore the effective Lagrangian expressed in terms of the infalling observer's time $\tau$ is given by
\be
  L_{eff}=-4\pi\sigma R^2\left[\sqrt{1+R_{\tau}^2}-R_{\tau}\sinh^{-1}(R_{\tau})-2\pi\sigma GR\right].
  \label{Leff_tau}
\ee

From Eq.(\ref{Leff_tau}) the generalized momentum $\tilde{\Pi}_R$ is derived to be
\be
  \tilde{\Pi}_R=4\pi\sigma R^2\sinh^{-1}(R_{\tau}).
  \label{Pi_tau}
\ee
From Eq.(\ref{Leff_tau}) and Eq.(\ref{Pi_tau}), the Hamiltonian in terms of $R_{\tau}$ is given by
\be
  H=4\pi\sigma R^2\left[\sqrt{1+R_{\tau}^2}-2\pi\sigma GR\right]
  \label{Ham_Rtau}
\ee
which is just Eq.(\ref{Mass_tau}) as expected. 

From Eq.(\ref{Ham_Rtau}) we can calculate $R_{\tau}$
\be
  R_{\tau}=\pm\sqrt{\left(\frac{\tilde{h}}{R^2}+2\pi\sigma GR\right)^2-1}
  \label{dRdtau}
\ee
where $\tilde{h}\equiv H/4\pi\sigma$. In general, Eq.(\ref{dRdtau}) cannot be solved analytically, at least not in very nice way. However, we can take some special cases to investigate the behavior of the solution to find the time dependence. 

As a first case we consider the zeroth order behavior near the horizon, i.e. $R\sim R_s$. In this region we note that we can write Eq.(\ref{dRdtau}) as
\be
  R_{\tau}=\pm\sqrt{\left(\frac{\tilde{h}}{R_s^2}+2\pi\sigma GR_s\right)^2-1}
  \label{dRdtau0}
\ee
hence $R_{\tau}$ is constant. Integrating Eq.(\ref{dRdtau0}) gives
\be
  R(\tau)=\tilde{R}_0-\tau\sqrt{\left(\frac{\tilde{h}}{R_s^2}+2\pi\sigma GR_s\right)^2-1}
  \label{R_tau_0}
\ee
where $\tilde{R}_0$ is the radius of the shell at $\tau=0$. 

As a second case we consider the case that $\tilde{h}/R^2>>2\pi\sigma GR>>1$. Therefore we can write Eq.(\ref{dRdtau}) as
\be
  R_{\tau}=-\frac{\tilde{h}}{R^2}.
  \label{dRdtauL}
\ee
Integrating Eq.(\ref{dRdtauL}) we then have the solution
\be
  R(\tau)=\left(\tilde{R}_0^3-3\tilde{h}\tau\right)^{1/3}.
  \label{RtauL}
\ee
Eq.(\ref{RtauL}) then gives that the time for an infalling observer to reach $R_s$ is
\be
  \tau=\frac{R_0^3-R_s^3}{3\tilde{h}}.
\ee

Here we make some comments on Eq.(\ref{R_tau_0}) and Eq.(\ref{RtauL}). These solutions imply that the infalling observer will reach $R_s$ in a finite amount of his/her proper time. This result  is expected from classical general relativity, since the observer is in a locally flat Minkowski reference frame. Therefore, there is no difficulty for the observer once he/she reaches the horizon, the horizon is just another locally flat point in space according to this observer. 

\begin{figure}[htbp]
\includegraphics{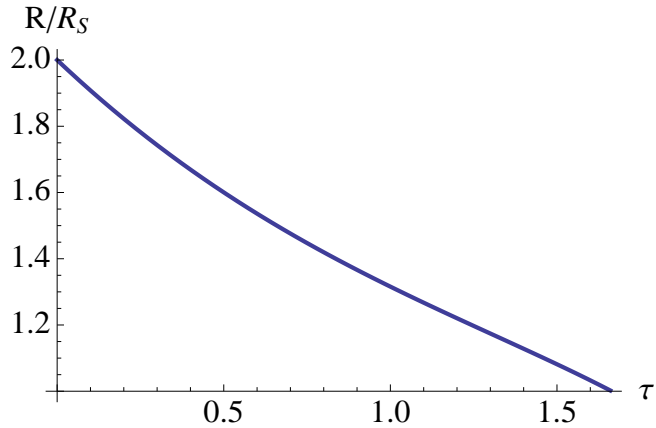}
\caption{Here we plot the numerical solution to Eq.(\ref{dRdtau}).}
\label{RtauFull}
\end{figure}

\begin{figure}[htbp]
\includegraphics{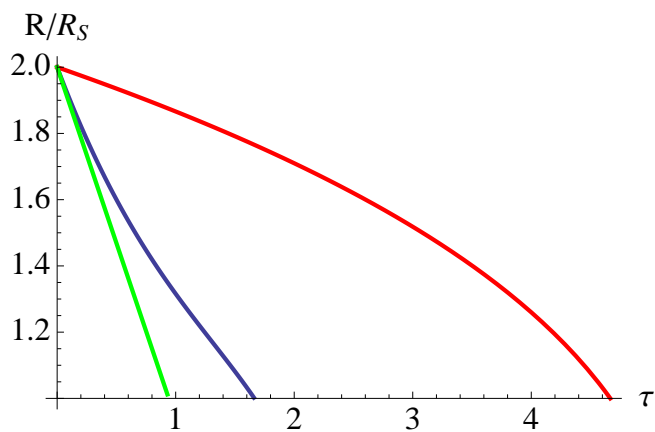}
\caption{Here we plot a comparison of the different approximations for the solution of Eq.(\ref{dRdtau}). The blue curve is the solution to Eq.(\ref{dRdtau}), the green curve is the solution to Eq.(\ref{dRdtau0}) and the red curve is the solution to Eq.(\ref{dRdtauL}).}
\label{RtauBoth}
\end{figure}

For consistency, in Figure \ref{RtauFull} we plot the numerical solution of Eq.(\ref{dRdtau}) for the parameters $\tilde{h}=1/2$ and $\sigma=0.1R_s$. Figure \ref{RtauFull} shows that the observer does in fact reach $R_s$ in a finite amount  of his proper time. Figure \ref{RtauBoth} compares the special cases discussed above with that of the numerical solution. Here the blue curve is the full solution of Eq.(\ref{dRdtau}), the green curve is the solution to Eq.(\ref{dRdtau0}) and the red curve is the solution to Eq.(\ref{dRdtauL}).

\begin{figure}[htbp]
\includegraphics{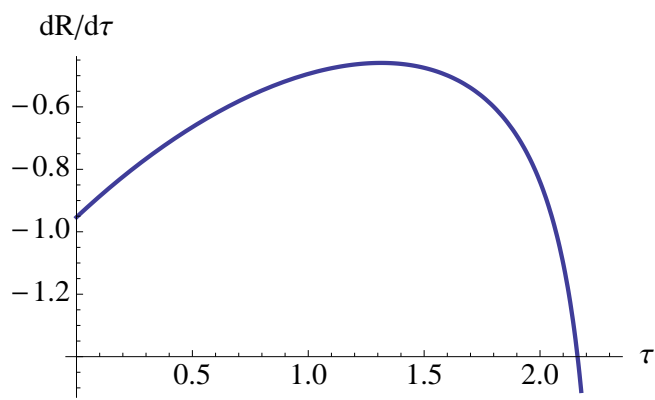}
\caption{Here we plot the corresponding numerical solution for Eq.(\ref{dRdtau}). Here we see that as $R\rightarrow R_s$ ($\tau=1.66$), the velocity of the domain wall is approximately constant. However, after the domain wall passes the Schwarzschild radius the velocity diverges as $R\rightarrow0$.}
\label{veloc_tau}
\end{figure}

In Figure \ref{veloc_tau} we plot the numerical solution for the velocity of the domain wall as seen by the infalling observer, from Eq.(\ref{dRdtau}). Since the domain wall crosses its own Schwarzschild radius at a time of $\tau=1.66$, Figure \ref{veloc_tau} shows that the velocity is infact approximately constant as $R\rightarrow R_s$. After the domain wall crosses the Schwarzschild radius, the velocity then increases and diverges as $R\rightarrow0$ (the classical singularity). 

\begin{figure}[htbp]
\includegraphics{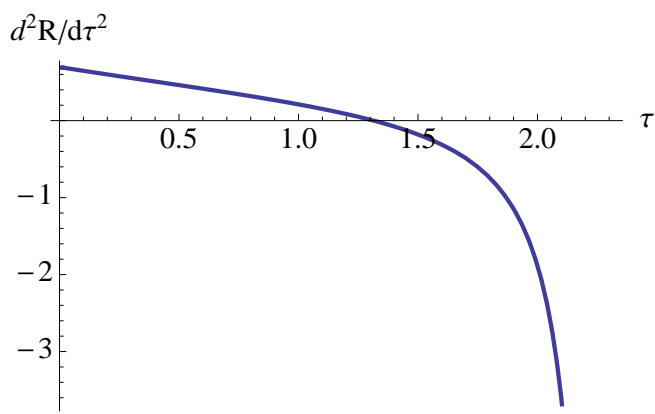}
\caption{Here we plot the corresponding numerical solution for the acceleration associated with Eq.(\ref{dRdtau}). Here we see that as $R\rightarrow R_s$ ($\tau=1.66$), the acceleration of the domain wall increases almost linearly. However, after the domain wall passes the Schwarzschild radius the acceleration diverges as $R\rightarrow0$.}
\label{accel_tau}
\end{figure}

In Figure \ref{accel_tau} we plot the numerical solution for the acceleration of the domain wall as seen by the infalling observer, from Eq.(\ref{dRdtau}). Since the domain wall crosses its own Schwarzschild radius at a time of $\tau=1.66$, Figure \ref{accel_tau} shows that the acceleration is increasing almost linearly as $R\rightarrow R_s$. After the domain wall crosses the Schwarzschild radius, the acceleration then increases and diverges as $R\rightarrow0$ (the classical singularity). Therefore, we can conclude that even though the observer is attached to the domain wall, he/she is not a truly free-falling observer (as stated at the beginning of this Section). The infalling Schwarzschild observer is an accelerated observer during the entire duration of the collapse.

\section{Comparing Asymptotic versus Infalling}

Here we wish to investigate the discrepancy of the observation between the infalling and asymptotic observers. 

\begin{figure}[htbp]
\includegraphics{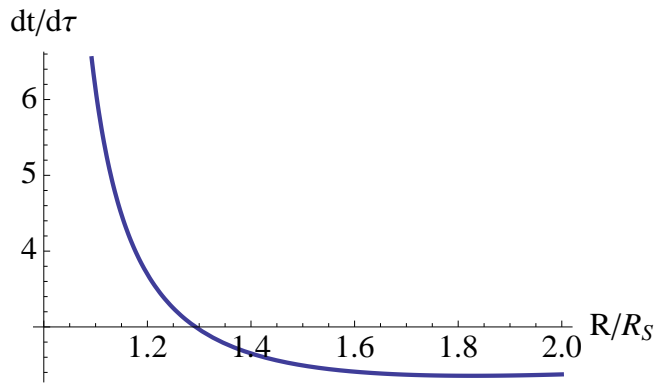}
\caption{Here we plot $dt/d\tau$ versus $R/R_s$. Here we can see that as $R\rightarrow R_s$, $dt/d\tau\rightarrow\infty$ as given in Eq.(\ref{grav_red}).}
\label{t_dot}
\end{figure}

\begin{figure}[htbp]
\includegraphics{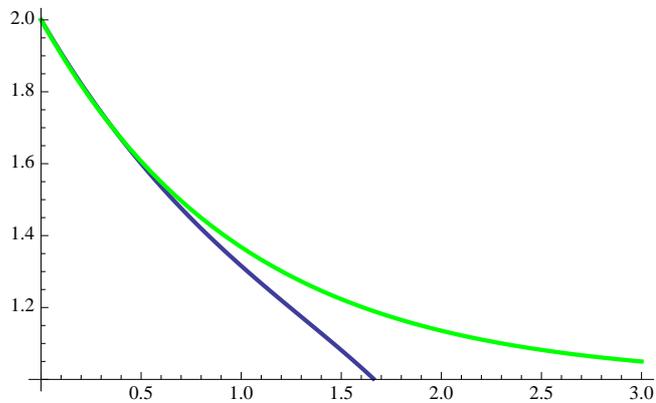}
\caption{Here we plot a comparison of the position of the domain wall relative to the Schwarzschild radius for both of the two different observers. Here the position of the domain wall as seen by the infalling observer is given in blue, while that of the asymptotic observer is given in green.}
\label{Rt_vs_Rtau}
\end{figure}

To understand this discrepancy we turn to Eq.(\ref{dtdtau}). As discussed in the previous section, in the limit $R\rightarrow R_s$, Eq.(\ref{dRdtau}) is approximately constant. This means that we can then write
\be
  \Delta t\approx \frac{1}{B}\sqrt{B+c}\Delta\tau.
\ee
Now, since $B\rightarrow0$ in this limit and the proper time taken to reach the Schwarzschild radius is finite, we can then see that 
\be
  \lim_{R\rightarrow R_s}\Delta t\rightarrow\infty.
  \label{grav_red}
\ee
This is a fairly crude approximating, thus in Figure \ref{t_dot} we plot $dt/dtau$ versus $R/R_s$. Figure \ref{t_dot} shows that as $R\rightarrow R_s$, $dt/d\tau$ does indeed diverge as given in Eq.(\ref{grav_red}). Therefore Eq.(\ref{grav_red}) can be thought of as the gravitational red-shift, which is the source of the discrepancy between the observations between the two observers. In Figure \ref{Rt_vs_Rtau} we show the the position of the domain wall as a function of time for both the asymptotic and infalling observers. Figure \ref{Rt_vs_Rtau} shows that initially the two observers are in agreement on where the domain wall is compared to the Schwarzschild radius. However, as the time increases (both asymptotic, $t$, and infalling, $\tau$) the discrepancy becomes more apparent. According to the asymptotic observer, the domain wall asymptotes to the Schwarzschild radius, while according to the infalling observer this happens in a finite amount of time (as stated earlier).

\section{Discussion}

In this section we investigated the classical equations of motion for the collapsing spherically symmetric domain wall. As discussed in the first section, the asymptotic observer sees the domain wall collapse to the Schwarzschild radius $R_s$ only as $t\rightarrow\infty$. This is not an unreasonable result since in classical General Relativity an asymptotic observer never sees a photon cross the Schwarzschild radius since it take an infinite amount of time $t$ to reach the horizon. Therefore one can easily believe the result here, since the time taken is due to the gravitational redshift of the photon. As the shell approaches $R_s$ the redshift will increase until it becomes infinite by the time it reaches the horizon.

In the second section, the infalling observer will see the shell collapse to the horizon in a finite amount of time. As discussed earlier, this is because the infalling observer is always in a locally flat Minkowski frame. Thus, the horizon is not a significant point for the observer, therefore there is no problem for him/her to pass right through and not even know it. This result is again expected from classical General Relativity. 

%% file: quantum.tex
\chapter{Quantum Treatment}
\label{ch:quantum}

In this chapter we wish to study the quantum equations of motion using the Functional Schr\"odinger equation Eq.(\ref{FSE}), again for the two different foliations of space-time. The idea here is that the quantum mechanical effects will change some of the difficulties that arise from the classical solutions. The ultimate goal is that examining these quantum mechanical effects will give us insight into the quantum mechanical nature of gravitational collapse and will possibly help guide us to be able to construct the appropriate theory of quantum gravity.

Some of the difficulties that arise from the classical solutions of gravitation collapse are discussed below. This is not meant to be an exhaustive list, but give the reader of an idea of the topics that we wish to address in this text. 

First we will discuss one difficulty faced by the asymptotic observer, the presence of the horizon. Why is the event horizon a difficult place for the asymptotic observer, while there is it is no problem for the infalling observer, since the horizon is nothing more than a coordinate singularity? Under the classical notion, as discussed in Chapter \ref{ch:Classical}, the presence of the coordinate singular creates an apparent gravitational time dilation, which makes the collapsing domain wall appear to stop. This effect is due to the divergence of the coordinate singularity when the observer sees the collapsing domain wall reach the Schwarzschild radius. There has been much discussion in this matter about how these quantum mechanical corrections can eliminate this effect. One such idea that has gained much attention over the past few years, we will discuss this process here. The basic idea here is that the upon quantizing the shell, the shell will now have quantum fluctuations in the position of the horizon. These fluctuations will then imply that the position of the horizon is no longer fixed, but will now be given by $R_s+\delta R_s$, where $\delta R_s$ represents the small fluctuations in the position of the horizon. These effects can then make the time as measured by the asymptotic observer finite (see for example Sec. 10.1.5 of Ref.\cite{FrolovNovikov})
\be
  \Delta t=\int_{R_s+\delta R_s}^{R_0}\frac{dr}{1-R_s/r}\sim R_s\ln\left(\frac{R_0-R_s}{\delta R_s}\right).
\ee
If this were correct, we would be able to observe black hole formation (due to the collapse) and other effects in finite time. Note however, that the fluctuations can go either way. In the case of $R_s-\delta R_s$, the result becomes infinite agin. 

Now we will discuss difficulties facing the infalling observer. For this observe there are really two important regimes: The regime in the region $R\sim R_s$ and the region $R\sim0$. For the region $R\sim R_s$, the concern is the exact opposite of that of the asymptotic observer. Will the quantization of the shell contradict the classical observation that the infalling observer sees the shell collapse to $R_s$ in a finite amount of time? For the region $R\sim0$, classically this represents the point of the classical singularity. Penrose and Hawking showed in Ref.\cite{PenroseHawking} that singularities are endemic in classical General Relativity. The question then arises whether these singularities are an intrinsic property of space-time or simply reflect our lack of the ultimate non-singular theory. The general belief is that quantization will rid gravitation of singularities. This is analogous to another theory where quantization got rid of the singularity, Electromagnetism. In atomic physics the singularity of the Coulomb potential, which has an identical $1/r$ behavior, was eliminated via quantization (see for example \cite{BogojevicStojkovic,Trodden,Borstnik,Shankaranarayanan}).

In this chapter we investigate these ideas. In later chapters we will investigate quantum mechanical corrects to addition aspects of gravitational collapse, those being thermodynamic.

\section{Asymptotic Observer}

First we will consider quantization of the shell from the view point of the asymptotic observer. This will be done by using Eq.(\ref{FSE}). For this section we outline the work done in Ref.\cite{Stojkovic}.

To utilize Eq.(\ref{FSE}) we need to use the Hamiltonian in Eq.(\ref{Ham Hor Pi_R}). However, we notice that Eq.(\ref{Ham Hor Pi_R}) has a squareroot in it. Therefore, we will consider the Hamiltonian squared
\be
  H^2=B\Pi_RB\Pi_R+B(4\pi\mu R^2)^2.
  \label{H2}
\ee
Before we proceed, we discuss the choice of ordering in the first term on the right hand side of Eq.(\ref{H2}). Since we are considering quantum mechanics, the distance $R$ and the conjugate momentum are now promoted to operators, which obey the standard commutation relations. Thus in general we would need to add terms to the squared Hamiltonian in Eq.(\ref{H2}) that depend on the commutator $[B,\Pi_R]$. However, in region of interest we find that the commutator is given by
\bd
  \left[B,\Pi_R\right]\sim\frac{1}{R_s}.
\ed
Estimating $H$ by the mass $M$ of the domain wall, as for the discussion in Chapter \ref{ch:model}, the terms due to the operater order ambiguity will be negligible provided
\bd
  M>>\frac{1}{R_s}\sim\frac{m_P^2}{M}
\ed
where $m_P$ is the Planck mass. Therefore we can ignore the ordering ambiguity and chose the ordering given in Eq.(\ref{H2}), provided that this limit is satisfied.

Now we apply the standard quantization procedure,
\bd
  \left[R,\Pi_R\right]=i.
\ed
We substitute
\be
  \Pi_R=-i\frac{\partial}{\partial R}
  \label{QPR}
\ee
into the squared Schr\"odinger equation
\be
  H^2\Psi=-\frac{\partial^2\Psi}{\partial t^2}.
  \label{Schrod2}
\ee
Inserting Eq.(\ref{H2}) into Eq.(\ref{Schrod2}) we then obtain
\be
  -B\frac{\partial}{\partial R}\left(B\frac{\partial\Psi}{\partial R}\right)+B(4\pi\mu R^2)^2\Psi=-\frac{\partial^2\Psi}{\partial t^2}.
  \label{Schrod R}
\ee

To find the wavefunction for the collapsing domain wall we need to solve Eq.(\ref{Schrod R}). To solve Eq.(\ref{Schrod R}) in terms of $R$ can be a formidable exercise in mathematics. However, we can simplify the matter by defining the tortoise coordinate
\be
  u=R+R_s\ln\left|\frac{R}{R_s}-1\right|.
  \label{u}
\ee
We can then see that Eq.(\ref{u}) then gives
\be
  B\Pi_R=-i\frac{\partial}{\partial u}
  \label{BPi_R}
\ee
where we used Eq.(\ref{QPR}). Using Eq.(\ref{BPi_R}) we can then rewrite Eq.(\ref{Schrod R}) as
\be
  \frac{\partial^2\Psi}{\partial t^2}-\frac{\partial^2\Psi}{\partial u^2}+B(4\pi\mu R^2)^2\Psi=0.
  \label{Massive Wave}
\ee
We can now identify Eq.(\ref{Massive Wave}) as just the massive wave equation in a Minkowski background with a mass term that depends on the position of the domain wall. We need to now solve Eq.(\ref{Massive Wave}) for $u$. To do so, we must first write the mass term in terms of $u$, rather than its present state of $R$. However, some care is needed since at $R=R_s$, $u$ is divergent, so we must take the appropriate branch. From Eq.(\ref{u}) we have that for the region $R\in(R_s,\infty)$ maps onto $u\in(-\infty,\infty)$ and $R\in(0,R_s)$ maps onto $u\in(0,-\infty)$. 

We are interested in the situation of a collapsing shell, hence the region $R\in(R_s,\infty)$. Thus we are interested in $u$ in the region $u\in(-\infty,\infty)$. We can solve Eq.(\ref{Massive Wave}) for the entire region, however, we are mostly concerned with the effect when $R\sim R_s$. In the region $R\sim R_s$, the logarithm in Eq.(\ref{u}) dominates, so we can write
\be
  R\sim R_s+R_se^{u/R_s}.
  \label{R(u)}
\ee
We look for wave-packet solutions propagating toward $R_s$, or in terms of $u$, $u\rightarrow-\infty$. Thus from Eq.(\ref{R(u)}) we have
\be
  B\sim e^{u/R_s}\rightarrow0.
\ee
This means that the last term in Eq.(\ref{Massive Wave}), the mass term, can be ignored in this region. 

In the region $R\sim R_s$, the dynamics of the wave-packet is simply given by the free wave equation, where any function of light-cone coordinates $(u\pm t)$ is a solution. To make this explicit, we consider a Gaussian wave-packet propagating toward the Schwarzschild radius
\be
  \Psi=\frac{1}{\sqrt{2\pi}s}e^{-(u+t)^2/2s^2}
  \label{wave pack t}
\ee
where $s$ is some chosen width of the wave packet in the $u$ coordinate. The width of the wave-packet remains fixed in the $u$-coordinate while it shrinks in the $R$ coordinate via the relation $dR=Bdu$, as follows from Eq.(\ref{u}). 

Let us consider some properties of Eq.(\ref{wave pack t}). First, we see that the wave-packet travels at the speed of light in the $u$ coordinate. This is expected since the Schr\"odinger equation takes the form of a massless wave equation in Minkowski space. Further, in the $u$ coordinate, the wave packet must travel out to $u=-\infty$ to get to the horizon, $R=R_s$. Thus we can conclude that the quantum domain wall does not collapse to $R_s$ in a finite amount of asymptotic time. 

Therefore one can conclude that the quantum mechanical effect, i.e. the quantization of the domain wall, does not smear out the presence of the coordinate singularity at the Schwarzschild radius. The asymptotic will not see the formation of the horizon in a finite amount of his/her time. Hence, the quantum solution does not alter the classical result found in Chapter \ref{ch:Classical}.

\section{Infalling observer}

Now we consider quantization of the domain wall from the view point of the infalling observer. This will be done using Eq.(\ref{Mass_tau}) as the Hamiltonian of the system. As discussed earlier, here we wish to solve the Functional Schr\"odinger equation in two different regions of interest. The first region is that near the Schwarzschild radius, $R_s$. The second is the region near the classical singularity, in the region $R\sim0$. For this section we summarize the work originally done in Ref.\cite{GreenStoj}.

\subsection{Near Horizon}

The exact Hamiltonian in terms of $R_{\tau}$ is again given by Eq.(\ref{Mass_tau}). From Eq.(\ref{Mass_tau}) we again see the presence of a square-root. However, in this case it is not as easy to remedy this as it was in the case of the asymptotic observer, since upon squaring the Hamiltonian will not get rid of the square root. To simplify the analysis we will require that $R_{\tau}$ is small. This is indeed a restriction to the special motion of the wall, since in general $R_{\tau}$ can be large near $R_s$ if the shell is falling from a very large distance. However, one may always choose initial conditions in such a way that the initial position of the shell $R(\tau=0)$ is very close to $R_s$. 

In the limit of small $R_{\tau}$, Eq.(\ref{Mass_tau}) simplifies to 
\be
  H=4\pi\sigma R_s^2\left[1+\frac{1}{2}R_{\tau}^2-2\pi\sigma GR_s\right].
  \label{Ham small Rtau}
\ee
Then again in the same limit, the conjugate momentum Eq.(\ref{Pi_tau}) simplifies to 
\be
  \tilde{\Pi}_R=4\pi\sigma R_s^2R_{\tau}.
  \label{Pi_tau small Rtau}
\ee
Ignoring the constant terms from the Hamiltonian Eq.(\ref{Ham small Rtau}) and using Eq.(\ref{Pi_tau small Rtau}) we can write the Hamiltonian as
\be
  H=\frac{\tilde{\Pi}_R}{8\pi\sigma R_s^2}.
\ee
Using the standard quantization procedure, we substitute
\be
  \tilde{\Pi}_{R}=-i\frac{\partial}{\partial R}
\ee
into the Schr\"odinger equation Eq.(\ref{FSE}), which yields
\be
  -\frac{1}{8\pi\sigma R_s^2}\frac{\partial^2\Psi}{\partial R^2}=i\frac{\partial\Psi}{\partial\tau}.
  \label{Rs_tau SE}
\ee

Investigating Eq.(\ref{Rs_tau SE}), we see that Eq.(\ref{Rs_tau SE}) is just the Schr\"odinger equation for a freely propagating ``particle" of mass $4\pi\sigma R_s^2$, as one can expect from this approximation. Since $R_s$ is only a finite distance away for an infalling observer we conclude that the wavefunction will collapse at $R_s$ in a finite amount of proper time. 

For the question on if the quantization of the domain wall will cause problems for the infalling observer, we can conclude that quantum effects do not alter the classical result. Hence a collapsing shell crosses its own Schwarzschild radius in a finite proper time. 

\subsection{Near the Origin}

Now we wish to investigate the quantization of the domain wall in the region of the classical singularity $R\sim0$.

The exact Hamiltonian in terms of $R_{\tau}$ is again given by Eq.(\ref{Mass_tau}), where $R_{\tau}$ is given by Eq.(\ref{dRdtau}). In the region near the classical singularity, i.e. in the limit $R\rightarrow0$, the classical expression for $R_{\tau}$ (keeping only the leading order term) becomes
\be
  R_{\tau}\approx-\frac{\tilde{h}}{R^2}
  \label{RtauN0}
\ee
where $\tilde{h}$ is defined in Chapter \ref{ch:Classical}. Eq.(\ref{RtauN0}) clearly shows that in this region the classical expression for $R_{\tau}$ diverges. Up to the leading term near the origin, Eq.(\ref{RtauN0}) implies that the Hamiltonian is
\be
  H=4\pi\sigma R^2R_{\tau}.
  \label{Hnear0}
\ee
Substituting the asymptotic behavior, Eq.(\ref{RtauN0}), in the expression for the generalized momentum Eq.(\ref{Pi_tau}) we have
\be
  \tilde{\Pi}_R=4\pi\sigma R^2\sinh^{-1}(R_{\tau}). 
  \label{Pi_tauN0}
\ee

From Eq.(\ref{Pi_tauN0}) we see that
\be
  \lim_{R\rightarrow0}\tilde{\Pi}_R=0.
\ee
This then gives
\be
  \lim_{R\rightarrow0}\frac{\tilde{\Pi}_R}{4\pi\sigma R^2}=-\infty.
\ee
This implies that $R_{\tau}$, which is defined as
\be
  R_{\tau}=\sinh\left(\frac{\tilde{\Pi}_R}{4\pi\sigma R^2}\right)
\ee
by virtue of Eq.(\ref{Pi_tau}), near the horizon becomes
\be
  R_{\tau}=\frac{1}{2}\exp\left(-\frac{\tilde{\Pi}_R}{4\pi\sigma R^2}\right).
  \label{Rtaue0}
\ee
Therefore substituting Eq.(\ref{Rtaue0}) into Eq.(\ref{Hnear0}) we have
\be
  2\pi\sigma R^2\exp\left(\frac{i}{4\pi\sigma R^2}\frac{\partial}{\partial R}\right)\Psi(R,\tau)=i\frac{\partial\Psi(R,\tau)}{\partial\tau}.
  \label{SchrodN0}
\ee

Let us consider some properties of Eq.(\ref{SchrodN0}). The differential operator in the exponent gives some unusual properties to the equation. First we note the presence of the $R^{-2}$ term. This implies that if we expand the exponent we can not stop the series after a finite number of terms, but instead need to include all orders of the expansion, we will make this explicit below. Thus, we need to include an infinite number of derivatives of the wavefunction $\Psi$ into the differential equation. An infinite number of derivatives of a certain function uniquely specifies the whole function. Thus, the value of (the derivative of) the function on the right hand side of Eq.(\ref{SchrodN0}) at one point depends on the values of the function at different points on the left hand side of the same equation. This is in strong contrast with ordinary local differential equations where the value of the function and certain finite number of its derivatives are related at the same point of space. This indicates that Eq.(\ref{SchrodN0}) describes physics which is not strictly local. 

Here we will make the non-locality of Eq.(\ref{SchrodN0}) explicit. To illustrate we will use the expression involving the $\sinh$ term, this is not necessary, however, it will make the explanation more clear. Note that we can rewrite $\sinh$ as
\bd
  \sinh(x)=\frac{e^x-e^{-x}}{2}.
\ed
By making a change of variable we can make this more explicit. If we introduce the new variable $v=R^3$, Eq.(\ref{SchrodN0}) becomes
\be
  \pi\sigma v^{2/3}\left[\exp\left(\frac{3i}{4\pi\sigma}\right)-\exp\left(-\frac{3i}{4\pi\sigma}\right)\right]\frac{\partial}{\partial v}\Psi(v,\tau)=i\frac{\partial\Psi(v,\tau)}{\partial\tau}.
  \label{Schrodv}
\ee
We can then see that the the differential operator in the exponents in Eq.(\ref{Schrodv}) are just a translation operator, which shifts the argument of the wavefunction by a non-infinitesimal amount of $3i/4\pi\sigma$. Since the wavefunction is complex in general, a shift by a complex value is not a problem. Therefore Eq.(\ref{Schrodv}) can be written as
\be
  \pi\sigma v^{2/3}\left[\Psi\left(v+\frac{3i}{4\pi\sigma},\tau\right)-\Psi\left(v-\frac{3i}{4\pi\sigma},\tau\right)\right]=i\frac{\partial\Psi(v,\tau)}{\partial\tau}.
  \label{ShiftSchrod}
\ee
Here we make an additional change in variable and define $v'=v-3i/4\pi\sigma$. Then we can rewrite Eq.(\ref{ShiftSchrod}) as
\be
  \pi\sigma\left(v'+\frac{3i}{4\pi\sigma}\right)^{2/3}\left[\Psi\left(v'+\frac{6i}{4\pi\sigma},\tau\right)-\Psi\left(v',\tau\right)\right]=i\frac{\partial\Psi(v,\tau)}{\partial\tau}.
\ee
To interpret this we rely on usual differential calculus. From calculus we have
\be
  f(x+\Delta x)-f(x)\approx f(x)+ \sum_{n=0}^{\infty}(\Delta x)^nf^{(n)}-f(x)=\sum_{n=0}^{\infty}(\Delta x)^nf^{(n)}\approx \Delta x\frac{df}{dx}.
  \label{deriv}
\ee
Here the last step assumes that $\Delta x$ is small so we are justified at keeping only the first term in the expansion. Now, performing the same procedure as in Eq.(\ref{deriv}) we can write
\bea
  \psi\left(v'+\Delta v',\tau\right)-\psi\left(v',\tau\right)&\approx&\psi(v',\tau)+\sum_{n=0}^{\infty}(\Delta v')^n\psi^{(n)}-\psi(v',\tau)\nonumber\\
      &=&\sum_{n=0}^{\infty}(\Delta v')^n\psi^{(n)}
\eea
where 
\be
  \Delta v'=-\frac{6i}{4\pi\sigma}.
  \label{Deltav}
\ee
However, here we cannot truncate the series after a finite number of derivatives since by virtue of Eq.(\ref{Deltav}), $\Delta v'$ is not a small shift, provided that $\sigma$ does not go to infinity. Therefore we must keep all orders of the derivative, since as in Eq.(\ref{deriv}) each derivative has higher powers of $\Delta v'$. 

An interesting thing to note here is that as $\sigma\rightarrow0$, the non-local effect becomes stronger and more predominant. A possible understanding of this is as follows. Outside of the domain wall, there exists a certain Hilbert space, while on the inside there exists a second Hilbert space. The transition from the first Hilbert space to the second is not necessarily a smooth transition. When the domain wall collapses to the singularity, the effect of each of these Hilbert spaces is now taken into account. When dealing with a massive domain wall, the warping of space-time is greater, however it exists over a larger distance. This makes the transition more smooth from point to point. However, for a small domain wall, the warping is less noticeable at larger distances. Therefore, the transition is more violent the closer one gets to the classical singularity in this case, since in both cases the warping diverges. 

What about the value of the wavefunction at the classical singularity? Eq.(\ref{ShiftSchrod}) shows that the wavefunction near the origin $\Psi(R\rightarrow0,\tau)$ is in fact related to the wavefunction at some distant point $\Psi(R\rightarrow(\frac{6i}{4\pi\sigma})^{1/3},\tau)$.  This also implies that the wavefunction describing the collapsing domain wall is non-singular at the origin. Indeed, in the limit $R\rightarrow0$, this equation becomes
\be
  \frac{\partial\Psi(R\rightarrow0,\tau)}{\partial\tau}=0
  \label{R-ind Schrod}
\ee
where we used the fact that the wavefunction at some finite $R$, i.e. $\Psi(R\rightarrow(\frac{6i}{4\pi\sigma})^{1/3},\tau)$, is finite. From Eq.(\ref{R-ind Schrod}) it then follows that $\Psi(R\rightarrow0)=const$. This gives strong indication that quantization of the domain wall may indeed rid gravity of the classical singularity.

\section{Discussion}

In this chapter we quantized the collapsing domain wall and investigated the quantum corrections. In the first section we did this with respect to the asymptotic observe to see if these fluctuations changed the classical observation that the domain wall takes an infinite amount of time to reach $R_s$. Upon quantizing the domain wall, we found that in this view point the classical scenario was not changed by the quantum fluctuations. 

In the second section we investigated the quantization of the domain wall from the view point of the infalling observer. Here we did this for two different points of interest, near the horizon $R\sim R_s$ and near the classical singularity $R\sim0$, respectively. In the region $R\sim R_s$, we found that upon quantizing the domain wall, the classical view point was again unchanged. The quantum fluctuations did not alter the fact that according to the infalling observer, the domain wall will collapse to $R_s$ in a finite amount of proper time. In the region $R\sim0$, we found some interesting properties of the wavefunction. First, we found that the physics of the wavefunction in this region are strongly non-local, meaning that the value of the wavefunction at $R\sim0$ depends on the value of the wavefunction some distance away from the classical singularity. This situation has been previously suggested in the context of the information loss paradox (see for example \cite{Lowe,Horowitz,Giddings}). As we pointed out in the section, what is interesting is that this non-local behavior becomes increasingly manifest in the limit that $\sigma\rightarrow0$, i.e. that the mass of the domain wall becomes very small. This may be a consequence of the non-separability of the Hilbert space between the outside of the domain wall and the singularity. In the massive domain wall scenario, the transition from the outside to the inside Hilbert space is smoother than in the case of the light domain wall. Secondly, we found that the wavefunction is in fact finite at the classical singularity, which implies that quantum fluctuations may rid gravity of the classical singularity.

%% file: Radiation.tex
\chapter{Radiation}
\label{ch:radiation}

In this chapter we wish to investigate one of the thermodynamic properties of gravitational collapse. Two of the most important thermodynamic properties of a black hole are the temperature (discussed in this chapter) and the entropy (discussed in Chapter \ref{ch:entropy}). As mentioned in Chapter \ref{ch:formal}, the benefit of the Functional Schr\"odinger equation is that this will allow us to evolve the system over time. This is in contrast with the usual method used to evaluate the thermodynamic properties of a black hole. 

The most widely used method of determining thermodynamic properties of a black hole is the so-called Bogolyubov method. The method here is as follows. One considers an initial asymptotically flat space-time, usually Minkowski, at the beginning of the gravitational collapse. The system is then allowed to evolve to a final asymptotically flat space-time, Schwarzschild in the context of a shell of matter only, with no knowledge of what happens in between. Then by matching the coefficients between these two space-times, the mismatch of these two vacua gives the number of produced particles. As mentioned, what happens in between the initial vacua and the final vacua is beyond the scope of the Bogolyubov method. 

Since the Functional Schr\"odinger equation allows one to find the time dependent wavefunction for the system, one can, in principle, ask the question of what happens during the evolution of the collapse. In this chapter we will investigate the time evolution of the radiation, in the form of the occupation developed in Chapter \ref{ch:number}, during the time of gravitational collapse. The occupation number will then allow us to fit the temperature of the radiation, and compare with that of the pre-formed black hole.

As discussed in Chapter \ref{ch:model}, we can consider the radiation given off during the collapse of the domain wall by considering a massless scalar field $\Phi$ that is coupled to the gravitational field. The action of the scalar field can then be written as in Eq.(\ref{gen_action})
\be
  S=\int d^4x\frac{1}{2}\sqrt{-g}g^{\mu\nu}\partial_{\mu}\Phi\partial_{\nu}\Phi.
  \label{rad_act}
\ee
We decompose the (spherically symmetric) scalar field into a complete set of real basis functions denoted by $\{f_k(r)\}$
\be
  \Phi=\sum_ka_k(t)f_k(r).
  \label{mode exp}
\ee
The exact form of the function $f_k(r)$ will not be important for us. We will, however, be interested in the wavefunction for the mode coefficients $\{a_k(t)\}$. Note, again here we use ``$t$" to be anytime coordinate of interest to us, not necessarily the asymptotic time.

From Eq.(\ref{Met_in}) and Eq.(\ref{Met_out}) we can see that the action for both the different foliations of space-time will consist of two parts, one from Eq.(\ref{Met_in}) and the second from Eq.(\ref{Met_out}). In both foliations, the asymptotic observer and the infalling observer, we will be interested in the region $R\sim R_s$, therefore we will explicitly write out Eq.(\ref{rad_act}) then take the limit to find the dominating contributions. 

From this action we can then find the Hamiltonian of the system using the usual methods. Substituting the Hamiltonian into Eq.(\ref{FSE}) we can find the time-dependent wavefunction of the system. As discussed in Chapter \ref{ch:number}, the time-dependent wavefunction allows us to find the occupation number $N$, which will allow us to fit the temperature of the radiation (which we will describe below).   

In 1975 Hawking showed that for a pre-existing static black hole, the black hole will radiate its mass away, see Ref.\cite{1975Hawking}. The radiation that is given off has a finite temperature, as viewed by an asymptotic observer, which is known as the Hawking temperature. Therefore, it will be instructive for us to compare our late time result with Hawking's original calculation. 

For the infalling observer, the calculation is also instructive to give us an idea of what the region is like for this observer. Some unanswered questions are: since the temperature as measured by the asymptotic observer is finite, what is the temperature at the horizon for the local observer? If the temperature is infinite at the horizon, as one would expect since the temperature for the asymptotic observer is finite, will the infalling observer burn up before he reaches the horizon? To answer this question, we will use two different foliations of space time, that of Schwarzschild and Eddington-Finkelstein, respectively.

\section{Asymptotic Observer}

Here we consider the radiation as measured by the asymptotic observer. For this section we summarize the work originally done in Ref.\cite{Stojkovic}.

As stated above, the action for the scalar field can be written in two parts
\be
  S=S_{in}+S_{out}
  \label{action_two}
\ee
where
\begin{align}
  S_{in}&=2\pi\int dt\int_0^{R(t)}drr^2\left[-\frac{(\partial_t\Phi)^2}{\dot{T}}+\dot{T}(\partial_r\Phi)^2\right]\label{S_t_in}\\
  S_{out}&=2\pi\int dt\int_{R(t)}^{\infty}drr^2\left[-\frac{(\partial_t\Phi)^2}{1-R_s/r}+\left(1-\frac{R_s}{r}\right)(\partial_r\Phi)^2\right]
  \label{S_t_out}
\end{align}
where $\dot{T}$ is given in Eq.(\ref{dTdt}), which with Eq.(\ref{dotRgen}), gives
\be
  \dot{T}=B\sqrt{1+(1-B)\frac{R^4}{h^2}}.
\ee
As mentioned above, we are interested in the $R\sim R_s$ behavior of the action. As $R\rightarrow R_s$, we see that $\dot{T}\sim B\rightarrow0$. Therefore the kinetic term in Eq.(\ref{S_t_in}) diverges as $(R-R_s)^{-1}$ in this limit, while the kinetic term in Eq.(\ref{S_t_out}) diverges logarithmically. Therefore the divergence of the kinetic term in Eq.(\ref{S_t_in}) dominates over that of the divergence of the kinetic term in Eq.(\ref{S_t_out}). The gradient term in Eq.(\ref{S_t_in}) vanishes in this limit, while the gradient term in Eq.(\ref{S_t_out}) becomes finite. Thus the gradient term in Eq.(\ref{S_t_out}) is dominant over that of the gradient term in Eq.(\ref{S_t_in}) in this limit. Hence the action can be written as
\be
  S\sim2\pi\int dt\left[-\frac{1}{B}\int_0^{R_s}drr^2(\partial_t\Phi)^2+\int_{R_s}^{\infty}drr^2\left(1-\frac{R_s}{r}\right)(\partial_r\Phi)^2\right]
  \label{Rad_act}
\ee
where we have changed the limits of integration to $R_s$ since this is the region of interest. This approximation is valid provided the contribution from $r\in(R_s,R(t))$ to the integrals remains subdominant, and also the time variation introduced by the true integration limit can be ignored. 

Now, using Eq.(\ref{mode exp}) we write the action in Eq.(\ref{Rad_act}) as
\be
  S=\int dt\left[-\frac{1}{2B}\dot{a}_k(t){\bf M}_{kk'}\dot{a}_{k'}(t)+\frac{1}{2}a_k(t){\bf N}_{kk'}a_{k'}(t)\right]
\ee
where ${\bf M}$ and ${\bf N}$ are matrices that are independent of $R(t)$ and are given by
\begin{align}
  {\bf M}_{kk'}&=4\pi\int_0^{R_s}drr^2f_k(r)f_{k'}(r)\label{M}\\
  {\bf N}_{kk'}&=4\pi\int_{R_s}^{\infty}drr^2\left(1-\frac{R_s}{r}\right)f'_k(r)f'_{k'}(r).
  \label{N}
\end{align}

Using the standard quantization procedure and Eq.(\ref{FSE}), the wavefunction $\psi(a_k,t)$ satisfies
\be
  \left[\left(1-\frac{R_s}{R}\right)\frac{1}{2}\Pi_k({\bf M}^{-1})_{kk'}\Pi_{k'}+\frac{1}{2}a_k(t){\bf N}_{kk'}a_{k'}(t)\right]\psi=i\frac{\partial\psi}{\partial t}
\ee
where
\be
  \Pi_k=-i\frac{\partial}{\partial a_k(t)}
\ee
is the momentum operator conjugate to $a_k(t)$.

The problem of radiation from the collapsing domain wall is equivalent to the problem of an infinite set of uncoupled harmonic oscillators whose masses go to infinity with time. We can see from Eq.(\ref{M}) and Eq.(\ref{N}) that the matrices ${\bf M}$ and ${\bf N}$ are hermitian. Therefore, it is possible to do a principal axis transformation to simultaneously diagonalize ${\bf M}$ and ${\bf N}$ (see Sec. 6-2 of Ref. \cite{Goldstein}) for example). Then for a single eigenmode, the Schr\"odinger equation takes the form
\be
  \left[-\left(1-\frac{R_s}{R}\right)\frac{1}{2m}\frac{\partial^2}{\partial b^2}+\frac{1}{2}Kb^2\right]\psi(b,t)=i\frac{\partial\psi(b,t)}{\partial t}
  \label{Pre-Rad_Schrod_t}
\ee
where $m$ and $K$ denote eigenvalues of ${\bf M}$ and ${\bf N}$, and $b$ is the eigenmode. 

Dividing Eq.(\ref{Pre-Rad_Schrod_t}) through by $B$, we can write in the standard form
\be
  \left[-\frac{1}{2m}\frac{\partial^2}{\partial b^2}+\frac{m}{2}\omega(\eta)^2b^2\right]\psi(b,\eta)=i\frac{\partial\psi(b,\eta)}{\partial\eta}
  \label{Rad_Schrod_t}
\ee
where
\be
  \eta=\int_0^tdt\left(1-\frac{R_s}{R}\right)
  \label{eta_t}
\ee
and
\be
  \omega^2(\eta)=\frac{K}{m}\frac{1}{1-R_s/R}\equiv\frac{\omega_0^2}{1-R_s/R}
  \label{omega_t}
\ee
where we have chosen to set $\eta(t=0)=0$. 

From Eq.(\ref{R(t)}) we can see that the classical late time behavior of the shell is given by $1-R_s/R\sim\exp(-t/R_s)$. For early times, the behavior depends on how the spherical domain wall was created and we are free to choose a behavior for $R(t)$ that is convenient for calculations and interpretation. The most convenient case to use is a static beginning. This can be obtained if we artificially take the collapse to stop at some time, $t_f$. Eventually we can then take $t_f\rightarrow\infty$, as given by Chapter \ref{ch:Classical}. We will then chose $R$ to be
\be
  1-\frac{R_s}{R}=\begin{cases} 1 & t\in(-\infty,0)\\
                   e^{-t/R_s}, &t\in(0,t_f)\\
                   e^{-t_f/R_s}, &t\in(t_f,\infty). \end{cases}
          \label{background_t}
\ee

With the choice of initial static space-time of the domain wall, the initial vacuum state for the modes is the simple harmonic oscillator ground state,
\be
  \psi(b,\eta=0)=\left(\frac{m\omega_0}{\pi}\right)^{1/4}e^{-m\omega_0b^2/2}.
  \label{HO_basis}
\ee
The exact solution for late times is given by Eq.(\ref{PedWave}) with initial conditions given by Eq.(\ref{IC}).

As discussed in Chapter \ref{ch:number} an observer with a detector will interpret the wavefunction of a given mode $b$ at late times in terms of simple harmonic oscillator states at the final frequency
\be
  \bar{\omega}=\omega_0e^{t_f/2R_s}
  \label{LateOmega_t}
\ee
where we have made use of Eq.(\ref{background_t}). 

The number of quanta in eigenmode $b$ can be evaluated from Eq.(\ref{OccNum}). By calculating $\dot{N}$ it can be checked that $N$ remains constant for $t<0$ and also $t>t_f$. Hence all the particle production occurs for $0< t<t_f$ and is a consequence of the gravitational collapse.

\begin{figure}[htbp]
\includegraphics{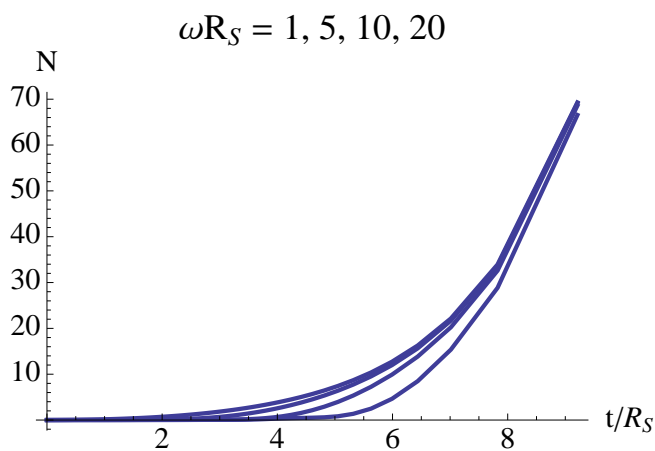}
\caption{$N$ versus $t/R_s$ for various fixed values of $\bar{\omega}R_s$. The curves are lower for higher $\bar{\omega}R_s$.}
\label{Nt_vs_t}
\end{figure}

Now we can take the limit $t_f\rightarrow\infty$. In this limit, $\rho$ remains finite but $\rho_{\eta}\rightarrow-\infty$ as $t>t_f\rightarrow\infty$, provided $\omega_0\not=0$ (see Appendix \ref{ch:rho_t} for details). However, we are interested in the behavior of $N$ for fixed frequency, $\bar{\omega}$. From Eq.(\ref{LateOmega_t}) in this limit implies $\omega_0\rightarrow0$. From the discussion in Appendix \ref{ch:rho_t}, we also know that $\rho\rightarrow\infty$ as $\omega_0\rightarrow0$. Hence we find
\be
  N(t,\bar{\omega})\sim\frac{\bar{\omega}\rho^2}{\sqrt{2}}\sim\frac{e^{t/(2R_s)}}{\sqrt{2}}, \hspace{2mm} t>t_f\rightarrow\infty.
  \label{LateOccNum_t}
\ee
Therefore the occupation number at any frequency diverges in the infinite time limit when backreaction is not taken into account. In Figure \ref{Nt_vs_t} we have plotted the occupation number $N$ versus $t/R_s$ for various values of $\bar{\omega}R_s$. Figure \ref{Nt_vs_t} confirms the late time behavior of the time dependence of the occupation number, Eq.(\ref{LateOccNum_t}).

\begin{figure}[htbp]
\includegraphics{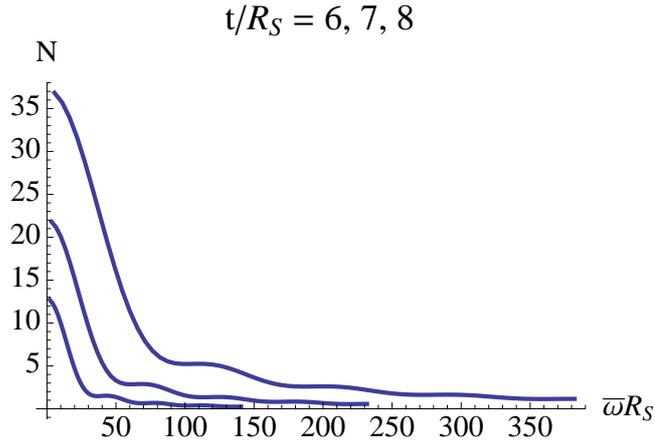}
\caption{$N$ versus $\bar{\omega}R_s$ for various fixed values of $t/R_s$. The occupation number at any frequency grows as $t/R_s$ increases.}
\label{Nt_vs_w}
\end{figure}

We have also numerically evaluated the spectrum of mode occupation numbers at any finite time and show the results in Figure \ref{Nt_vs_w} for several different values of $t/R_s$. Figure \ref{Nt_vs_w} shows that as the asymptotic observer's time increases, the occupation number of larger values of $\bar{\omega}R_s$ increases. This is consistent with Eq.(\ref{omega_t}), since as $t\rightarrow\infty$, $\bar{\omega}\rightarrow\infty$.

To find the temperature of the radiation, we compare the curve in Figure \ref{Nt_vs_w} with the occupation numbers for the Planck distribution
\be
  N_P(\omega)=\frac{1}{e^{\beta\omega}-1}
  \label{OccPlanck}
\ee
where $\beta$ is the inverse temperature. We can see that the spectrum of occupation numbers is non-thermal. As an example, there is no singularity in $N$ at $\omega=0$ at finite time. However, as $t\rightarrow\infty$, the peak at $\omega=0$ does diverge and the distribution becomes more and more thermal for these times. There are also oscillations in $N$.

\begin{figure}[htbp]
\includegraphics{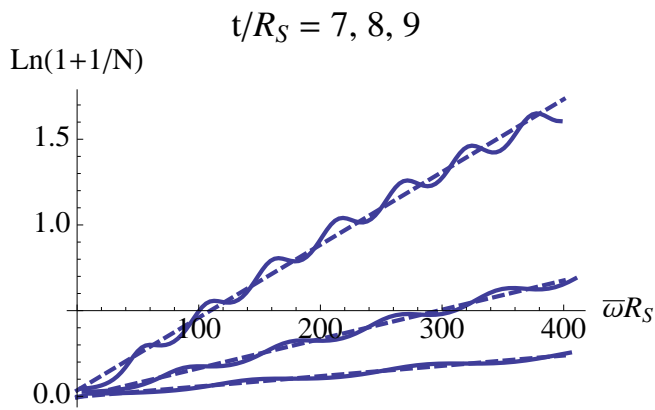}
\caption{$\ln(1+1/N)$ versus $\bar{\omega}R_s$ for various values of $t/R_s$.}
\label{LnNt_vs_w}
\end{figure}

We now wish to fit the temperature of the radiation. From Eq.(\ref{OccPlanck}), we can find the inverse temperature to be
\be
  \beta=\frac{\ln(1+1/N_P)}{\bar{\omega}R_s}=T^{-1}.
\ee 
In Figure \ref{LnNt_vs_w} we plot $\ln(1+1/N)$ versus $\bar{\omega}R_s$ for various values of $t/R_s$. Here we see that for smaller values of $t/R_s$ the spectrum for $\beta$ is non-thermal. For example, Figure \ref{LnNt_vs_w} shows a thermal-like distribution for only small values of $\bar{\omega}R_s$ for $t/R_s=2$, while the larger values or $\bar{\omega}R_s$ are not yet thermally induced. If one fitted the slope of $\beta$ for this particular time, the only relevant region is that between $0<\bar{\omega}R_s<200$. We can also see that the fluctuations of $\beta$ are large. However, as $t/R_s$ increases more and more values of $\bar{\omega}R_s$ are thermally induced, hence one can fit a larger region. The fluctuations for larger values of $t/R_s$ become much smaller, until they become almost completely non-existent. Another feature which occurs as $t/R_s$ increases is that the slope of $\beta$ goes to zero, which would imply that the temperature of the radiation in fact goes to infinity, not to a finite number as predicted by Hawking.

\begin{figure}[htbp]
\includegraphics{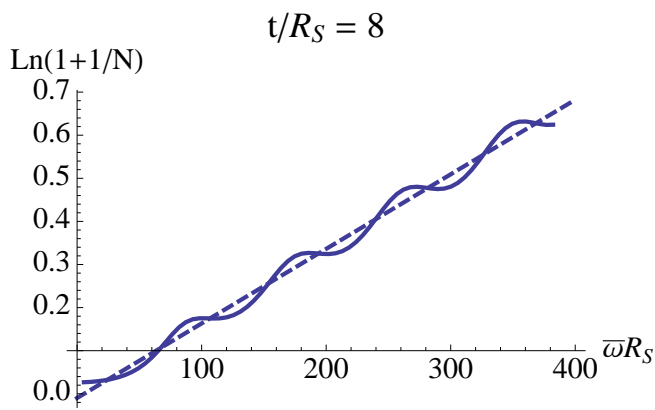}
\caption{$\ln(1+1/N)$ versus $\bar{\omega}R_s$ for $t/R_s=8$. The dashed line shows $\ln(1+1/N_P)$ versus $\bar{\omega}R_s$ where $N_P$ is a Planck distribution. The slope gives $\beta^{-1}$ and the temperature is given in Eq.(\ref{HawkTemp}).}
\label{LnNt_vs_w8}
\end{figure}

However, from Eq.(\ref{Rad_Schrod_t}) we see that the time derivative of the wavefunction on the right-hand side is with respect to $\eta$, not with respect to $t$, and $\omega$ is the mode frequency with respect to $\eta$ as well. Eq.(\ref{eta_t}) tells us that the frequency in $t$ is $(1-R_s/R)$ times the frequency in $\eta$, so at the final time $t_f$, this implies 
\be
  \omega^{(t)}=e^{-t_f/R_s}\bar{\omega}
\ee
where the superscript $(t)$ on $\omega$ refers to the fact that this frequency is with respect to time $t$. 
However, since we are interested in the temperature in time $t$, we must also rescale the temperature in the same manner as the frequency. So the temperature seen by the asymptotic observer is
\be
  T=e^{-t_f/R_s}\beta^{-1}(t_f).
\ee
Fitting a thermal spectrum to the collapsed spectrum of Figure \ref{Nt_vs_w}, as shown in Figure \ref{LnNt_vs_w8}, we obtain
\be
  T\approx\frac{0.19}{R_s}=2.4T_H
  \label{HawkTemp}
\ee
where $T_H=1/4\pi R_s$ is the Hawking temperature. Since there is ambiguity in fitting the non-thermal spectrum by a thermal distribution, we can only say that the constant temperature, $T$, and the Hawking temperature are of comparable magnitude. 

\section{Infalling Observer}

Here we consider the radiation as measured by the infalling observer. To do so we will consider two different foliations of space time. As seen in Chapter \ref{ch:Classical}, the acceleration of the Schwarzschild observer becomes divergent as the observer crosses the horizon. Therefore it is important to switch to another observer whose acceleration is no longer divergent upon crossing the horizon. For this observer, we will work in Eddington-Finkelstein coordinates. In this section we summarize the work originally done in Ref.\cite{Greenwood}.

First we will consider the Schwarzschild observer, after which we will consider the Eddington-Finkelstein observer.

\subsection{Schwarzschild Coordinates}

The action can again be written in two parts, see Eq.(\ref{action_two}), where
\begin{align}
  S_{in}=&2\pi\int d\tau\int_0^{R(\tau)}drr^2\left[-\frac{1}{\sqrt{1+R_{\tau}^2}}(\partial_{\tau}\Phi)^2+\sqrt{1+R_{\tau}^2}(\partial_r\Phi)^2\right]\\
  S_{out}=&2\pi\int d\tau\int_{R(\tau)}^{\infty}drr^2\left[-\frac{B}{\sqrt{B+R_{\tau}^2}}\frac{(\partial_{\tau}\Phi)^2}{1-R_s/r}+\frac{\sqrt{B+R_{\tau}^2}}{B}\left(1-\frac{R_s}{r}\right)(\partial_r\Phi)^2\right].
\end{align}
The most interesting things happen when the shell approaches the Schwarzschild radius. From Eq.(\ref{dRdtau}) we see that $R_{\tau}$ is constant in the limit when $R\rightarrow R_s$. Therefore the kinetic term for $S_{in}$ is roughly constant. The kinetic term in $S_{out}$ goes to zero as $R\rightarrow R_s$, so the $S_{in}$ kinetic term is dominant. Similarly the potential term in $S_{in}$ goes to a constant while the potential term in $S_{out}$ becomes very large, so the potential term in $S_{out}$ dominates. Therefore in the region $R\sim R_s$ we can write the action as
\be
  S\approx\int d\tau\left[-\int_0^{R_s}drr^2\frac{1}{\sqrt{1+R_{\tau}^2}}(\partial_{\tau}\Phi)^2+\int_{R_s}^{\infty}drr^2\frac{|R_{\tau}|}{B}\left(1-\frac{R_s}{r}\right)(\partial_r\Phi)^2\right]
  \label{rad_act_tau}
\ee
where we have changed the limits of integration from $R(\tau)$ to $R_s$ since this is the region of interest. 

Using the expansion in modes, Eq.(\ref{mode exp}), we can write the action as
\be
  S=\int d\tau\left[-\frac{1}{2}\frac{1}{\sqrt{1+R_{\tau}^2}}\dot{a}_k(\tau){\bf A}_{kk'}\dot{a}_{k'}(\tau)+\frac{|R_{\tau}|}{2B}a_k(\tau){\bf C}_{kk'}a_{k'}(\tau)\right]
\ee
where $\dot{a}=da/d\tau$, and ${\bf A}$ and ${\bf C}$ are matrices that are independent of $R(\tau)$ and are given by
\begin{align}
  {\bf A}_{kk'}&=4\pi\int_0^{R_s}drr^2f_k(r)f_{k'}(r)\\
  {\bf C}_{kk'}&=4\pi\int_{R_s}^{\infty}drr^2\left(1-\frac{R_s}{r}\right)f'_k(r)f'_{k'}(r).
  \label{AC}
\end{align}

From the action Eq.(\ref{rad_act_tau}) we can find the Hamiltonian, and according to the standard quantization procedure, the wave function $\psi(a_k,\tau)$ must satisfy the Functional Schr\"odinger equation. We can write the Schr\"odinger equation as
\be
  i\frac{\partial\psi}{\partial\tau}=\left[\frac{1}{2}\sqrt{1+R_{\tau}^2}\Pi_k({\bf A}^{-1})_{kk'}\Pi_{k'}+\frac{|R_{\tau}|}{2B}a_k(\tau){\bf C}_{kk'}a_{k'}(\tau)\right]\psi
\ee
where 
\be
  \Pi_k=-i\frac{\partial}{\partial a_k(\tau)}
\ee
is the momentum operator conjugate to $a_k(\tau)$.

Again, the problem of radiation from the collapsing domain wall for the infalling observer is equivalent to the problem of an infinite set of uncoupled harmonic oscillators with time dependent mass and frequency. Following the principal axis transformation used in the section above, the single eigenmode Schr\"odinger equation take the form
\be
  \left[-\frac{1}{2m}\sqrt{1+R_{\tau}^2}\frac{\partial^2}{\partial b^2}+\frac{|R_{\tau}|}{2B}Kb^2\right]\psi(b,\tau)=i\frac{\partial\psi(b,\tau)}{\partial\tau}
  \label{Schrod_rad_tau}
\ee
where $m$ and $K$ denote eigenvalues of ${\bf A}$ and ${\bf C}$, and $b$ is the eigenmode. 

Re-writing Eq.(\ref{Schrod_rad_tau}) in the standard form we obtain
\be
  \left[-\frac{1}{2m}\frac{\partial^2}{\partial b^2}+\frac{m}{2}\omega^2(\eta)b^2\right]\psi(b,\eta)=i\frac{\partial\psi(b,\eta)}{\partial\eta}
  \label{Schrod_Rad_tau}
\ee
where
\be
  \omega^2(\eta)=\frac{K}{m}\frac{|R_{\tau}|}{B\sqrt{1+R_{\tau}^2}}\equiv\omega_0^2\frac{|R_{\tau}|}{B\sqrt{1+R_{\tau}^2}}
\ee
and
\be
  \eta=\int d\tau'\sqrt{1+R_{\tau}^2}
  \label{eta_tau}
\ee
where we defined $\omega_0^2\equiv K/m$.

To proceed further, we will use the classical background of the collapsing domain wall Eq.(\ref{R_tau_0}). The initial vacuum state for the modes is the simple harmonic oscillator ground state,
\be
  \psi(b,\eta=0)=\left(\frac{m\omega_0}{\pi}\right)^{1/4}e^{-m\omega_0b^2/2}.
\ee
The exact solution for late times is given by Eq.(\ref{PedWave}) with initial conditions given by Eq.(\ref{IC}).

As discussed in Chapter \ref{ch:number} an observer with a detector will interpret the wavefunction of a given mode $b$ at late times in terms of simple harmonic oscillator states at the final frequency $\bar{\omega}$. 

The number of quanta in eigenmode $b$ can be evaluated from Eq.(\ref{OccNum}). By calculating $N_{\tau}$ it can be checked that $N$ remains constant for $\tau<0$ and also $\tau>\tau_f$. Hence all the particle production occurs for $0< \tau<\tau_f$ and is a consequence of the gravitational collapse.

Now we can take the limit $\tau_f\rightarrow\tau_c$. In this limit, $\rho$ remains finite but $\rho_{\eta}\rightarrow-\infty$ as $\tau>\tau_f\rightarrow\tau_c$, provided $\omega_0\not=0$ (see Appendix \ref{ch:rho(tau)} for details). However, we are interested in the behavior of $N$ for fixed frequency, $\bar{\omega}$. From the discussion in Appendix \ref{ch:rho(tau)}, we also know that $\rho\rightarrow\infty$ as $\omega_0\rightarrow0$. Therefore the occupation number at any frequency diverges in the infinite time limit when backreaction is not taken into account. 

\begin{figure}[htbp]
\includegraphics{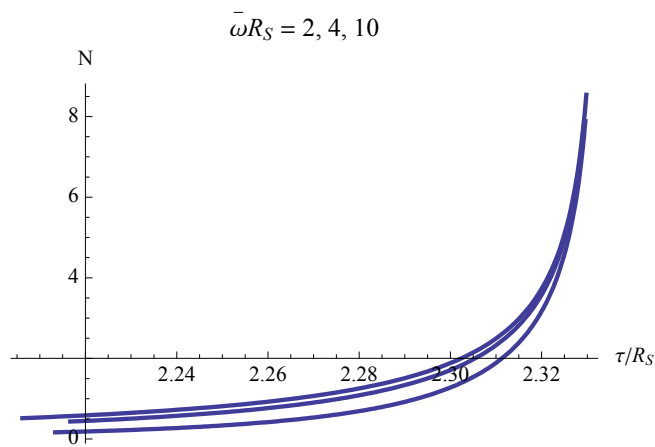}
\caption{The occupation number $N$ as a function of proper time $\tau/R_s$ for various fixed values of particle frequencies $\bar{\omega}R_s$. The curves are lower for higher values of $\bar{\omega}R_s$. The occupation number diverges as the infalling observer approaches $R_s$, which happens as $\tau\rightarrow\tau_c$.}
\label{Ntau_vs_tau}
\end{figure}

In Figure \ref{Ntau_vs_tau} we plot the occupation number of produced particles as a function of time (for several fixed frequencies $\bar{\omega}R_s$). The amount of proper time needed for the shell to reach $R_s$ can be obtained by integrating Eq.(\ref{dRdtau}). For $\sigma=0.01R_s^{-3}$ this critical proper time is $\tau_c=7/3R_s$. Figure \ref{Ntau_vs_tau} shows that, as the infalling observer approaches $R_s$, the occupation number increases and diverges exactly at $R_s$. The same conclusion as found by analyzing the occupation number $N$ as a function of $\rho$ and $\rho_{\tau}$ (see Appendix \ref{ch:rho(tau)}). This is in agreement with what one would expect in the absence of backreaction. Hawking showed, see Ref.\cite{1975Hawking}, that the flux of particles at late times diverges for a fixed background, i.e. fixed mass of the object. Here, from Eq.(\ref{Mass_tau}), we are treating the mass of the domain wall as a constant of motion. This means that we keep adding energy to the domain wall during the time of collapse, despite the the loss of mass due to the radiation. For the asymptotic observer it takes an infinite amount of his time for the domain wall to collapse to $R_s$, see Chapter \ref{ch:Classical} for discussion of this. However, this infinite time interval corresponds to a finite amount of time for the infalling observer's time. Thus, one may conclude that the infalling observer has to encounter the infinite number of particles produced during this finite amount of time before he reaches $R_s$. 

\begin{figure}[htbp]
\includegraphics{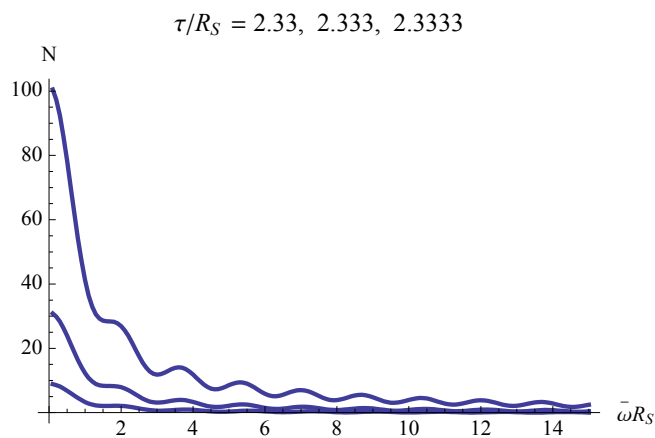}
\caption{The occupation number $N$ as a function of frequency $\bar{\omega}R_s$ for various fixed values of proper time $\tau/R_s$. The occupation number increases for larger values of $\tau/R_s$ as $\tau\rightarrow\tau_c$.}
\label{Ntau_vs_w}
\end{figure}

We have also numerically evaluated the spectrum of mode occupation numbers at any finite time and show the results in Figure \ref{Ntau_vs_w} for several different values of $\tau/R_s$. Figure \ref{Ntau_vs_w} shows that as the infalling observer time increases, the occupation number of larger values of $\bar{\omega}R_s$ increases. 

To find the temperature of the radiation, we again compare the curve in Figure \ref{Ntau_vs_w} with the occupation numbers for the Planck distribution, which is given by Eq.(\ref{OccPlanck}), where $\beta$ is again the inverse temperature. We can see that the spectrum of occupation numbers is non-thermal. As an example, there is no singularity in $N$ at $\omega=0$ at finite time. However, as $\tau\rightarrow\tau_c$, the peak at $\omega=0$ does diverge and the distribution becomes more and more thermal for these times. There are also oscillations in $N$, which are not present in the Planck distribution.

\begin{figure}[htbp]
\includegraphics{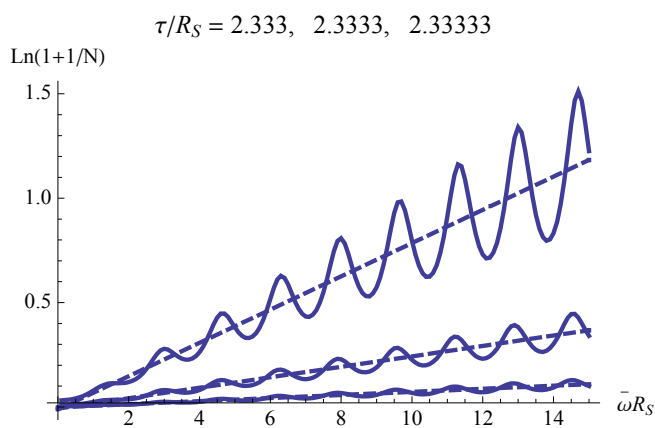}
\caption{Plot of $\ln(1+1/N)$ as a function of frequency $\bar{\omega}R_s$ for various fixed values of proper time $\tau/R_s$. The slope of the best fit line is $\beta$, which is the inverse temperature. The non-thermal features disappear and the temperature diverges as the Schwarzschild radius is approached, i.e. $\tau\rightarrow\tau_c$.}
\label{LnNtau_vs_w}
\end{figure}

We now wish to fit the temperature of the radiation, however, from Eq.(\ref{Schrod_Rad_tau}) we see that the time derivative of the wavefunction on the right-hand side is with respect to $\eta$, not with respect to $\tau$, and $\omega$ is the mode frequency with respect to $\eta$ as well. Eq.(\ref{eta_tau}) tells us that the frequency in $\tau$ is $\sqrt{1+R_{\tau}^2}$ times the frequency in $\eta$. However, recall from Chapter \ref{ch:Classical} Eq.(\ref{dRdtau}) tells us that as $R\rightarrow R_s$, $R_{\tau}$ is in fact a constant. Therefore, $\eta$ and $\tau$, for the case of the infalling observer, only differ by a constant amount. Hence, without loss of generality, we can ignore this shift by a constant amount, since the general features of the temperature will be the same. From Eq.(\ref{OccPlanck}), we can find the inverse temperature to be
\be
  \beta=\frac{\ln(1+1/N_P)}{\bar{\omega}R_s}=T^{-1}.
\ee
In Figure \ref{LnNtau_vs_w} we fit a thermal spectrum to the collapsed spectrum of Figure \ref{Ntau_vs_w}.  Several important features of the Hawking-like radiation can be taken from from this plot. First, the non-thermal features of the radiation are apparent. However, the departure from thermality (the fluctuations) are larger for earlier times, hence larger frequencies. This observation was first argued in Ref.\cite{Stojkovic}. Second, as $\tau\rightarrow\tau_c$ and the infalling observer approaches $R_s$, the radiation becomes more and more thermal even at large frequencies. Third, at $\tau=\tau_c$, i.e. $R=R_s$, the radiation becomes purely thermal. At this point, the black hole is formed and the radiation becomes thermal, as known from various studies of quantum radiation from a pre-existing black hole. Finally, it is apparent that the slope of $\ln(1+1/N)$ versus $\bar{\omega}R_s$ is decreasing as the infalling observer approaches $R_s$. Exactly at $R_s$, the slope of the curve is zero, indicating that the temperature of the radiation is infinite. This is not surprising since, as it is well known, the asymptotic observer in the nearly flat asymptotic region will register Hawking radiation with a finite temperature (see previous section). When the temperature is blue-shifted back to $R_s$, it clearly diverges. 

\subsection{Infalling Eddington-Finkelstein Coordinates}

Now we consider the collapse from the point of view of an infalling Eddington-Finkelstein observer. This is a different space-time foliation than that in Schwarzschild coordinates, and we expect crucially different results. In particular, since the metric is not divergent at the horizon, we do not expect infinite temperature there.  

For this purpose, we define the ingoing null coordinate $v$ as
\be
  v=t+r^*
\ee
where $r^*$ is the tortoise coordinate. We can then rewrite Eq.(\ref{Met_out}) as
\be
  ds^2=-\left(1-\frac{R_s}{r}\right)dv^2+2dvdr+r^2d\Omega^2, \hspace{2mm} r>R(v).
  \label{out_Met}
\ee
where the trajectory of the collapsing wall is $r=R(v)$.
The interior metric is the same as in Eq.(\ref{Met_in}). The interior time coordinate, $T$, is related to the ingoing null coordinate, $v$, via the proper time on the shell, $\tau$. The relations are
\be
  \frac{dT}{d\tau}=\sqrt{1+\left(\frac{dR}{d\tau}\right)^2}
  \label{dTdtau}
\ee
and
\be
  \frac{dv}{d\tau}=\frac{1}{B}\left(\frac{dR}{d\tau}-\sqrt{B+\left(\frac{dR}{d\tau}\right)^2}\right)
  \label{dvdtau}
\ee
where
\be
  B\equiv 1-\frac{R_s}{R}.
\ee

Consider again a massless scalar field $\Phi$ which propagates in the background of the collapsing shell. The action for the scalar field is
\be
  S=\int d^4x\sqrt{-g}\frac{1}{2}g^{\mu\nu}\partial_{\mu}\Phi\partial_{\nu}\Phi,
  \label{action}
\ee
where $g^{\mu\nu}$ is the background metric given by Eqs.(\ref{Met_in}) and (\ref{out_Met}). Decomposing the (spherically symmetric) scalar field into a complete set of real basis functions denoted by $\{f_k(r)\}$
\be
  \Phi=\sum_ka_k(v)f_k(r)
  \label{mode_ex}
\ee
we can find a complete set of independent eigenmodes $\{b_k\}$ for which the Hamiltonian is a sum of terms.

Since the metric inside and outside of the shell have different forms, we again split the action into two parts
\be
  S_{in}=2\pi \int dT\int_0^{R(v)}drr^2\left[-(\partial_T\Phi)^2+(\partial_r\Phi)^2\right],
  \label{S_in}
\ee
\begin{align}
  S_{out}=2\pi\int dv\int_{R(v)}^{\infty}drr^2&\Big{[}\partial_v\Phi\partial_r\Phi+\partial_r\Phi\partial_v\Phi\nonumber\\
  &+\left(1-\frac{R_s}{r}\right)(\partial_r\Phi)^2\Big{]}.\label{S_out}
\end{align}
We are again interested in the near horizon behavior of the radiation, i.e. as $R\rightarrow R_s$. In this limit we can write Eq.(\ref{dvdtau}) as
\be
  \frac{dv}{d\tau}\approx-\frac{1}{2R_{\tau}}
\ee
where $R_{\tau}=dR/d\tau$. Then with the help of Eq.(\ref{dTdtau}) we can write Eq.(\ref{S_in}) as
\begin{align}
  S_{in}=2\pi\int dv\int_0^{R(v)}drr^2&\Big{[}-\frac{1}{2}\frac{1}{\sqrt{R_v/2(R_v/2+1)}}(\partial_v\Phi)^2\nonumber\\
  &+2\sqrt{R_v/2(R_v/2+1)}(\partial_r\Phi)^2\Big{]}\label{S_in_v}
\end{align}
where $R_v=dR/dv$. Obviously, the action is not singular as $R(v) \rightarrow R_s$, unlike the Schwarzschild case. From Eqs.(\ref{S_out}) and (\ref{S_in_v}) we can write the total action as
\begin{align}
  S\approx&2\pi\int dv\Big{[}-\int_0^{R_s}drr^2\frac{1}{2}\frac{1}{\sqrt{R_v/2(R_v/2+1)}}(\partial_v\Phi)^2\nonumber\\
  &+\int_{R_s}^{\infty}drr^2\partial_v\Phi\partial_r\Phi+\int_{R_s}^{\infty}drr^2\partial_r\Phi\partial_v\Phi\nonumber\\
  &+\int_{R_s}^{\infty}drr^2\left(1-\frac{R_s}{r}\right)(\partial_r\Phi)^2\Big{]}
\end{align}
where we have changed the limits of integration from $R(v)$ to $R_s$ since this is the region of interest.

Now using the expansion in modes Eq.(\ref{mode_ex}), we can rewrite the action as
\begin{align}
  S\approx\int dv&\Big{[}-\frac{1}{2}\frac{1}{\sqrt{R_v/2(R_v/2+1)}}\dot{a}_k{\bf A}_{kk'}\dot{a}_{k'}\nonumber\\
  &+\frac{1}{2}\dot{a}_k{\bf Y}_{kk'}a_{k'}+\frac{1}{2}a_k{\bf Y}_{kk'}^{-1}\dot{a}_{k'}+\frac{1}{2}a_k{\bf C}_{kk'}a_{k'}\Big{]}
\end{align}
where $\dot{a}=da/dv$, and ${\bf A}$, ${\bf Y}$ and ${\bf C}$ are matrices that are independent of $R(v)$ and are given by
\begin{align}
  {\bf A}_{kk'}=2\pi\int_0^{R_s}drr^2f_k(r)f_{k'}(r),\\
  {\bf Y}_{kk'}=4\pi\int_{R_s}^{\infty}drr^2f_k(r)f'_{k'}(r),\\
  {\bf C}_{kk'}=8\pi\int_{R_s}^{\infty}drr^2\left(1-\frac{R_s}{r}\right)f'_k(r)f'_{k'}(r).
\end{align}
However if we take that the matrices are symmetric and real, we can see that ${\bf Y}={\bf Y}^{-1}$, so we can write the action as
\begin{align}
  S\approx\int dv&\Big{[}-\frac{1}{2}\frac{1}{\sqrt{R_v/2(R_v/2+1)}}\dot{a}_k{\bf A}_{kk'}\dot{a}_{k'}\nonumber\\
  &+\frac{1}{2}{\bf Y}_{kk'}\left(\dot{a}_ka_{k'}+a_k\dot{a}_{k'}\right)+\frac{1}{2}a_k{\bf C}_{kk'}a_{k'}\Big{]}.\label{Action}
\end{align}

From the action Eq.(\ref{Action}) we can find the Hamiltonian, and according to the standard quantization procedure, the wave function $\psi(a_k,v)$ must satisfy
\be
  i\frac{\partial\psi}{\partial v}=H\psi,
\ee
or
\begin{align}
  i\frac{\partial\psi}{\partial v}=&\Big{[}\frac{1}{2}\sqrt{R_v/2(R_v/2+1)}\Pi_k({\bf A^{-1}})_{kk'}\Pi_{k'}\nonumber\\
  &+\frac{1}{2}a_k\left(\sqrt{R_v/2(R_v/2+1)}{\bf Y}^2_{kk'}({\bf A^{-1}})_{kk'}+{\bf C}_{kk'}\right)a_{k'}\nonumber\\
  &+\frac{1}{2}\sqrt{R_v/2(R_v/2+1)}\Pi_k{\bf Y}_{kk'}({\bf A^{-1}})_{kk'}a_{k'}\Big{]}\psi
\end{align}
where
\be
  \Pi_k=-i\frac{\partial}{\partial a_k}
\ee
is the momentum operator conjugate to $a_k$. Using the momentum operator conjugate to $a_k$, we can rewrite the Schr\"odinger equation as
\begin{align}
  i\frac{\partial\psi}{\partial v}=&\Big{[}\frac{1}{2}\sqrt{R_v/2(R_v/2+1)}\Pi_k({\bf A^{-1}})_{kk'}\Pi_{k'}\nonumber\\
  &+\frac{1}{2}a_k\left(\sqrt{R_v/2(R_v/2+1)}{\bf Y}^2_{kk'}({\bf A^{-1}})_{kk'}+{\bf C}_{kk'}\right)a_{k'}\nonumber\\
  &-i\frac{1}{2}\sqrt{R_v/2(R_v/2+1)}{\bf Y}_{kk'}({\bf A^{-1}})_{kk'}\delta_{kk'}\Big{]}\psi
\end{align}
where $\delta_{kk'}$ is the Kronecker delta function.

So the problem of radiation from the collapsing domain wall for the infalling observer is equivalent to the problem of solving an infinite set of decoupled damped harmonic oscillators with time-dependent frequency. Since ${\bf A}$, ${\bf Y}$ and ${\bf C}$ are symmetric and real, it is possible to simultaneously diagonalize them using the principle axis transformation. Then for a single eigenmode, the Schr\"odinger equation takes the form
 \begin{align}
  i\frac{\partial\psi}{\partial v}=&\Big{[}-\frac{1}{2m}\sqrt{R_v/2(R_v/2+1)}\frac{\partial^2}{\partial b^2}\nonumber\\
  &+\frac{1}{2}\left(\sqrt{R_v/2(R_v/2+1)}\frac{y^2}{m}+K\right)b^2\nonumber\\
  &-i\frac{y}{2m}\sqrt{R_v/2(R_v/2+1)}\Big{]}\psi
  \label{schrod}
\end{align}
where $m$, $y$ and $K$ denote eigenvalues of ${\bf A}$, ${\bf Y}$ and ${\bf C}$, and $b$ is the eigenmode.

Re-writing Eq.(\ref{schrod}) in the standard form we obtain
\be
  \left[-\frac{1}{2m}\frac{\partial^2}{\partial b^2}+\frac{m}{2}\omega^2(\eta)-i\frac{y}{2m}\right]\psi(b,\eta)=i\frac{\partial\psi(b,\eta)}{\partial\eta}
  \label{Schrod}
\ee
where
\begin{align}
  \omega^2(\eta)&=\frac{y^2}{m^2}+\frac{K}{m}\frac{1}{\sqrt{R_v/2(R_v/2+1)}}\nonumber\\
        &\equiv\frac{y^2}{m^2}+\frac{\omega_0^2}{\sqrt{R_v/2(R_v/2+1)}}
\end{align}
and
\be
  \eta=\int dv'\sqrt{R_v/2(R_v/2+1)}
\ee
where we defined $\omega_0^2\equiv K/m$. To find solutions to equation Eq.(\ref{Schrod}) we use the ansatz
\be
  \psi(b,\eta)=e^{-y\eta/2m}\phi(b,\eta).
  \label{ansatz}
\ee
This leads to the equation for $\phi(b,\eta)$
\be
  -\frac{1}{2m}\frac{\partial^2\phi}{\partial b^2}+\frac{m\omega^2}{2}b^2\phi=i\frac{\partial\phi}{\partial\eta}.
\ee
The exact solution for late times is given by Eq.(\ref{PedWave}) with initial conditions given by Eq.(\ref{IC}).
Then Eq.(\ref{ansatz}) gives
\be
  \psi=e^{-y\eta/2m}\phi(b,\eta)
  \label{psi}
\ee
where $\phi$ given in Eq.(\ref{PedWave}).

As discussed in Chapter \ref{ch:number} an observer with a detector will interpret the wavefunction of a given mode $b$ at late times in terms of simple harmonic oscillator states at the final frequency $\bar{\omega}$.

\begin{figure}[htbp]
\includegraphics{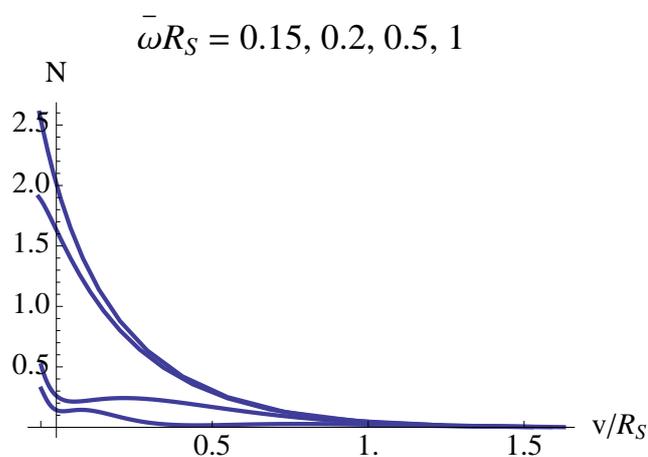}
\caption{Here we plot $N$ versus $v/R_s$ for various fixed values of $\bar{\omega} R_s$. The curves are lower for higher values of $\bar{\omega} R_s$. }
\label{N_versus_v}
\end{figure}

In Fig.~\ref{N_versus_v} we plot $N$ versus $v/R_s$ for various fixed values of $\bar{\omega}R_s$. We can see that the occupation number at any frequency increases as $v/R_s$ decreases. Thus more particles are created as the shell reaches and crosses the horizon. However, the number of created particles does not diverge as $R(v) \rightarrow R_s$.

\begin{figure}[htbp]
\includegraphics{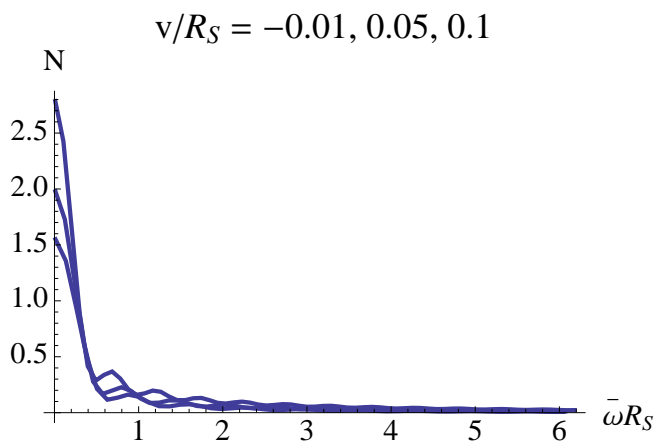}
\caption{Here we plot $N$ versus $\bar{\omega} R_s$ for various fixed values of $v/R_s$. The occupation number at any frequency grows as the collapse progresses  (i.e. $v/R_s$ decreases) but in never diverges.}
\label{N_versus_w}
\end{figure}

We then numerically evaluate the spectrum of mode occupation numbers at any finite time and show the results in Fig.~\ref{N_versus_w} for several values of $v/R_s$. The first sign of non-thermality is the fact that the occupation number is non-divergent at $\bar{\omega}=0$, as opposed to the thermal Planck distribution in Eq.(\ref{OccPlanck}).

\begin{figure}[htbp]
\includegraphics{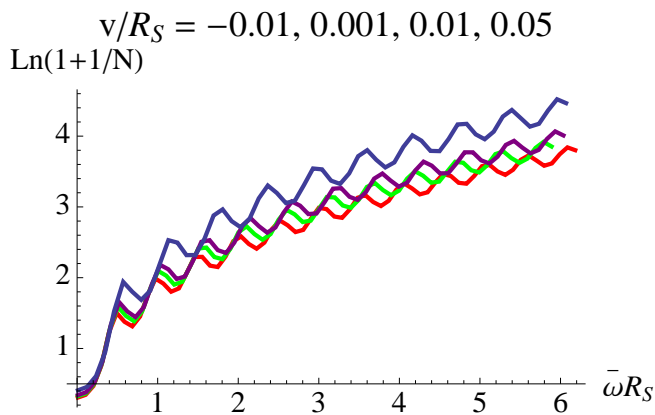}
\caption{Here we plot $\ln(1+1/N)$ versus $\bar{\omega} R_s$ for various fixed values of $v/R_s$. The curves are lower and display more fluctuations as $v/R_s$ decreases.}
\label{LnN_versus_w}
\end{figure}

In Fig.~\ref{LnN_versus_w} we plot $\ln(1+1/N)$ versus $\bar{\omega}R_s$ for various values of $v/R_s$. As $v/R_s$ decreases (as the shell is collapsing), the curves decrease. A thermal spectrum should gives us a straight line, however, we see that is not the case here. The best one can do is to fit the low frequency part of the spectrum and get the temperature in that regime. In our case we get $T=(0.17R_s)^{-1}$. Unlike the case of Schwarzschild coordinates, where the spectrum becomes thermal in the whole frequency range, in Eddington-Finkelstein coordinates the spectrum never becomes thermal in the high frequency range. Another feature is apparent in Fig.~\ref{LnN_versus_w}. As the collapse progresses, the fluctuations in the spectrum become more violent. This is indicative of the shell approaching the actual singularity at $R=0$ which is the region of strong gravitational fields.

\section{Discussion}

In this chapter we investigated the Hawking-like radiation produced during the time of gravitational collapse for both the asymptotic observer and the infalling observer. The occupation number of the radiation was then used to fit the temperature of the radiation as the shell approaches $R_s$. When considering Schwarzschild coordinates, in both cases the resulting analysis lead to the same conclusions: First, that the spectrum of the occupation display non-thermal characteristics during the time of collapse. This non-thermality is seen by a non-divergent occupation number when $\bar{\omega}=0$ and in oscillations about thermality. Second, the spectrum becomes more and more thermal as the domain wall approaches $R_s$, corresponding to large $\bar{\omega}$ values. Finally, the spectrum becomes purely thermal when the domain wall reaches $R_s$. When consider Eddington-Finkelstein coordinates for the infalling observer, we find that the spectrum never becomes thermal in the high frequency range.

In the case of the asymptotic observer, upon fitting the temperature, we find that the temperature of the radiation is on the order of the Hawking temperature. This value is not exactly the Hawking temperature for two reasons. First, when fitting the temperature we use a best fit approximation for the slope of $\beta$. However, there is ambiguity for choosing the best fit approximation, thus the true slope of $\beta$ may be different from the one chosen. Second, we are fitting the temperature numerically. There is always an inherent approximation used we using numerical methods, therefore our calculation is inherently ambiguous. 

In the case of the infalling observer, upon fitting the temperature in Schwarzschild coordinates, we find that when the shell reaches $R_s$ the temperature of the radiation becomes divergent. This would seem to imply that the local temperature measured by the observer is then infinite, meaning that the observer will burn up before he/she reaches $R_s$. However, this is not necessarily the case. It has been argued in Ref.\cite{Fulling}, where a simple $1+1$ model was studied, that the local vacuum polarization will cancel out the divergent temperature energy density due to the radiation. Therefore, the true local value of the stress-energy tensor is small in the region $R\sim R_s$. A simple reason for this divergent temperature is that the Schwarzschild observer is actually an accelerated observer, so to truly investigate the local temperature one must consider a truly freely falling observer, i.e. the Eddington-Finkelstein coordinates. In this case, upon fitting the temperature, we find that the local temperature is in fact finite.  

%% file: Entropy.tex
\chapter{Entropy}
\label{ch:entropy}

In 1972 Bekenstein argued that a black hole of mass $M$ has an entropy proportional to its surface area, see Ref.\cite{Bekenstein}. Further calculations by Gibbons and Hawking showed that the entropy of a black hole is always a constant, despite the type of metric which is used, see Ref.\cite{GibbonsHawking}. They showed that the expression for the entropy is given by 
\be
  S_{BH}=\frac{A_{hor}}{4}=\pi R_s^2
  \label{SBH}
\ee
where $A_{hor}$ is the surface area of the event horizon and $R_s$ is the Schwarzschild radius for a black hole which contains only mass.

The typical method for calculating the entropy of the black hole is to first calculate the temperature of the black hole using the so-called Bogolyubov method. Here, one considers that the system starts in an asymptotically flat metric (typically Minkowski), then the system evolves to a new asymptotically flat metric (in the case of just mass, the typically final metric is that of Schwarzschild). One then matches the coefficients between the two asymptotically flat spaces at the beginning and end of the gravitational collapse. The mismatch of these two vacua gives the number of particles produced during the collapse. What happens in between is then beyond the scope of the Bogolyubov method, since the method is generally independent of time. Therefore the time-evolution of the thermodynamics properties of the collapse cannot be investigated in the context of the Bogoyubov method. 

Here we will investigate the time-evolution of a spherically symmetric infinitely shell of collapsing matter in the context of the Functional Schr\"odinger formalism. Since the Functional Schr\"odinger formalism depends on the observer's degrees of freedom, one can introduce the ``observer" time into the quantum mechanical processes, with the use of the Wheeler-de Witt equation, in the form of the Schr\"odinger equation, see Chapter \ref{ch:formal}. To study the case of gravitational collapse, one can then choose the classical Hamiltonian of the collapsing object, then employ the standard quantization condition. The wavefunctional is then dependent on the observer time chosen, hence one can view the quantum mechanical processes of a given system under any foliation of space-time that one chooses. The benefit of using the Functional Schr\"odinger formalism is that, in principle, one can solve the time-dependent wavefunctional equation exactly, as discussed in the previous chapters. Therefore the Functional Schr\"odinger formalism goes beyond the approximations of the Bogolyubov method, since the system is allowed to evolve over time, which allows one to investigate the intermediate regime during the collapse. Since the wavefunctional contains all the information of the system, one can, in principle, study the time evolution of the thermodynamical processes of the system. Of current interest is the time-evolution of the entropy of a collapsing gravitational object. We will do so from the view point of a stationary asymptotic observer, since this is the more relevant question. In this chapter we summarize the work originally done in Ref.\cite{EG_ent}.

\section{Partition Function}

To study the entropy of the system, we will first develop the partition function for the system. In order to study the time-evolution of the entropy we shall employ the so-called Liouville-von Neumann approach, which was developed to study equilibrium and non-equilibrium quantum processes (see Ref.\cite{Thermal}). The Liouville-von Neumann approach is a canonical method which unifies the Liouville-von Neumann equation and the Functional Schr\"odinger equation. This approach utilizes the invariant operator approach developed by Lewis and Riesenfeld (see Ref.\cite{Lewis}, Chapter \ref{ch:number} and Appendix \ref{ch:Invariant}), which allows one to exactly solve time-indepedent and time-dependent quantum systems.  The Liouville-von Neumann approach has been employed for several different situations ranging from Condensed matter physics to Cosmology, see for example see Ref.\cite{Kim}. The basic assumption of the Liouville-von Neuman approach is that non-equilibrium processes are consequences of underlying microscopic processes which are well described by quantum theory. The details about the collapse will depend on the particular foliation of space-time used to study the system. From the point of view of an infalling observer, in order to calculate the backreaction and local effect around the event horizon it is important to choose a state that is non-singular at the horizon. In this region, the vacuum of choice is the Unruh vacuum (see Refs.\cite{Davies,Unruh}). However, discussed above, we are interested in the view point of the asymptotic observer.

Using the Liouville-von Neumann approach, and following the procedure used in Ref.\cite{Kim}, we can write the partition function as
\be
  Z=\textrm{Tr}\left[e^{-\beta I}\right]
  \label{Z}
\ee
where $I$ is any operator which satisfies the equation
\be
  \frac{dI}{dt}=\frac{\partial I}{\partial t}-i\left[I,H\right]=0
   \label{Heisenberg}
\ee
 and $\beta$ is a free parameter. Here we note that Eq.(\ref{Heisenberg}) is just the Heisenberg equation of motion for the operator $I$, see Ref.\cite{Sakurai}, where the total time derivative of the operator is zero. In the case that the total derivative is equal to zero in the Heisenberg, this case is known as the Liouville-von Neumann equation, see Ref.\cite{Lewis}.
 
 From Ref.\cite{Pedrosa}, we can write the invariant operator $I$ as
\be
  I=\frac{1}{2}\left[\sqrt{\frac{b}{\rho}}+\left(\pi_b\rho-m\rho_{\eta}b\right)^2\right].
  \label{I}
\ee
Here we note that the invariant operator $I$ is time dependent since $\rho$ is time dependent (see Eq.(\ref{gen_rho})). Using Eq.(\ref{I}) we can therefore write the partition function, Eq.(\ref{Z}), as
\be
  Z=\textrm{Tr}\exp\left[-\beta\frac{1}{2}\left[\sqrt{\frac{b}{\rho}}+\left(\pi_b\rho-m\rho_{\eta}b\right)^2\right]\right].
\ee
In this form we can see that the partition function is time dependent since the invariant operator $I$ is time dependent by virtue of Eq.(\ref{I}).

We note that we can rewrite the invariant operator in a more suggestive manner by writing Eq.(\ref{I}) as
\bea
  I&=&\left(\frac{1}{\sqrt{2}}\right)^2\left[\left(\frac{b}{\rho}\right)^{1/4}-i\left(\pi_b\rho-m\rho_{\eta}b\right)\right]\left[\left(\frac{b}{\rho}\right)^{1/4}+i\left(\pi_b\rho-m\rho_{\eta}b\right)\right]\nonumber\\
    &\equiv&n(t)+\frac{1}{2}
\eea
where
\be
  n(t)=a^{\dagger}(t)a(t)
  \label{n_t}
\ee
and
\be
  a(t)\equiv\frac{1}{\sqrt{2}}\left[\left(\frac{b}{\rho}\right)^{1/4}+i\left(\pi_b\rho-m\rho_{\eta}b\right)\right].
\ee
Here $n(t)$ is the time dependent number of states. Hence, the invariant operator $I$ takes on the form of a time-dependent harmonic oscillator Hamiltonian, where the number operator is time dependent. 

For a physical meaning of the partition function, we need to act the invariant operator on a quantum state. In the Heisenberg picture, the quantum states span a particular Hilbert space. A convienient basis in this Hilbert space is the so-called Fock space representation, see Ref.\cite{Birrell}. This basis is an eigenstate of the Number operator, Eq.(\ref{n_t}). Thus at a particular time $t$, one has in the Fock space representation
\be
  n(t)\big{|}n,t\rangle=n\big{|}n,t\rangle. 
\ee
Thus, in this space we can then write the partition function as
\bea
  Z&=&\textrm{Tr}\exp\left[-\beta\omega_0\left(n+\frac{1}{2}\right)\right]\nonumber\\
   &=&\frac{1}{2\sinh\left(\frac{\beta\omega_0}{2}\right)}.
   \label{part func}
\eea
At first glance, one would be tempted to say that the partition function in Eq.(\ref{part func}) is not time-dependent since the partition function now only depends on the initial frequency of the induced scalar field. However, recall that $\beta$ is free parameter which we can choose. Here we discuss our choice in the free parameter $\beta$. 

In Refs.\cite{Stojkovic,Greenwood} one can define the occupation number for a frequency $\bar{\omega}$, Eq.(\ref{OccNum}). Then by fitting the number of particles created as the usual Planck distribution Eq.(\ref{OccPlanck}), one can then in principle fit the temperature of the radiation. Here, we then choose to define $\beta$ as 
\be
  \beta=\frac{\partial\ln\left(1+1/N\right)}{\partial\bar{\omega}}.
  \label{beta}
\ee
This implies that all of the time dependence of the system is encoded into the temperature of the system.

Therefore we can see that Eq.(\ref{part func}) is just the standard entropy for a time-independent harmonic oscillator, however, the temperature here is time-dependent. Thus we recover the time-dependence of the partition function. Since the partition function is time-dependent, therefore the entropy is also time-dependent. 

\section{Entropy}

In terms of the partition function, the thermodynamic definition of entropy is given by, see for example Ref.\cite{Landau},
\be
  S=\ln Z-\beta\frac{\partial\ln Z}{\partial\beta}.
\ee
Using Eq.(\ref{part func}), we can then write the entropy of the system as
\be
  S=-\ln\left(1-e^{-\beta\omega_0}\right)+\beta\frac{e^{-\beta\omega_0}}{1-e^{-\beta\omega_0}}.
\ee
Therefore, this is again just the entropy of the usual time-independent harmonic oscillator. From Eq.(\ref{beta}) it follows that the temperature is time-dependent. 

To be able to calculate the entropy of the domain wall we will consider the entropy of the entire system, i.e. the domain wall and radiation, and the radiation alone. We will assume that the total entropy is a linear equation in the entropy of the domain wall and the entropy of the radiation. Thus we will write the total entropy as
\be
  S_{SR}=S_S+S_R,
\ee
where the subscripts $SR$ stands for domain wall and radiation, $S$ for just domain wall and $R$ radiation only, respectively. Then by subtracting these two quantities one can then determine the entropy of the domain wall
\be
  S_S=S_{SR}-S_R.
  \label{S_shell}
\ee
In Chapter \ref{ch:radiation} we considered the wavefunction and occupation number of the radiation only system. To proceed further, we must now consider the wavefunction and occupation number for the entire system, $SR$.

\subsection{Entire System}

To find the wavefunction and occupation number for the entire system, we first note that from Eq.(\ref{Ham_Rdot}) we can approximate the Hamiltonian of the domain wall as
\be
  H_{wall}\approx-B\Pi_R.
  \label{HW_t}
\ee
Then using Eq.(\ref{HW_t}) and Eq.(\ref{Pre-Rad_Schrod_t}) we can write the Hamiltonian of the entire system as
\be
  H=H_{wall}+H_b=-B\Pi_R+B\frac{\Pi_b^2}{2m}+\frac{K}{2}b^2
  \label{tot Ham}
\ee
where $\Pi_R$ is given in Chapter \ref{ch:Classical} and $\Pi_b$ is given by
\be
  \Pi_b=-i\frac{\partial}{\partial b}
\ee
The wavefunction for the entire system is then a function of $b$, $R$, and $t$, which we can write as
\be
  \Psi=\Psi(b,R,t).
  \label{psi_b_R_t}
\ee

Substituting Eq.(\ref{tot Ham}) into Eq.(\ref{FSE}), we can then write the Functional Schr\"odinger equation as
\be
  iB\frac{\partial\Psi}{\partial R}-\frac{B}{2m}\frac{\partial^2\Psi}{\partial b^2}+\frac{K}{2}b^2\Psi=i\frac{\partial\Psi}{\partial t}.
  \label{Schrod1}
\ee
To solve Eq.(\ref{Schrod1}) we will use the semiclassical case, i.e. we will use the classical background for the collapsing shell. Since the distance of the shell only depends on the time, see Eq.(\ref{Rdot}), we can then write
\begin{equation*}
  iB\frac{dt}{dR}\frac{\partial\Psi}{\partial t}-\frac{B}{2m}\frac{\partial^2\Psi}{\partial b^2}+\frac{K}{2}b^2\Psi=i\frac{\partial\Psi}{\partial t}.
  \label{Schrod2}
\end{equation*}
Hence, we are eliminating the $R$ dependence from Eq.(\ref{psi_b_R_t}), so $\Psi(b,R,t)\rightarrow\Psi(b,t)$. Rewriting gives
\be
  -\frac{B}{2m}\frac{\partial^2\Psi}{\partial b^2}+\frac{K}{2}b^2\Psi=i\frac{\partial\Psi}{\partial t}\left(1-B\frac{dt}{dR}\right).
  \label{Schrod3}
\ee
Making use of Eq.(\ref{Rdot}), i.e. $dt/dR=-B$, this becomes
\be
  -\frac{B}{2m}\frac{\partial^2\Psi}{\partial b^2}+\frac{K}{2}b^2\Psi=2i\frac{\partial\Psi}{\partial t}.
  \label{ent Schrod}
\ee

We now rewrite Eq.(\ref{ent Schrod}) in the standard form
\be
  \left[-\frac{1}{2m}\frac{\partial^2}{\partial b^2}+\frac{m}{2}\omega^2(\tilde{\eta})b^2\right]\psi(b,\tilde{\eta})=i\frac{\partial\psi(b,\tilde{\eta})}{\partial \tilde{\eta}}
  \label{Ent Schrod}
\ee
where
\be
  \tilde{\eta}=\frac{1}{2}\int_0^tdt'\left(1-\frac{R_s}{R}\right)
  \label{tilde_eta}
\ee
and
\be
  \omega^2(\tilde{\eta})=\frac{K}{m}\frac{1}{1-R_s/R}\equiv\frac{\omega_0^2}{1-R_s/R}.
\ee
Here we have chosen to set $\tilde{\eta}(t=0)=0$.

The solution to Eq.(\ref{Ent Schrod}) is given by Eq.(\ref{PedWave}), as discussed in Chapter \ref{ch:radiation}. We can then find the occupation number $N$ for the entire system, Eq.(\ref{OccNum}). 

Here we will make some quick comments regarding the occupation number. We can see that from Eqs.(\ref{Ent Schrod}) and (\ref{Rad_Schrod_t}), the Schr\"odinger equations for the entire system and radiation only are of the same form. Hence one would expect that there is no difference between the occupation number for the entire system and the radiation only. However, the time parameters $\tilde{\eta}$ and $\eta$, given in Eqs.(\ref{tilde_eta}) and (\ref{eta_t}), are different. Hence the occupation numbers of the two systems will evolve differently, which leads to different temperatures in each of the two systems. Therefore, the entropy of each system will be different.

\section{Analysis}

\begin{figure}[htbp]
\includegraphics{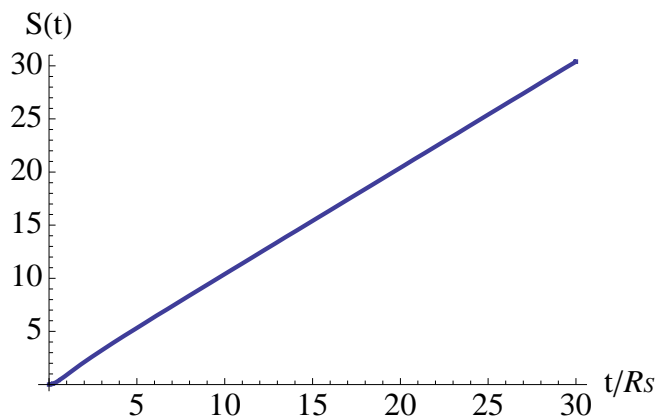}
\caption{We plot the entropy of the entire system as a function of asymptotic observer time $t$.}
\label{EntEntire}
\end{figure}

First we consider the entropy of the entire system. In Figure \ref{EntEntire} we plot the entropy of the entire system as a function of dimensionless time $t/R_s$. Figure \ref{EntEntire} shows that the system starts with an initial entropy of zero. This is expected since initially there is only one degree of freedom, meaning that $S=\ln(1)=0$. Here we have normalized the initial entropy of the shell to be zero. To justify this normalization, consider a solar mass black hole. Under the usual Bekenstein-Hawking entropy, the order of magnitude estimate of the entropy of a solar mass black hole is $S_{BH}\approx10^{75}$. Now consider that the shell is actually made up of protons. The initial entropy of the shell then is approximately $S_{S,0}\approx10^{57}$. Comparing the entropy of the final black hole versus the initial entropy of the shell, the entropy of the final black hole is much much greater than that of the initial entropy of the shell, thus the initial entropy of the shell only contributes a negligible amount of entropy to the entropy of the final black hole. Thus our normalization of the initial entropy of the shell to zero is justified. As $t/R_s$ increases, initially the entropy increases rapidly, then settles down to increase approximately linearly. Due to the linear increase, we see that as $t/R_s$ goes to infinity, the entropy will then diverge. This is again expected since as the asymptotic time goes to infinity, the number of particles that are produced diverges (see Ref.\cite{Stojkovic}). This is a consequence of the fact that we keep the background fixed (i.e. $R_s$ is a constant). In reality, $R_s$ should decrease over time since the radiation is taking away mass and energy from the system. Therefore as $t/R_s$ goes to infinity, the entropy of the entire system as measured by the asymptotic observer diverges as $R\rightarrow R_s$. 

This is consistent with the results found in Refs.\cite{Saida}. Here the authors consider the time-dependent non-equilibrium evolution of a black hole as well as the incorporation of the given off radiation. Here one can see that the entropy of the system diverges as the time goes to infinity. 

The results of Figure \ref{EntEntire} are consistent with the generalized second law of black hole thermodynamics. The generalized second law states that, see for example Ref.\cite{Jacobson} and references there in
\be
  \delta(S_{out}+A/4)\geq0
  \label{GSL}
\ee
where here, $S_{out}=S_R$, $A/4=S_S$ and $S_{out}+A/4=S_{SR}$, respectively. Eq.(\ref{GSL}) simply states that the total entropy of the system must constantly be increasing as in agreement with thermodynamics entropy Ref.\cite{Landau}. As stated above, a realistic model for gravitational collapse will have that the Schwarzschild radius $R_s$ will decrease over time, since the domain wall is losing mass. Eq.(\ref{GSL}) allows for this result as long as the entropy increase of the radiation compensates for the loss in entropy of the collapsing domain wall.

\begin{figure}[htbp]
\includegraphics{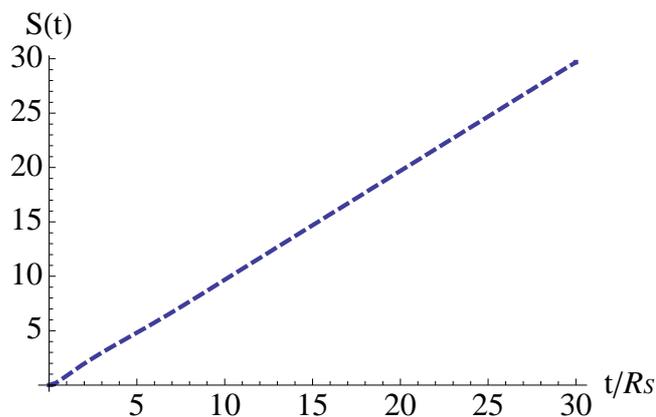}
\caption{We plot the entropy of the particles created during the collapse as a function of asymptotic time $t$.}
\label{EntR}
\end{figure}

Now we consider the radiation only. Considering just the particles which are created, i.e. the radiation, during the collapse, we can then plot the entropy as a function or the rescaled asymptotic time $t/R_s$, see Figure \ref{EntR}. Figure \ref{EntR} shows initially the entropy of the system is zero. Again, this is expected since initially the domain wall is in vacuum, meaning that there are no particles produced. Therefore the only degree of freedom is that of the domain wall, this then gives that the initial entropy must be zero. As the asymptotic observer time increases, initially there is rapid increase in the entropy, but again, the entropy then increases linearly as the asymptotic observer time increases. As in the case of the entire system, as the time measured by the asymptotic observer goes to infinity, the entropy of the particles created during the time of collapse diverges. This is expected since the number of particles which are created during the time of collapse diverges as $R\rightarrow R_s$, hence as the domain wall approaches the horizon the number of particles created during the collapse diverges. This result again is in agreement with the generalized second law of black hole thermodynamics, Eq.(\ref{GSL}).

\begin{figure}[htbp]
\includegraphics{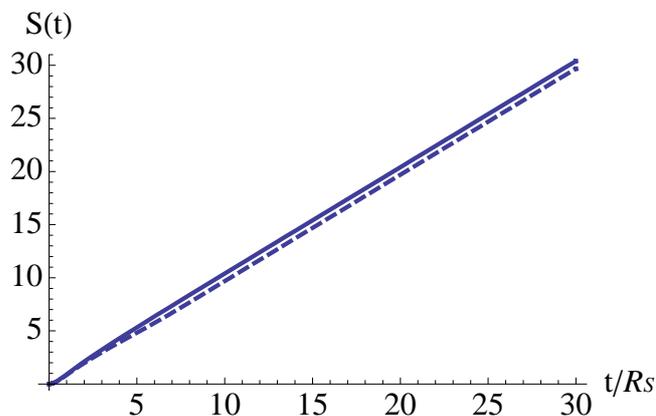}
\caption{We plot the entropy as a function of asymptotic observer time $t$ for both the entire system and the particles created during the time of collapse.}
\label{EntB}
\end{figure}

In Figure \ref{EntB} we plot the entropy as a function of the rescaled asymptotic observer time $t/R_s$ of both the entire system and the particles created during the time of collapse. Figure \ref{EntB} shows that except for the initial increase in the entropy, for later asymptotic observer time, the slopes of the entropy versus time are approximately equal. Therefore, one can expect that the entropy of the domain wall is approximately constant for late times. 

\begin{figure}[htbp]
\includegraphics{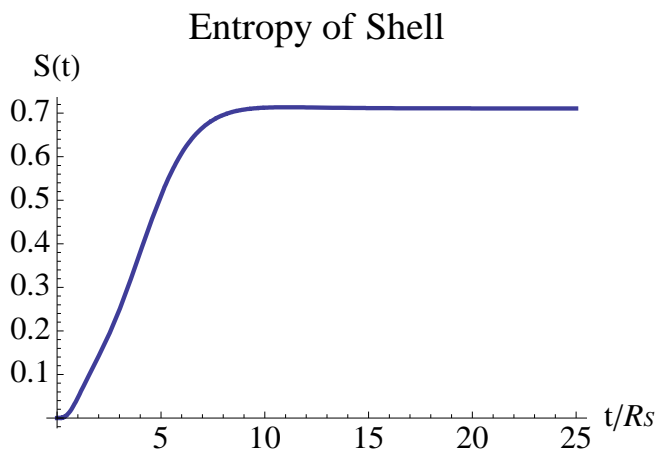}
\caption{We plot the entropy of the shell as a function of asymptotic observer time $t$.}
\label{EntShell}
\end{figure}

As stated earlier, what is of interest is the entropy of the collapsing domain wall, since this will collapse to form a black hole. To find the entropy of the domain wall, we can take the entropy of the entire system and subtract off the entropy of the particles produced (since these are the only relevant objects which contribute to the entropy), see Eq.(\ref{S_shell}). The result is then given in Figure \ref{EntShell}. Figure \ref{EntShell} shows that initially the entropy of the domain wall is zero. As stated above, this is expected since initially there is only one degree of freedom. As asymptotic time increases, the entropy of the domain wall rapidly increases. However, for late times, the entropy of the domain wall goes to a constant. As stated above, this is expected since the late time entropies for entire system and for the particles created during collapse are approximately parallel. However, as discussed earlier, one would expect that in a realistic model the entropy of the domain wall should in fact decrease over time since $R_s$ is decreasing because the domain wall is losing mass. The entropy here, however, is constant since we are assuming that the mass is approximately the Hamiltonian of the system, which is a constant of motion, see Chapters \ref{ch:Classical} and \ref{ch:model}. This means that since we are holding the mass of the domain wall constant, we need to keep adding energy to the system to counter act the loss of mass from the Hawking radiation. Therefore one can expect that the entropy of the domain wall must be a constant for late times.

In reality, radiation takes mass away from the system, so the entropy of the domain wall will go to zero as $R_s$ goes to zero. This means that after the black hole disappears, all the entropy will go into the entropy of the radiation, which is in agreement with the generalized second law of black hole thermodynamics.

From Figure \ref{EntShell}, we see that our numerical value for the late time entropy of the domain wall is
\begin{equation*}
  S\approx0.7 R_s^2.
\end{equation*}
Comparing with Eq.(\ref{SBH}), we can view this discrepancy as a shift in the Schwarzschild radius $R_s$. In order to get the theoretical value for the entropy, Eq.(\ref{SBH}), we see that we would require $R_s\rightarrow2.11R_s$. This is an understandable numerical error, which implies that our numerical solution is of the same order as the Hawking-Bekenstein entropy.

Another interesting thing to note is that Figure \ref{EntShell} tells us that change in entropy occurs for early times, then gets frozen as time increases. From the plot we see that the change in entropy occurs during the time range $0\leq t/R_s<7.5$. At first sight this seems to be an arbitrary value for the entropy of the domain wall to stop increasing. However, from Eq.(\ref{R(t)}) one can see that this time is not an arbitrary value. 

\begin{figure}[htbp]
\includegraphics{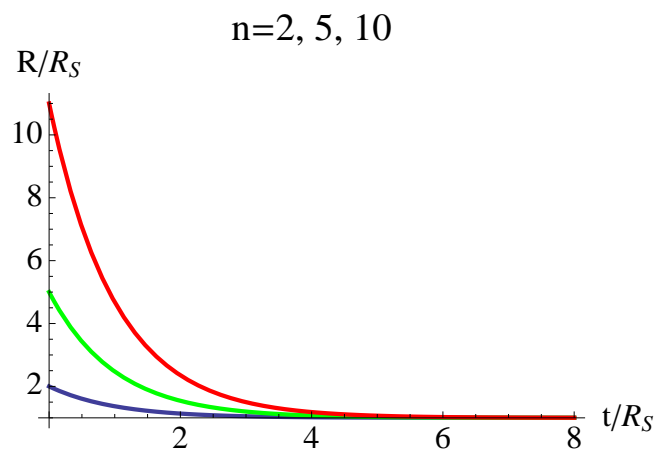}
\caption{We plot $R/R_s$ versus $t/R_s$ for various values of $n$. Here the blue curve corresponds to $n=2$, the green curve corresponds to $n=5$ and the red curve corresponds to $n=10$.}
\label{Position_n}
\end{figure}

\begin{figure}[htbp]
\includegraphics{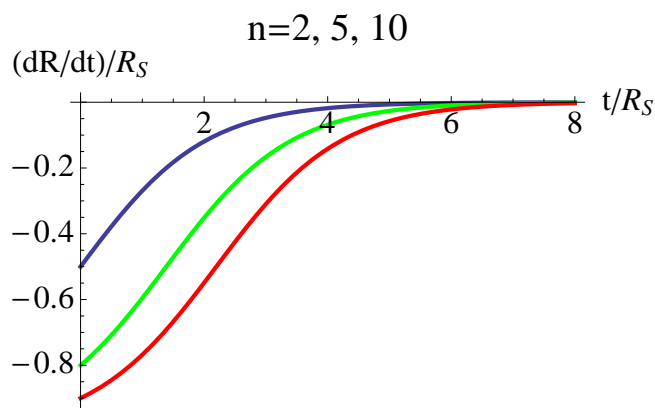}
\caption{We plot $\dot{R}/R_s$ versus $t/R_s$ for various values of $n$. Here the blue curve corresponds to $n=2$, the green curve corresponds to $n=5$ and the red curve corresponds to $n=10$.}
\label{velocity_n}
\end{figure}

To see this, let us first consider Eq.(\ref{R(t)}) and make the requirement that $R_0=nR_s$, where $n$ is some integer. Then we can write Eq.(\ref{R(t)}) as
\bd
  R(t)=R_s\left(1+(n-1)e^{-t/R_s}\right).
\ed
For illustration purposes let's restrict the value of $n$ to be $n\leq10$, which is a restriction that the domain wall starts off at a position ten times it's Schwarzschild radius. In Figure \ref{Position_n} we plot $R/R_s$ versus $t/R_s$ for various values of $n$. For each value of $n$ chosen, we see that the value $R/R_s\approx1$ occurs for $t/R_s\approx7$. In the case of $n=10$, we see that $R=1.005R_s$, while the value is less than that for smaller values of $n$. Hence, the time $t/R_s=7.5$ seems to be a universal time when the domain wall is almost to the Schwarzschild radius. From Eq.(\ref{Rdot}) we see that by this time we have
\bd
  \dot{R}=-B\approx0.
\ed
Hence in this time limit, the velocity of the domain wall is approximately zero, meaning that as far as the asymptotic observer is concerned the domain wall has stopped moving and there are no more dynamics. This can be seen in Figure \ref{velocity_n}, where we plot the corresponding velocities for the same values of $n$. Figure \ref{velocity_n} also shows that the time $t/R_s=7.5$ corresponds to a universal time of when the different velocities go approximately to zero. Recall from Chapter \ref{ch:Classical} that it takes an infinite amount of time for the domain wall to reach the horizon, so from $t/R_s=7.5$ to infinity the entropy is constant since all the dynamics are essentially done and the shell is approximately stationary for the observer. Hence the volume of the spherically symmetric domain wall becomes essentially constant by the time $t/R_s=7.5$.

Second, we can show that the entire system and the induced radiation come into thermal equilibrium at this time. In Figure \ref{beta} we plot $\beta$ versus $t/R_s$ for the entire system (continuous curve) and the induced radiation (dashed curve). Figure \ref{beta} shows that for the time $t/R_s\approx7.5$ the values of the two $\beta$'s become approximately equal, meaning that the entire system and the induced radiation are now at the same temperature. Therefore the system is now in thermal equilibrium, meaning that there is no more change in entropy of the domain wall as $t/R_s$ increases. Further more, the fluctuations (departure from thermality) in $\beta$ become very small at this time, as discussed in Refs.\cite{Stojkovic,Greenwood}.

\begin{figure}[htbp]
\includegraphics{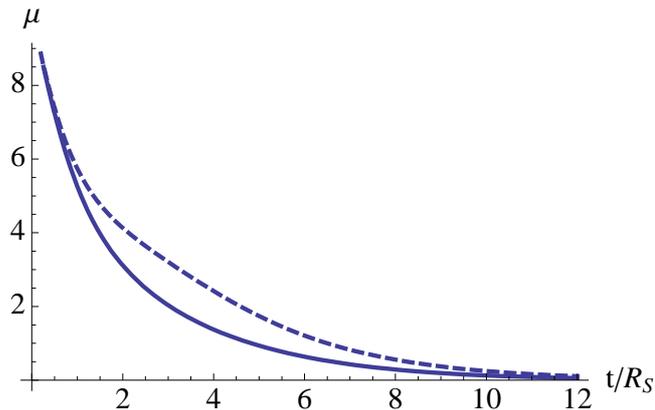}
\caption{We plot $\mu$ versus $t/R_s$. The solid line corresponds to the entire system while the dashed line corresponds to the induced radiation only. Here we see that as $t/R_s$ increases, the chemical potential for each goes to zero.}
\label{ChemPot}
\end{figure}

Finally we can evaluate the the chemical potential for both the entire system and for the induced radiation. From definition we can write the chemical potential as
\be
  \mu=\frac{\partial S}{\partial N}.
\ee
In Figure \ref{ChemPot} we plot the chemical potential for both the entire system and for the induced radiation. We can see that as $t/R_s$ increases the chemical potential of the entire system and the induced radiation goes to zero. This means that the dispersion of particles goes to zero and the system goes into equilibrium. 

During the dynamical process, the entropy increases almost linearly. If one applies a best-fit line, we see that the entropy oscillates about the best line. These oscillations may be attributed to several different circumstances. First, the oscillations may be caused by the non-thermal property of the radiation (see Ref.\cite{Stojkovic}). Secondly, these oscillations may be a manifestation of the error associated with the numerical calculations. Lastly, the oscillations may be an artifact of expanding the calculations beyond the region of validity, since we are using the near horizon approximation. Hence for values large compared to $R_s$, we cannot completely trust our result.

\section{Discussion}

Here we have shown that the entropy of the collapsing domain wall and the entropy of the radiation given off during the time of collapse are in agreement with the generalized second law of black hole thermodynamics. The results of Figure \ref{EntEntire} are clearly in agreement with Eq.(\ref{GSL}). The results of Figure \ref{EntShell} are in agreement with the results of Hawking and Gibbons, Eq.(\ref{SBH}), that the entropy of the black hole is in fact finite and proportional to the area of the event horizon. 

Note, here we do not discuss or explain the origin of Eq.(\ref{SBH}), we merely verify that our model gives the correct result. The origin of Eq.(\ref{SBH}) is still not understood, however, many attempts have been made to make sense of this result (see for example Refs.\cite{Strominger,tHooft,Susskind}). However, the answer to this question may lie in understanding the entanglement nature between the particles inside and outside of the event horizon, see for example Refs.\cite{Sorkin,Bombelli,Frolov}.

%% file: BR.tex
\chapter{Back Reaction}
\label{ch:BR}

In this section we make some general comments on how one can include the effect of back reaction for an infalling observer during gravitational collapse. In this section we do not completely solve the equations of motion for the included back reaction, we merely set up the situation and make some comments about it. 

To incorporate back reaction into gravitational collapse, one must consider the entire Hamiltonian, as in Chapter \ref{ch:entropy}, as well as the interaction Hamiltonian between the domain wall and the induced radiation. 

Thus the total Hamiltonian is given by
\begin{align}
  H=&H_{Wall}+H_{Rad}+H_{Int}\nonumber\\
   =&4\pi \sigma R^2 [ \sqrt{1+R_\tau^2} - 2\pi G\sigma R] +\sum_{modes}\left[\sqrt{1+R_{\tau}^2}\frac{\Pi_b}{2m}+\frac{|R_{\tau}|}{2B}Kb^2\right]+T_{\mu\nu}S^{\mu\nu}
\end{align}
where $H_{Int}=T_{\mu\nu}S^{\mu\nu}$ is the interaction Hamiltonian and $T_{\mu\nu}$ and $S^{\mu\nu}$ are the energy-momentum tensors for the radiation and the domain wall, respectively, which are given by
\begin{align}
  T_{\mu\nu}=&\int d^4x\left[\frac{1}{2}g_{\mu\nu}g^{\alpha\beta}\partial_{\alpha}\Phi\partial_{\beta}\Phi-\partial_{\mu}\Phi\partial_{\nu}\Phi\right]\nonumber\\
                     =&2\pi^2\int dt\int drr^2\left[\frac{1}{2}g_{\mu\nu}g^{\alpha\beta}\partial_{\alpha}\Phi\partial_{\beta}\Phi-\partial_{\mu}\Phi\partial_{\nu}\Phi\right]
   \label{T_rad}
\end{align}
and
\be
  S^{\mu\nu}\sqrt{-g}=\sigma\int d^3\xi\gamma^{ab}\partial_aX^{\mu}\partial_bX^{\nu}\delta^{(4)}(X^{\sigma}-X^{\sigma}(\xi^a)).
  \label{S_shell}
\ee
From the expansion of the scalar field in Eq.(\ref{Phi_Exp}), we can see that the stress-energy tensor for the scalar field takes on the form 
\be
  (T_{\mu\nu})=\left[ \begin{array}{cccc}
     T_{00} & T_{01} & 0& 0 \\ T_{10} & T_{11} &0 &0 \\ 0 & 0 & T_{22} & 0 \\ 0 & 0 & 0 & T_{33} 
  \end{array} \right].
\ee
While from Eq.(\ref{S_shell}) we see that the stress-energy tensor for the domain wall takes the form,
\be
  (S^{\mu\nu})=\left[ \begin{array}{cccc}
     S^{00} & 0 & 0& 0 \\ 0 & S^{11} &0 &0 \\ 0 & 0 & S^{22} & 0 \\ 0 & 0 & 0 & 0 
  \end{array} \right].
\ee
Hence we see that the interaction Hamiltonian doesn't contain any off-diagonal terms.

\section{Stress-Energy Tensor}

Here we develop the stress-energy tensor for the radiation and the domain wall, respectively. First we will discuss the stress-energy tensor for the induced radiation. Second we will discuss the stress-energy tensor for the domain wall.

\subsection{Radiation Stress-Energy Tensor}

Here we examine the stress-energy tensor for the radiation. From the discussion in Chapter \ref{ch:radiation} we can write the stress-energy tensor as
\begin{align}
  T_{\mu\nu}=&4\pi\int d\tau\Big{[}\frac{1}{2}g_{\mu\nu}\Big{(}-\int_0^{R(\tau)} drr^2\frac{1}{\sqrt{1+R_{\tau}^2}}(\partial_{\tau}\Phi)^2-\int_{R(\tau)}^{\infty} drr^2\frac{B}{\sqrt{B+R_{\tau}^2}}\frac{(\partial_{\tau}\Phi)^2}{1-R_s/r}\nonumber\\
    &+\int_0^{R(\tau)} drr^2\sqrt{1+R_{\tau}^2}(\partial_r\Phi)^2+\int_{R(\tau)}^{\infty} drr^2\frac{\sqrt{B+R_{\tau}^2}}{B}\left(1-\frac{R_s}{r}\right)(\partial_r\Phi)^2\Big{)}\nonumber\\
     &-\int_0^{\infty} drr^2\partial_{\mu}\Phi\partial_{\nu}\Phi\Big{]}
\end{align}
Now using the metric we can write the individual terms, which are given as
\begin{align}
  T_{00}=&4\pi\int d\tau\Big{[}-\frac{1}{2}\Big{(}-\int_0^{R(\tau)} drr^2\left(1-\frac{R_s}{r}\right)\frac{1}{\sqrt{1+R_{\tau}^2}}(\partial_{\tau}\Phi)^2\nonumber\\
    &-\int_{R(\tau)}^{\infty} drr^2\frac{B}{\sqrt{B+R_{\tau}^2}}(\partial_{\tau}\Phi)^2+\int_0^{R(\tau)} drr^2\left(1-\frac{R_s}{r}\right)\sqrt{1+R_{\tau}^2}(\partial_r\Phi)^2\nonumber\\
    &+\int_{R(\tau)}^{\infty} drr^2\frac{\sqrt{B+R_{\tau}^2}}{B}\left(1-\frac{R_s}{r}\right)^2(\partial_r\Phi)^2\Big{)}-\int_0^{\infty} drr^2(\partial_{\tau}\Phi)^2\Big{]},
    \label{T00}
\end{align}
\begin{align}
  T_{01}=-4\pi\int d\tau\int drr^2\partial_{\tau}\Phi\partial_r\Phi,
  \label{T01}
\end{align}
\be
  T_{10}=-4\pi\int d\tau\int drr^2\partial_r\Phi\partial_{\tau}\Phi,
  \label{T10}
\ee
\begin{align}
  T_{11}=&4\pi\int d\tau\Big{[}\frac{1}{2}\Big{(}-\int_0^{R(\tau)} drr^2\frac{1}{\sqrt{1+R_{\tau}^2}}\frac{(\partial_{\tau}\Phi)^2}{1-R_s/r}\nonumber\\
    &-\int_{R(\tau)}^{\infty} drr^2\frac{B}{\sqrt{B+R_{\tau}^2}}\frac{(\partial_{\tau}\Phi)^2}{(1-R_s/r)^2}+\int_0^{R(\tau)} drr^2\sqrt{1+R_{\tau}^2}\frac{(\partial_r\Phi)^2}{1-R_s/r}\nonumber\\
    &+\int_{R(\tau)}^{\infty} drr^2\frac{\sqrt{B+R_{\tau}^2}}{B}(\partial_r\Phi)^2\Big{)}-\int_0^{\infty} drr^2(\partial_{r}\Phi)^2\Big{]},
    \label{T11}
\end{align}
\begin{align}
  T_{22}=&4\pi\int d\tau\Big{[}\frac{1}{2}\Big{(}-\int_0^{R(\tau)} drr^4\frac{1}{\sqrt{1+R_{\tau}^2}}(\partial_{\tau}\Phi)^2-\int_{R(\tau)}^{\infty} drr^4\frac{B}{\sqrt{B+R_{\tau}^2}}\frac{(\partial_{\tau}\Phi)^2}{1-R_s/r}\nonumber\\
    &+\int_0^{R(\tau)} drr^4\sqrt{1+R_{\tau}^2}(\partial_r\Phi)^2+\int_{R(\tau)}^{\infty} drr^4\frac{\sqrt{B+R_{\tau}^2}}{B}\left(1-\frac{R_s}{r}\right)(\partial_r\Phi)^2\Big{)}\Big{]},
    \label{T22}
\end{align}
and
\begin{align}
  T_{33}=&\frac{8}{3}\pi\int d\tau\Big{[}\frac{1}{2}\Big{(}-\int_0^{R(\tau)} drr^4\frac{1}{\sqrt{1+R_{\tau}^2}}(\partial_{\tau}\Phi)^2-\int_{R(\tau)}^{\infty} drr^4\frac{B}{\sqrt{B+R_{\tau}^2}}\frac{(\partial_{\tau}\Phi)^2}{1-R_s/r}\nonumber\\
    &+\int_0^{R(\tau)} drr^4\sqrt{1+R_{\tau}^2}(\partial_r\Phi)^2+\int_{R(\tau)}^{\infty} drr^4\frac{\sqrt{B+R_{\tau}^2}}{B}\left(1-\frac{R_s}{r}\right)(\partial_r\Phi)^2\Big{)}\Big{]}.
    \label{T33}
\end{align}
Here note that from Eqs.(\ref{T22}) and (\ref{T33}) show that $T_{33}=(2/3)T_{22}$. 

For a full analysis of the stress-energy tensor we will look in the near the horizon limit. Ideally we would like to extend this analysis to the near singularity limit as well. However, we are working in Schwarzschild coordinates, which we cannot extend to the near singularity limit due to the fact that the observer is being constantly accelerated (see Chapter \ref{ch:Classical}). We will then exam the behavior of the stress-energy tensor near the horizon, i.e. in the region $R\sim R_s$.

Of interest is the behavior of the stress-energy tensor near the horizon. To investigate the effect of the radiation we will change the limit of integration from $R(\tau)$ to $R_s$, allowing us to find the dominate terms in this regime. From Eq.(\ref{T00}) we see that in this limit and with the expansion in modes we have
\bead
  T_{00}&=&4\pi\int d\tau\Big{[}-\frac{1}{2}\Big{(}-\int_0^{R_s} drr^2\left(1-\frac{R_s}{r}\right)\frac{1}{\sqrt{1+R_{\tau}^2}}(\partial_{\tau}\Phi)^2\\
      &&+\int_{R_s}^{\infty} drr^2\frac{\sqrt{B+R_{\tau}^2}}{B}\left(1-\frac{R_s}{r}\right)^2(\partial_r\Phi)^2\Big{)}-\int_0^{\infty} drr^2(\partial_{\tau}\Phi)^2\Big{]}\\
       &=&\int d\tau\left[\frac{1}{2}\frac{1}{\sqrt{1+R_{\tau}^2}}\dot{a}_k\tilde{{\bf A}}_{kk'}\dot{a}_{k'}-\frac{1}{2}\frac{\sqrt{B+R_{\tau}^2}}{B}a_k\tilde{{\bf C}}_{kk'}a_{k'}-\frac{1}{2}\dot{a}_k\tilde{{\bf D}}_{kk'}\dot{a}_{k'}\right],
\eead
from Eq.(\ref{T01})
\bd
  T_{01}=-\int d\tau\dot{a}_k\tilde{{\bf E}}_{kk'}a_{k'},
\ed
from Eq.(\ref{T10})
\bd
  T_{10}=-\int d\tau a_k\tilde{{\bf E}}^{-1}_{kk'}\dot{a}_{k'},
\ed
from Eq.(\ref{T11})
\bead
  T_{11}&=&4\pi\int d\tau\Big{[}\frac{1}{2}\Big{(}-\int_0^{R_s} drr^2\frac{1}{\sqrt{1+R_{\tau}^2}}\frac{(\partial_{\tau}\Phi)^2}{1-R_s/r}\\
      &&+\int_{R_s}^{\infty} drr^2\frac{\sqrt{B+R_{\tau}^2}}{B}(\partial_r\Phi)^2\Big{)}-\int_0^{\infty} drr^2(\partial_{r}\Phi)^2\Big{]}\\
       &=&\int d\tau\left[-\frac{1}{2}\frac{1}{\sqrt{1+R_{\tau}^2}}\dot{a}_k\tilde{{\bf F}}_{kk'}\dot{a}_{k'}+\frac{1}{2}\frac{\sqrt{B+R_{\tau}^2}}{B}a_k\tilde{{\bf G}}_{kk'}a_{k'}-\frac{1}{2}a_k\tilde{{\bf H}}_{kk'}a_{k'}\right],
\eead
and from Eq.(\ref{T22}) we have
\bead  
    T_{22}&=&4\pi\int d\tau\left[\frac{1}{2}\Big{(}-\int_0^{R_s} drr^4\frac{1}{\sqrt{1+R_{\tau}^2}}(\partial_{\tau}\Phi)^2+\int_{R_s}^{\infty} drr^4\frac{\sqrt{B+R_{\tau}^2}}{B}\left(1-\frac{R_s}{r}\right)(\partial_r\Phi)^2\Big{)}\right]\\
    &=&\int d\tau\left[-\frac{1}{2}\frac{1}{\sqrt{1+R_{\tau}^2}}\dot{a}_k\tilde{{\bf J}}_{kk'}\dot{a}_{k'}+\frac{1}{2}\frac{\sqrt{B+R_{\tau}^2}}{B}a_k\tilde{{\bf K}}_{kk'}a_{k'}\right],
\eead
where the matrices are defined by
\begin{align}
  \tilde{{\bf A}}_{kk'}=&4\pi\int_0^{R_s}drr^2\left(1-\frac{R_s}{r}\right)f_kf_{k'},\\
  \tilde{{\bf C}}_{kk'}=&4\pi\int_{R_s}^{\infty}drr^2\left(1-\frac{R_s}{r}\right)^2f'_kf'_{k'},\\
  \tilde{{\bf D}}_{kk'}=&8\pi\int_0^{\infty}drr^2f'_kf_{k'},\\
  \tilde{{\bf E}}_{kk'}=&4\pi\int_0^{\infty}drr^2f_kf'_{k'},\\
  \tilde{{\bf F}}_{kk'}=&4\pi\int_0^{R_s}drr^2\left(1-\frac{R_s}{r}\right)^{-1}f_kf_{k'},\\
  \tilde{{\bf G}}_{kk'}=&4\pi\int_{R_s}^{\infty}drr^2f'_kf'_{k'},\\
  \tilde{{\bf H}}_{kk'}=&8\pi\int_0^{\infty}drr^2f'_kf'_{k'},\\
  \tilde{{\bf J}}_{kk'}=&4\pi\int_0^{R_s}drr^4f_kf_{k'},\\
  \tilde{{\bf K}}_{kk'}=&4\pi\int_{R_s}^{\infty}drr^4\left(1-\frac{R_s}{r}\right)f'_kf'_{k'}.
  \label{Matrices}
\end{align}
To further investigate the problem, for a moment let us assume that the basis functions are planewaves. This is a valid approximation in the asymptotic regime, however, this will give us some insight into the problem here. For the basis functions as plane waves we have
\bd
  f_k=e^{ikr},
\ed
however since we are requiring real basis functions then we will take the real part of this. Therefore we can write
\bd
  Re\left(\int_0^{\infty}drr^2f_k'f_{k'}\right)=Re\left(\int_0^{\infty}drr^2e^{i(k+k')r}\right).
\ed
Performing the integral over $r$ we then have
\bd
  Re\left(\int_0^{\infty}drr^2f_k'f_{k'}\right)=\delta(k+k')
\ed
which is finite. Hence the other terms in Eqs.(\ref{T00}), (\ref{T11}), (\ref{T22}) and (\ref{T33}) are dominate due to the divergences of these terms. Therefore we can ignore these extra terms, so we then have
\bead
  T_{00}&=&\int d\tau\left[\frac{1}{2}\sqrt{1+R_{\tau}^2}\dot{a}_k(\tilde{{\bf A}}^{-1})_{kk'}\dot{a}_{k'}-\frac{1}{2}\frac{\sqrt{B+R_{\tau}^2}}{B}a_k\tilde{{\bf C}}_{kk'}a_{k'}\right],\\
  T_{01}&=&-\int d\tau\dot{a}_k\tilde{{\bf E}}_{kk'}a_{k'},\\
  T_{10}&=&-\int d\tau a_k\tilde{{\bf E}}^{-1}_{kk'}\dot{a}_{k'},\\
  T_{11}&=&\int d\tau\left[-\frac{1}{2}\sqrt{1+R_{\tau}^2}\dot{a}_k(\tilde{{\bf F}}^{-1})_{kk'}\dot{a}_{k'}+\frac{1}{2}\frac{\sqrt{B+R_{\tau}^2}}{B}a_k\tilde{{\bf G}}_{kk'}a_{k'}\right],\\
  T_{22}&=&\int d\tau\left[-\frac{1}{2}\sqrt{1+R_{\tau}^2}\dot{a}_k(\tilde{{\bf J}}^{-1})_{kk'}\dot{a}_{k'}+\frac{1}{2}\frac{\sqrt{B+R_{\tau}^2}}{B}a_k\tilde{{\bf K}}_{kk'}a_{k'}\right],
\eead
which is of the same form as $H_{rad}$. We can see that the matrices in Eq.(\ref{Matrices}) are just multiples of the matrices in Eq.(\ref{AC}). Therefore we can see that the eigenvalues of Eq.(\ref{Matrices}) are multiples of those of Eq.(\ref{AC}), hence we can simultaneously diagonalize the matrices as we did in Chapter \ref{ch:radiation}. Finally we can write
\begin{align}
  T_{00}=&\int d\tau\left[-\frac{1}{2}\frac{\sqrt{1+R_{\tau}^2}}{nm}\frac{\partial^2}{\partial b^2}-\frac{nK}{2}\frac{\sqrt{B+R_{\tau}^2}}{B}b^2\right],\label{T_00n}\\
  T_{01}=&-\int d\tau\dot{a}_k\tilde{{\bf E}}_{kk'}a_{k'},\label{T_01n}\\
  T_{10}=&-\int d\tau a_k\tilde{{\bf E}}^{-1}_{kk'}\dot{a}_{k'},\label{T_10n}\\
  T_{11}=&\int d\tau\left[\frac{1}{2}\frac{n\sqrt{1+R_{\tau}^2}}{m}\frac{\partial^2}{\partial b^2}+\frac{K}{2n}\frac{\sqrt{B+R_{\tau}^2}}{B}b^2\right],\label{T_11n}\\
  T_{22}=&\int d\tau\left[\frac{1}{2}\frac{\sqrt{1+R_{\tau}^2}}{m}\frac{\partial^2}{\partial b^2}+\frac{K}{2}\frac{\sqrt{B+R_{\tau}^2}}{B}b^2\right]\label{T_22n},
\end{align}
where as in Chapter \ref{ch:radiation}, $m$ and $K$ are eigenvalues, $b$ are the eigenmodes and $n$ is a constant multiple. Here we note that $T_{00}$, $T_{11}$ and $T_{22}$ have the structure of a Harmonic oscillator. 

We now calculate the expectation value of the stress-energy tensor. To do this we consider 
\be
  \langle T_{\mu\nu}\rangle=\langle0|T_{\mu\nu}|0\rangle
\ee
where $|0\rangle$ is the vacuum state. Since the components of the stress-energy tensor have the structure of a harmonic oscillator we take that the vacuum state is the ground state of the harmonic oscillator (see Chapter \ref{ch:radiation}). The ground state of the harmonic oscillator is given by Eq.(\ref{HO_basis}), thus the expectation value is 
\be
  \langle T_{\mu\nu}\rangle=\int db\left(\frac{m\omega_0}{\pi}\right)^{1/2}e^{-m\omega_0b^2/2}T_{\mu\nu}e^{-m\omega_0b^2/2}.
\ee
From the structure of Eqs.(\ref{T_00n})-(\ref{T_22n}), we can see that there is a kinetic term and a potential term which all have the same dependence on the eigenmode $b$. So, using Eq.(\ref{HO_basis}) we can write
\bead
  \text{Kinetic Term}&=&\int db\left(e^{-m\omega_0b^2/2}\frac{\partial^2}{\partial b^2}e^{-m\omega_0b^2/2}\right)\\
         &=&-\frac{\sqrt{m\omega_0\pi}}{4},
\eead 
and
\bead
  \text{Potential Term}&=&\int db\left(e^{-m\omega_0b^2/2}b^2e^{-m\omega_0b^2/2}\right)\\
        &=&\frac{\sqrt{\pi}}{4(m\omega_0)^{3/2}}
\eead
where we used the fact that there are an infinite number of eigenmodes $b$ (hence the integrals are from zero to infinity). The individual components are then,
\begin{align}
  \langle T_{00}\rangle=&\int d\tau\left[\frac{1}{2}\frac{\sqrt{1+R_{\tau}^2}}{nm}\left(\frac{\sqrt{m\omega_0}}{4}\right)-\frac{nK}{2}\frac{\sqrt{B+R_{\tau}^2}}{B}\left(\frac{\sqrt{\pi}}{4(m\omega_0)^{3/2}}\right)\right],\label{T_00nh}\\
  \langle T_{11}\rangle=&\int d\tau\left[-\frac{1}{2}\frac{n\sqrt{1+R_{\tau}^2}}{m}\left(\frac{\sqrt{m\omega_0}}{4}\right)-\frac{K}{2n}\frac{\sqrt{B+R_{\tau}^2}}{B}\left(\frac{\sqrt{\pi}}{4(m\omega_0)^{3/2}}\right)\right],\label{T_11nh}\\
  \langle T_{22}\rangle=&\int d\tau\left[-\frac{1}{2}\frac{\sqrt{1+R_{\tau}^2}}{m}\left(\frac{\sqrt{m\omega_0}}{4}\right)-\frac{K}{2}\frac{\sqrt{B+R_{\tau}^2}}{B}\left(\frac{\sqrt{\pi}}{4(m\omega_0)^{3/2}}\right)\right]\label{T_22nh}.
\end{align}

Investigating Eqs.(\ref{T_00nh})-(\ref{T_22nh}) we can see that as $R\rightarrow R_s$ the potential term diverges such as in the Hamiltonian of the induced radiation. Therefore we can conclude that the components of the stress-energy tensor are divergent, however, not due to the usual reasons. Typically this divergence of the stress-energy tensor is associated with the divergence in the frequency $\omega$ (see for example Refs.\cite{Davies,Unruh,Birrell}). To get around this divergence, one usually either applies a cut-off for the allowed frequency or applies a renormalization technique that makes the stress-energy tensor finite. Here we can see that this process is not needed since the divergence is not due to the frequency (since we never specify the basis functions), however the divergence is due to the metric itself.

The divergence in Eqs.(\ref{T_00nh})-(\ref{T_22nh}) in the regime $R\sim R_s$ is due to the $B$ term in the potential term. As stated earlier, Chapters \ref{ch:Classical} and \ref{ch:radiation}, this is due to the fact that we are using Schwarzschild coordinates. The Schwarzschild observer is in an accelerated reference frame, which causes the divergence. Therefore, as we saw in Chapter \ref{ch:radiation}, to study the question of backreaction the more appropriate observer to use would be a truly free-falling observer such as an Eddington-Finkelstein observer. However, we will not investigate such an observer here.

\subsection{Stress-Energy of the domain wall}

From Eq.(\ref{S_shell}) we can write the determinant of the induced metric as
\begin{align}
  \sqrt{-\gamma}&=R^2\sin\theta\sqrt{B-\frac{\dot{R}^2}{B}}\nonumber\\
         &=R^2\sin\theta\sqrt{B-\frac{BR_{\tau}^2}{B+R_{\tau}^2}}
\end{align}
where we used Eq.(\ref{dtdtau}). The stress-energy tensor for the domain can then be written as
\be
  S^{\mu\nu}=-\sigma\gamma^{ab}
\ee
where $\gamma^{ab}$ is again the induced metric on the surface of the domain wall. This is expected from Eq.(\ref{Stress-Energy}) in the case of the domain wall ($\sigma=\eta$).

\section{Quantum Hamiltonian}

Here we wish to find an appropriate way to take into account the fact that the mass of the domain wall is changing, due to the fact that the radiation is taking mass away from the system. To do this we will follow a technique that was first introduced in Ref.\cite{Vach}.

Eq.(\ref{Mass_tau}) tells us that the mass of the domain wall is approximately the Hamiltonian, therefore we can write the Schwarzschild radius as
\be
  R_s\rightarrow2GH_{Wall}.
  \label{RsH}
\ee
The factor $B$ in the Hamiltonian for the radiation contains the energy of the wall via the Schwarzschild radius. So we then have
\be
  B=1-\frac{2GH_{Wall}}{R}.
  \label{BH}
\ee

In the near horizon limit Eq.(\ref{Mass_tau}) can be written as
\be
  M = 4\pi \sigma R_s^2 [ \sqrt{1+R_\tau^2} - 2\pi G\sigma R_s] =H_{wall}.
\ee
Assuming that the velocity at the horizon is small and dropping the constant terms we can write this as
\be
  H_{wall}=2\pi \sigma R_s^2R_{\tau}^2.
\ee
Using Eq.(\ref{RsH}) we can then rewrite this as
\begin{align}
  H_{wall}&=\frac{1}{8\pi\sigma G^2R_{\tau}^2}\\
     &=\left(\frac{\Pi_R^2}{16\pi\sigma G^2}\right)^{1/3}
\end{align}
where in the second line we used Eq.(\ref{Pi_tau}). 

The total Hamiltonian in terms of a single mode then becomes
\begin{align}
  H=&\left(\frac{\Pi_R^2}{8\pi\sigma G^2}\right)^{1/3}+\left(1+\left(\frac{1}{1024\pi\sigma G^2\Pi_R}\right)^{2/3}\right)\frac{\Pi^2_b}{2m}\nonumber\\
     &+\frac{RKb^2}{2\left((16\pi\sigma G^2)^{1/3}R-2G(\Pi^2_R)^{1/3}\right)(\Pi_R^2)^{2/3}}+T_{\mu\nu}S^{\mu\nu}.
     \label{Ham_tot_BR}
\end{align}
Here we note an unusual property of Eq.(\ref{Ham_tot_BR}), the appearance of the fractional derivatives. In general, fractional derivatives are non-local, that is, one cannot say that the fractional derivative at a point $x$ of a function $f$ depends only on the graph of $f$ very near $x$, see for example Ref.\cite{Wiki}. Therefore it is expected that the theory of fractional derivatives involves some sort of boundary conditions, involving information further out. The most general definition of the fractional derivative is
\be
  _aD^q_t=\begin{cases} \frac{d^q}{dx^q}, & Re(q)>0\\
          1, & Re(a)=0\\
          \int_a^t(dx)^{-q}, & Re(a)<0. \end{cases}
\ee
Here the first case is defined as
\be
  \frac{d^q}{dx^q}x^k=\frac{k!}{(k-a)!}x^{k-q}=\frac{\Gamma(k+1)}{\Gamma(k-q+1)}x^{k-q},
\ee
while the third case is defined as
\be
  \int_a^t(dx)^{|q|}f(x)=\frac{1}{\Gamma(q)}\int_0^x(x-t)^{|q|-1}f(t)dt.
\ee
Hence Eq.(\ref{Ham_tot_BR}) is an differential-integral equation.

The study of the behavior backreaction is therefore very complicated. However, the interesting thing to point out here is that, similarly to the investigation of the quantum mechanical effects studied in Chapter \ref{ch:quantum}, the presence of the non-locality again emerges. However, in Chapter \ref{ch:quantum} the non-locality was only present when investigating the near classical singularity regime. Here, the non-local effect is even present in the near horizon regime.

\section{Discussion}

In this chapter we investigated the stress-energy tensor for the radiation given off during collapse as well as investigated a way to include the loss of mass during this collapse. We found some interesting properties of these two quantities.

First for the stress-energy tensor for the radiation, we found that the expectation value for the stress-energy tensor is in fact infinite. However, this is not due to the usual difficulties. Generally when one investigates the stress-energy tensor, the infinities arise from the basis function. Traditionally one assumes a plane-wave basis function for the radiation, and the divergence is therefore due to the frequency of the basis function. To avoid these infinities, one usually institutes a cut-off frequency. Here, we do not have this problem. This is due to the fact that we never actually specify our basis functions, hence we do not have the problem of infinities in the basis function. The divergence in this case is due to the presence of the $B^{-1}$ in the potential term. As $R\rightarrow R_s$, $B\rightarrow0$ which causes the divergence. As stated in Chapters \ref{ch:Classical} and \ref{ch:radiation} this is due to the fact that we are using Schwarzschild coordinates, where the observer is being accelerated. 

To include the loss of mass into the Hamiltonian of the system, we used the technique originally developed in Ref.\cite{Vach}. Here one uses the approximation that the Hamiltonian of the domain wall is approximately the mass of the domain wall. Therefore one can replace the mass in the Schwarzschild radius by the Hamiltonian of the wall. Using this, we can then rewrite the total Hamiltonian as in Eq.(\ref{Ham_tot_BR}). The interesting thing here is that the Hamiltonian is now in terms of fraction, not whole or partial, derivatives. By definition, fractional derivatives are not strictly local quantities and will either give a differential or integral equation depending on the sign of the fractional derivative. As in Chapter \ref{ch:quantum}, we recover the non-locality of the quantum effects during gravitational collapse. Unlike in Chapter \ref{ch:quantum}, these effects are now manifest even near the horizon. 

%% file: Invariant.tex
\chapter{Invariant Method and the Schr\"odinger Equation}
\label{ch:Invariant}

In this section we breifly review the invariant operator method developed by Lewis and Reisenfeld in Ref.\cite{Lewis} as a solution to the time-dependent Schr\"odinger Equation.

Consider a system whose Hamiltonian operator $H(t)$ is an explicit function of time, and assume the existence of another explicitly time-dependent non-trivial Hermitian operator $I(t)$, which is invariant. To say that $I(t)$ is invariant means that $I(t)$ satisfies the Liouville-von Neumann equation
\be
  \frac{dI}{dt}=\frac{\partial I}{\partial t}+\frac{1}{i}\left[I,H\right]=0
  \label{dIdt}
\ee
and since $I(t)$ is Hermitian we have
\be
  I^{\dagger}=I.
\ee
Here we will consider the analysis for a state vector $\big{|}\psi\rangle$, however, in general this also works for a wavefunction since $\psi(x)=\langle x\big{|}\psi\rangle$. We can then write the Schr\"odinger equation as
\be
  H(t)\big{|}\psi\rangle=i\frac{\partial}{\partial t}\big{|}\psi\rangle.
\ee
By operating with the left-hand side of Eq.(\ref{dIdt}) on the state vector and using the Schr\"odinger equation, we obtain the relation
\be
  i\frac{\partial}{\partial t}\left(I\big{|}\psi\rangle\right)=H\left(I\big{|}\psi\rangle\right),
\ee
which implies that the action of the invariant operator on a Schr\"odinger state vector produces another solution of the Schr\"odinger equation. In general, this result is valid for any invariant, even if the invariant involves the operation of time differentiation. However, for our purposes, we shall consider invariants which do not involve time differentiation. This choice allows one to derive simple and explicit rules for choosing the phases of the eigenstates of $I(t)$ such that these states themselves satisfy the Schr\"odginer equation. 

Assume that the invariant is one of a complete set of commuting observables, so that there is a complete set of eigenstates of $I$. Denote the eigenvalues of $I$ by $\lambda$, and the orthonormal eigenstates associated with a given $\lambda$ by $\big|\lambda,\kappa\rangle$, where $\kappa$ represents all of the quantum numbers other than $\lambda$ that are necessary for specifying the eigenstates:
\begin{align}
  I(t)\big|\lambda,\kappa\rangle=&\lambda\big|\lambda,\kappa\rangle\label{actionI}\\
  \langle\lambda',\kappa'\big|\lambda,\kappa\rangle=&\delta_{\lambda'\lambda}\delta_{\kappa'\kappa}.
\end{align}
Since the invariant is Hermitian, the eigenvalues $\lambda$ are real. They are also time-independent as we shall now see. By differentiating Eq.(\ref{actionI}) with respect to time, we obtain
\be
  \frac{\partial I}{\partial t}\big|\lambda,\kappa\rangle+I\frac{\partial}{\partial t}\big|\lambda,\kappa\rangle=\frac{\partial\lambda}{\partial t}\big|\lambda,\kappa\rangle+\lambda\frac{\partial}{\partial t}\big|\lambda,\kappa\rangle.
  \label{dIdt_full}
\ee
Using Eq.(\ref{dIdt}) we can write
\be
  i\frac{\partial I}{\partial t}\big|\lambda,\kappa\rangle+IH\big|\lambda,\kappa\rangle-\lambda H\big|\lambda,\kappa\rangle=0.
  \label{dIdt_action}
\ee
The scalar product of Eq.(\ref{dIdt_action}) with a state $\big|\lambda',\kappa'\rangle$ is 
\be
  i\langle\lambda',\kappa'\big{|}\frac{\partial I}{\partial t}\big|\lambda,\kappa\rangle+(\lambda'-\lambda)\langle\lambda',\kappa'\big{|}H\big|\lambda,\kappa\rangle=0
  \label{eigen}
\ee
which then implies 
\be
  \langle\lambda',\kappa'\big{|}\frac{\partial I}{\partial t}\big|\lambda,\kappa\rangle=0.
\ee
Taking the scalar product of Eq.(\ref{dIdt_full}) with $\big|\lambda,\kappa\rangle$, we obtain
\be
  \frac{\partial\lambda}{\partial t}=\langle\lambda,\kappa\big{|}\frac{\partial I}{\partial t}\big|\lambda,\kappa\rangle=0.
  \label{dlambdadt}
\ee
Since the eigenvalues are time-independent, it is clear that the eigenstates must be time-dependent.

To investigate the connection between the eigenstates of $I$ and the solutions so the Schr\"odinger equation, we first write the equation of motion of $\big|\lambda,\kappa\rangle$ starting from Eq.(\ref{dIdt_full}) and using Eq.(\ref{dlambdadt}):
\be
  (\lambda-I)\frac{\partial}{\partial t}\big|\lambda,\kappa\rangle=\frac{\partial I}{\partial t}\big|\lambda,\kappa\rangle.
\ee
By taking the scalar production with $\big|\lambda',\kappa'\rangle$ and using Eq.(\ref{eigen}) to eliminate 
\bd
  \langle\lambda',\kappa'\big{|}\frac{\partial I}{\partial t}\big|\lambda,\kappa\rangle
\ed
we get
\be
  i(\lambda-\lambda')\langle\lambda',\kappa'\big{|}\frac{\partial}{\partial t}\big|\lambda,\kappa\rangle=(\lambda-\lambda')\langle\lambda',\kappa'\big{|}H\big|\lambda,\kappa\rangle.
  \label{eigen_relation}
\ee
From this, for $\lambda'\not=\lambda$, we infer
\be
  i\langle\lambda',\kappa'\big{|}\frac{\partial}{\partial t}\big|\lambda,\kappa\rangle=\langle\lambda',\kappa'\big{|}H\big|\lambda,\kappa\rangle.
  \label{partial_H}
\ee
Eq.(\ref{eigen_relation}) does not imply
\bd
  i\langle\lambda',\kappa'\big{|}\frac{\partial}{\partial t}\big|\lambda,\kappa\rangle=\langle\lambda',\kappa'\big{|}H\big|\lambda,\kappa\rangle.
\ed
If Eq.(\ref{partial_H}) held for $\lambda'=\lambda$ as well as for $\lambda'\not=\lambda$, then we would immediately deduce that $\big|\lambda,\kappa\rangle$ satisfies the Schr\"odinger equation, that is $\big|\lambda,\kappa\rangle$ is a special case of $\big{|}\psi\rangle$.

Note that the phase of $\big|\lambda,\kappa\rangle$ has not been fixed by our definitions. We are still free to multiply $\big|\lambda,\kappa\rangle$ by an arbitrary time-dependent phase factor. Thus, we can define a new set of eigenvectors of $I(t)$ related to our initial set by a time-dependent gauge transformation 
\be
  \big|\lambda,\kappa\rangle_{\alpha}=e^{i\alpha_{\lambda\kappa}(t)}\big|\lambda,\kappa\rangle,
\ee
where the $\alpha_{\lambda\kappa}(t)$ are arbitrary real functions of time. Because $I(t)$ is assumed not to contain time-derivative operators, the $\big|\lambda,\kappa\rangle_{\alpha}$ are orthonormal eigenstates of $I(t)$ just as are the $\big|\lambda,\kappa\rangle$. For $\lambda'\not=\lambda$, Eq.(\ref{eigen_relation}) also holds for matrix elements taken with respect to the new eigenstates. Each of the new eigenstates will statisfy the Schr\"odinger if we choose the phases $\alpha_{\lambda\kappa}(t)$ such that Eq.(\ref{eigen_relation}) holds for $\lambda'=\lambda$. This requirement is equivalent to the following first-order differential equation for the $\alpha_{\lambda\kappa}(t)$:
\be
  \delta_{\kappa\kappa'}\frac{d\alpha_{\lambda\kappa}}{dt}=\langle\lambda,\kappa'\big{|}i\frac{\partial}{\partial t}-H\big|\lambda,\kappa\rangle.
\ee
Since each of the new set of eigenstates of $I(t)$, $\big|\lambda,\kappa\rangle_{\alpha}$, satisfies the Schr\"odinger equation, the general solution is 
\be
  \big{|}t\rangle=\sum_{\lambda,\kappa}c_{\lambda\kappa}e^{i\alpha_{\lambda\kappa}(t)}\big|\lambda,\kappa;t\rangle,
  \label{t_state}
\ee
where the $c_{\lambda\kappa}$ are time-independent coefficients. All of the state vectors with which we have dealt so far are time-dependent, while in Eq.(\ref{t_state}) we modified the notation to indicate the dependence on time explicitly. The Schro\"odinger state vector is now denoted by $\big{|}t\rangle$ and the eigenstates of the invariant by $\big|\lambda,\kappa;t\rangle$.

Assume that in the remote past the Hamiltonian $H(t)$ is a constant operator $H(-\infty)$ having a complete, orthonormal set of time-independent eigenstates $\big{|}n;i\rangle$, $n$ being a label for all relevant quantum numbers and $i$ standing for ``initial state." Similiarly, assume that in the remote future, the Hamiltonian is a constant operator $H(\infty)$ and it possesses time-independent eigenstates $\big{|}m;f\rangle$, $m$ labeling the quantum numbers and $f$ standing for ``final state." The explicit time variation of $H(t)$ for intermediate times is arbitrary except for piecewise continuity; in particular, we do not exclude the possibility of variations rapid enough to render an analysis in terms of quasistationary states of $H(t)$ impossible. 

We want to calculate the transition amplitude $T(n\rightarrow m)$ connecting an initial state $\big{|}n;i\rangle$ to a final state $\big{|}m;f\rangle$. Thus we consider the case in which the Schr\"odinger state vector $\big{|}-\infty\rangle$ corresponds to an eigenstate $\big{|}n;i\rangle$. The superposition coefficients of Eq.(\ref{t_state}) for this problem are given by
\be
  c_{\lambda\kappa}=e^{-i\alpha_{\lambda\kappa}(-\infty)}\langle\lambda,\kappa;-\infty\big|n;i\rangle
\ee
from which we obtain
\be
  \big|t\rangle=\sum_{\lambda,\kappa}\exp\left(i\left[\alpha_{\lambda\kappa}(t)-\alpha_{\lambda\kappa}(-\infty)\right]\right)\big|\lambda,\kappa;t\rangle\langle\lambda,\kappa;-\infty\big|n;i\rangle.
\ee
The transition amplitude is therefore given by
\begin{align}
  T(n\rightarrow m)&=\langle m;f\big|\infty\rangle\nonumber\\
      &=\sum_{\lambda,\kappa}\exp\left(i\left[\alpha_{\lambda\kappa}(\infty)-\alpha_{\lambda\kappa}(-\infty)\right]\right)\langle m;f\big|\lambda,\kappa;\infty\rangle\langle\lambda,\kappa;-\infty\big|n;i\rangle.
      \label{trans_amp}
\end{align}

The properties of $I(t)$ apply equally well to any operator that is an invariant corresponding to a given $H(t)$. In general, for a system of $f$ degrees of freedom, there is an infinite family of such invariants, the members of which are functions of a set of $f$ independent invariants. Two such invariants will, in general, have different eigenstates, different time derivatives, and different commutators with the Hamiltonian. However, one must arrive at the same physical results no matter what invariant we use and, therefore, the choice of which particular invariant to use may be made on the basis of mathematical convenience. Here we demonstrate that the physical result is independent of the choice of invariant, we give a direct proof that a transition amplitude, such as in Eq.(\ref{trans_amp}) is indeed independent of our choice of invariant.

Suppose that we have two complete orthonormal sets of states, $\big{|}v;t\rangle$ and $\big{|}w;t\rangle$, all of which satisfy the time-dependent Schr\"odinger equation; and suppose that the states $\big|v;t\rangle$ are eigenstates of one set of operators, whose eigenvalues are labeled by $v$, and that the states $\big|w;t\rangle$ are eigenstates of a different set of operators, whose eigenvalues are labeled by $w$. The transition amplitude $T(n\rightarrow m)$ can be expressed as
\be
  T(n\rightarrow m)=\sum_v\langle m;f\big|v;\infty\rangle\langle v;-\infty\big|n;i\rangle
  \label{trans_amp1}
\ee 
or as
\be
  T(n\rightarrow m)=\sum_w\langle m;f\big|w;\infty\rangle\langle w;-\infty\big|n;i\rangle.
  \label{trans_amp2}
\ee
We want to show directly that these two expressions are the same. The completeness of the states $\big|w;t\rangle$ requires
\be
  \big|v;t\rangle=\sum_w\big|w;t\rangle\langle w;t\big|v;t\rangle.
\ee
Operating on this equation with $(i(\partial/\partial t)-H)$, and using the facts that all of the states satisfy the Schr\"odinger equation and that the states $\big|w;t\rangle$ are orthogonal, we obtain
\be
  \frac{\partial}{\partial t}\langle w;t\big|v;t\rangle=0.
  \label{diff_t_state}
\ee
Thus the quantity $\langle w;t\big|v;t\rangle$ is independent of time. We now use the completeness of the state $\big|v;t\rangle$ and $\big|w;t\rangle$, Eq.(\ref{diff_t_state}), and the orthonormality of the states $\big|w;t\rangle$ to rewrite Eq.(\ref{trans_amp1}) as
\begin{align}
  T(n\rightarrow m)=&\sum_{v,w,w'}\langle m;f\big|w;\infty\rangle\langle w;\infty\big|v;\infty\rangle\langle v;-\infty\big|w';-\infty\rangle\langle w';-\infty\big|n;i\rangle\nonumber\\
  =&\sum_{v,w,w'}\langle m;f\big|w;\infty\rangle\langle w;-\infty\big|v;-\infty\rangle\langle v;-\infty\big|w';-\infty\rangle\langle w';-\infty\big|n;i\rangle\nonumber\\
  =&\sum_w\langle m;f\big|w;\infty\rangle\langle w;-\infty\big|n;i\rangle.
\end{align}
Thus, Eqs.(\ref{trans_amp1}) and (\ref{trans_amp2}) are the same, as asserted.

Suppose for simplicity that the eigenstates of $I$ are nondegenerate, so that the eigenvalue of $I$ is the only quantum number required for describing the system. When this is so, as it is in our discussion of the time-dependent harmonic oscillator, then it is particularly convenient to choose an invariant having the property that it becomes time-independent as $t\rightarrow-\infty$ so that the commutator $[I(-\infty),H(-\infty)]$ vanishes. Then the normalized eigenvectors of $H(-\infty)$ and $I(-\infty)$ are identical to within constant phase factors. Consequently, we may choose the initial state $|n;i\rangle$ simply to be a eigenstate of $I(-\infty)$, say $|\lambda;-\infty\rangle$. Eq.(\ref{trans_amp}) then reduces to 
\be
  T(n\rightarrow m)=\exp\left(i\left[\alpha_n(\infty)-\alpha_n(-\infty)\right]\right)\langle m;f|\lambda_n;\infty\rangle,
\ee
and the transition probability is given by
\begin{align}
  P_{nm}&=\left|T(n\rightarrow m)\right|^2\nonumber\\
      &=\left|\langle m;f|\lambda_n;\infty\rangle\right|^2.
      \label{P_mn}
\end{align}
As $t\rightarrow\infty$, the invariant operator $I(t)$ in general remains time-dependent and does not commute with the Hamiltonian. Therefore, the state $|\lambda_m;\infty\rangle$ in Eq.(\ref{P_mn}) is a superposition of eigenstates of $H(\infty)$; this is another expression of the fact that energy is not conserved in our system. 

From the structure of Eq.(\ref{trans_amp}), it is apparent that we may express the transition amplitude as a matrix element of an $S$ matrix by writing 
\bd
  S=\sum_{\lambda,\kappa}e^{i\alpha_{\lambda\kappa}(\infty)}|\lambda,\kappa;\infty\rangle\langle\lambda,\kappa;-\infty|e^{-\alpha_{\lambda\kappa(-\infty)}},
\ed
\be
  T(n\rightarrow m)=\langle m;f|S|n;i\rangle.
\ee
It is easily verified that this operator is unitary:
\be
  S^{\dagger}S=SS^{\dagger}=1.
\ee
In the special case that the Hamiltonian operators in the remote past and distant future are identical, $H(-\infty)=H(\infty)$, so that the initial and final states are the same set, we may define an elastic scattering operator $R$ in the standard fashion:
\be
  S=1+2\pi iR.
\ee
The operator $R$ describes the nondiagonal transitions just as $S$ does, but subtracts a noninteracting part from the diagonal amplitudes so that $\langle n|R|n\rangle$ represents a ``forward reaction amplitude" from the state $|n\rangle$ to the same state. The unitarity of the $S$ matrix implies
\be
  \sum_m\left|\langle m|R|n\rangle\right|^2=\frac{1}{\pi}Im(\langle n|R|n\rangle),
\ee
which is a statement of the optical theorem: the total reaction probability is proportional to the imaginary part of the forward reaction amplitude. 

%% file: Gauss_Codazzi.tex
\chapter{Gauss Codazzi}
\label{ch:GC}

\section{The Gauss-Codazzi Formalism}

Here we wish to solve Einstein's equations in the presence of stress-energy sources confined to three-dimensional time-like hypersurfaces for a general metric. Following the methods used by Ipser and Sikivie, Ref.\cite{Ipser}, we shall use the Gauss-Codazzi formalism.

The Gauss-Codazzi equations relate the four-dimensional geometry of the overall global space-time to their projection onto a three-dimensional hypersurface embedded within the original four-dimensional space-time. This is done by investigating the intrinsic and extrinsic curvature of the three-dimensional time-like hypersurface. The Gauss-Codazzi formalism allows one to find the equations of motion for a collapsing domain wall in a very systematic way. To find the equations of motion, one needs to specify the metric (and associated energy-momentum tensor) only. 

In this chapter we wish to develop the Gauss-Codazzi formalism for a general metric where the only initial requirement is that the coefficients of the metric depend on position and time only. We will then arrive a final equation which depends on the coefficient (and derivatives of), as well as its associated energy-momentum tensor, which will allow us to find the equations of motion for the collapsing domain wall once the metric is completely specified. After we develop the general equations, we will compare our result with that found in the literature for two different specified metrics: the Schwarzschild and Reissner-Nordstr\"om metrics, respectively. The Schwarzschild and Reisner-Norstr\"om metric coefficients both depend on position only, hence these are an example of a special case of the general method we are working with here.

\subsection{The Equations}

Here we follow the technologies developed in Ref.\cite{Ipser}. Let $S$ denote a three-dimensional time-like hypersurface containing stress-energy and let $\xi^a$ be its unit spacelike normal ($\xi_a\xi^a=1$). The three-metric intrinsic to the hypersurface $S$ is 
\be
  h_{ab}=g_{ag}-\xi_a\xi_b
\ee
where $g_{ab}$ is the four-metric of the space-time. Here $h_{ab}$ is known as the projected tensor for the hypersurface $S$, see Ref.\cite{Carroll}. This is due to the fact that, when acting $h_{ab}$ on a vector $v^a$, it will project it tangent to the hypersurface, hence orthogonal to $\xi^a$, 
\begin{align*}
  (h_{ab}v^a)\xi^b&=g_{ab}v^a\xi^b-\xi_a\xi_bv^a\xi^b\\
      &=v^a\xi_a-v^a\xi_a\\
       &=0.
\end{align*}
Let $\nabla_a$ denote the covariant derivative associated with $g_{ab}$ and let 
\be
  D_a=h_a{}^b\nabla_b,
  \label{D_a}
\ee
hence $D_a$ is the covariant derivative on the induced three-dimensional hypersurface. The extrinsic curvature of $S$, denoted by $\pi_{ab}$, is defind by
\be
  \pi_{ab}\equiv D_a\xi_b=\pi_{ba}.
  \label{pi}
\ee
The extrinsic curvature depends on how the hypersurface is embedded in the full four-dimensional space-time. The extrinsic curvature is used to differentiate different topologies. For example, intrinsic geometry of a cylinder and a torus can be flat, however, we know the exterior geometry of each is different. This different topology is given in the extrinsic curvature, which will tell us that we are actually on a torus or a cylinder. 

The contracted forms of the first and second Gauss-Codazzi equations are then given by
\bea
  ^3R+\pi_{ab}\pi^{ab}-\pi^2&=&-2G_{ab}\xi^a\xi^b\label{Gauss}\\
  h_{ab}D_c\pi^{ab}-D_a\pi&=&G_{bc}h^b{}_a\xi^c\label{Codazzi}.
\eea
Here $^3R$ is the Ricci scalar curvature of the three-geometry $h_{ab}$ of $S$, $\pi$ is the trace of the extrinsic curvature, and $G_a{}^b$ is the Einstein tensor in four-dimensional space-time.

Here we will be working with infinitely thin domain walls. The stress-energy tensor $T_{ab}$ of four-dimensional space-time then is assumed to have a $\delta$-function singularity on $S$. This in turn implies that the extrinsic curvature has a jump discontinuity across $S$, since the extrinsic curvature is analogous to the gradient of the Newtonian gravitational potential. Therefore we can introduce
\be
  \gamma_{ab}\equiv\pi_{+ab}-\pi_{-ab}
\ee
which is the difference between the exterior and interior extrinsic curvatures, and
\be
  S_{ab}\equiv\int dlT_{ab},
\ee
where $l$ is the proper distance through $S$ in the direction of the normal $\xi^a$, and where the subscripts $\pm$ refer to values just off the surface on the side determined by the direction of $\pm\xi^a$. Hence the direction for, say $+\xi^a$ will be in the direction of the exterior geometry of the domain wall, while $-\xi^a$ will denote the direction of the interior geometry of the domain wall. As we shall discuss below, these geometries will be different for the case of the spherically symmetric domain wall. Using Einstein's and the Gauss-Codazzi equations, one can show that (see Ref.\cite{Wheeler})
\be
  S_{ab}=-\frac{1}{8\pi G_N}\left(\gamma_{ab}-h_{ab}\gamma_c{}^c\right).
  \label{action}
\ee
We can also introduce the ``average" extrinsic curvature
\be
  \tilde{\pi}_{ab}=\frac{1}{2}\left(\pi_{+ab}+\pi_{-ab}\right)
  \label{tilde_pi}
\ee
which will be important later.

\subsection{The Surface Stress-Energy Tensor}

Here we restrict ourselves to sources for which the stress energy tensor is given by, see Ref.\cite{Ipser}
\be
  S^{ab}=\sigma u^au^b-\eta\left(h^{ab}+u^au^b\right)
  \label{Stress-Energy}
\ee
which is the material sources consisting of a perfect fluid. In Eq.(\ref{Stress-Energy}) $u^a$ is the four-velocity of any observer whose world line lies within $S$ and who sees no energy flux in his local frame, and where $\sigma$ is the energy per unit area and $\eta$ is the tension measured by the observer. For a dust wall it is well known that $\eta=0$, while for a domain wall $\eta=\sigma$. For a domain wall Eq.(\ref{Stress-Energy}) reduces to 
\be
  S^{ab}=-\sigma h^{ab}.
\ee
We also note that the four-velocity $u^a$ is a time-like unit vector orthogonal to the space-like unit normal $\xi^a$, i.e.,
\begin{equation*}
  u_au^a=-1, \hspace{2mm} \xi_au^a=0, \hspace{2mm} \xi_a\xi^a=+1.
\end{equation*}

\subsection{Attractive Energy}

Here we derive equations for an observer who is hovering just above the surface $S$ on either side. Let the vector field $u^a$ be extended off $S$ in a smooth fashion. The acceleration
\bea
  u^a\nabla_au^b&=&(h^b{}_c+\xi^b\zeta_c)u^a\nabla_au^c\nonumber\\
     &=&h^b{}_cu^a\nabla_au^c-\xi^bu^au^c\pi_{ab}
\eea
has a jump discontinuity across $S$ since the extrinsic curvature has such a discontinuity. The perpendicular components of the accelerations of observers hovering just off $S$ on either side satisfy
\begin{align}
  \xi_bu^a\nabla_au^b\Big{|}_++\xi_bu^a\nabla_au^b\Big{|}_-=&-2u^au^b\tilde{\pi}_{ab}\nonumber\\
     =&-2\frac{\eta}{\sigma}(h^{ab}+u^au^b)\tilde{\pi}_{ab}-2\frac{1}{\sigma}S^{ab}\tilde{\pi}_{ab}
     \label{perp1}
\end{align}
and
\bea
  \xi_bu^a\nabla_au^b\Big{|}_+-\xi_bu^a\nabla_au^b\Big{|}_-&=&-u^au^b\gamma_{ab}\nonumber\\
     &=&4\pi G_n(\sigma-2\eta).
     \label{perp2}
\eea
Here we comment on the precense of the second term on the right hand side of Eq.(\ref{perp1}). This term takes into account the contributions to the energy-tensor $T_{ab}$ which are present in the vacuo on opposite sides of $S$. For example, if there is only mass present, then $T_{ab}$ vanishes off the shell, hence the second term is zero. In the case of charge present, then $T_{ab}$ does not vanish, then the contribution to $T_{ab}$ outside can be taken from the Maxwell tensor. 

\section{Spherical Walls}

In this section we shall obtain the asymptotically flat solutions to Einstein's equations for spherically symmetric domain walls with an arbitrary metric. Here we will consider two cases. First we will consider the case where the metric coefficients only depend on the radial position of the domain wall. Second, we will consider the case where the metric coefficients depend on both the radial position of the domain wall and the time.

\subsection{Radial dependence only}

For a spherical shell of stress-energy, let the unit normal $\xi_+$ point in the outward radial direction. It is well known that asymptotic flatness and spherical symmetry requires that the interior geometry is flat (Birkhoff's theorem). For the external geometry we will choose an arbitrary metric. First we shall consider the case where the coefficients only depend on position. Hence,
\begin{align}
  (ds^2)_+=&-e^{v(r)}dt^2+e^{u(r)}dr^2+r^2d\Omega^2\nonumber\\
       =&-A(r)dt^2+B(r)^2dr^2+r^2d\Omega^2 \hspace{2mm} \text{for $r>R(t)$}
       \label{out_metric}
\end{align}
and
\be
  (ds^2)_-=-dT^2+dr^2+r^2d\Omega^2 \hspace{2mm} \text{for $r<R(t)$}
  \label{in_metric}
\ee
where 
\be
  d\Omega^2=d\theta^2+\sin^2\theta d\phi^2.
\ee
Here the equation of the wall is 
\be
  r=R(t).
  \label{wallEq}
\ee
One finds for the components of $u^a$ and $\xi^a$ $(a=t$ or $T,r,\theta,\phi$, in that order$)$
\begin{align}
  (u^a{}_+)&=(\beta A(r)^{-1},R_{\tau},0,0), \hspace{2mm} (u^a{}_-)=(\alpha,R_{\tau},0,0),\nonumber\\
  (\xi^a{}_+)&=(R_{\tau}A^{-1},\beta(AB)^{-1},0,0), \hspace{2mm} (\xi^a{}_-)=(R_{\tau},\alpha,0,0).
  \label{u_xi}
\end{align}
Here $R_{\tau}=dR/d\tau$, where $\tau$ is the propertime of an observer moving with four-velocity $u^a$ at the wall, and 
\begin{align}
  \alpha\equiv&T_{\tau}=\sqrt{1+R_{\tau}^2},\\
  \beta\equiv&At_{\tau}=\sqrt{A(r)+A(r)B(r)R_{\tau}^2}.
  \label{beta1}
\end{align}
However, here we should comment that the condition that $\xi$ is of unit normal, this then implies the condition that 
\be
  B(r)=\frac{1}{A(r)}.
\ee 
Therefore we can rewrite Eq.(\ref{beta1}) as
\begin{align}
  \alpha\equiv&T_{\tau}=\sqrt{1+R_{\tau}^2},\label{alpha}\\
  \beta\equiv&=\sqrt{A(r)+R_{\tau}^2}.
  \label{beta2}
\end{align}
These expressions and the definitions Eqs.(\ref{D_a}), (\ref{pi}) and (\ref{tilde_pi}) imply that
\be
  (h^{ab}+u^au^b)\tilde{\pi}_{ab}=(\xi^r{}_++\xi^r{}_-)\frac{1}{R},
\ee
and
\begin{align}
  \xi_bu^a\nabla_au^b\Big{|}_+&=\frac{1}{\beta}\left[R_{\tau\tau}+\frac{A'}{2}\right]\nonumber\\
  \xi_bu^a\nabla_au^b\Big{|}_-&=\frac{1}{\alpha}R_{\tau\tau}
  \label{acceleration}
\end{align}
where
\be
  A'=\frac{dA(r)}{dr}\Big{|}_{r=R(t)}.
  \label{A_prime}
\ee
Substituting into Eqs.(\ref{perp1}) and (\ref{perp2}) then yields the equations of motion
\begin{align}
  (\alpha+\beta)R_{\tau\tau}=&-2\frac{\eta}{\sigma}\frac{\alpha\beta(\alpha+\beta)}{R}-\frac{\alpha A'}{2}-2\frac{\alpha\beta}{\sigma}S^{ab}\tilde{\pi}_{ab}\label{plus}\\
  (\alpha-\beta)R_{\tau\tau}=&4\pi\alpha\beta G(\sigma-2\eta)-\frac{\alpha A'}{2}.\label{minus}
\end{align}
Taking the ratio of Eqs.(\ref{plus}) and (\ref{minus}) allows us to eliminate $R_{\tau\tau}$ from the expression, so we then find
\be
  \sigma(\sigma-2\eta)+\frac{(1-A)\eta}{2\pi(\alpha+\beta)GR}-\frac{A'\sigma}{4\pi(\alpha+\beta)G}+\frac{(1-A)S^{ab}\tilde{\pi}_{ab}}{4\pi(\alpha+\beta)^2G}=0
  \label{EOM}
\ee

Here we make some general comments on Eqs.(\ref{plus}) and (\ref{minus}). First, in the absence of stress-energy outside the domain wall, $R_{\tau\tau}$ is always negative provided $\eta\geq0$. Hence a spherical domain wall with, say only mass, with $\eta\geq0$ will always collapse to a black hole, regardless of its size. Second, in the presence of stress-energy outside the domain wall, $R_{\tau\tau}$ is always positive provided that the source term is small compared to the other terms. However, if the source term is large compared to the other terms, $R_{\tau\tau}$ can become positive at some point. This means that the collapsing object will turn around and begin to expand. 

Eq.(\ref{EOM}) allows us to find the equations of motion for a specific geometry, provided that the coefficient $A=A(r)$, i.e. is only a function of position. In the next section we demonstrate the findings in Eq.(\ref{EOM}) for two specific cases found in the literature. This will allow us to demonstrate ease of the general form of the equations of motion.

\subsection{Radial and Time dependence}

In this section we will write the exterior metric, Eq.(\ref{out_metric}), as
\begin{align}
  (ds^2)_+=&-e^{v(r,t)}dt^2+e^{u(r,t)}dr^2+r^2d\Omega^2\nonumber\\
       =&-A(r,t)dt^2+B(r,t)^2dr^2+r^2d\Omega^2 \hspace{2mm} \text{for $r>R(t)$}
       \label{out_metric_t}
\end{align}
where we will maintain that the interior metric is still given by the Minkowski line element. We will again take that the equation of the wall is given by Eq.(\ref{wallEq}), this then gives that the components of $u^a$ and $\xi^a$ are unchanged in form from Eq.(\ref{u_xi}). Note however that one does have to make the change from $A(r)$ and $B(r)$ to $A(r,t)$ and $B(r,t)$, respectively. As in the case of radial dependence only, the condition that $\xi^a$ is a normalized space-like vector, we again have the condition that 
\be
  B(r,t)=\frac{1}{A(r,t)}.
\ee
Therefore we define $\alpha$ and $\beta$ in the same manner as in the case with only radial dependence, using the suitable substitution.

We can then find that the acceleration outside and inside the domain wall are given by
\begin{align}
  \xi_bu^a\nabla_au^b\Big{|}_+=&\frac{1}{\beta}\left[R_{\tau\tau}+\frac{A'}{2}\right]\nonumber\\
       &+\frac{\dot{A}\dot{R}}{2A^3\beta}\left[A^2+2\dot{R}^2(A-\beta)-3A\beta\right]\nonumber\\
  \xi_bu^a\nabla_au^b\Big{|}_-=&\frac{1}{\alpha}R_{\tau\tau}
  \label{acceleration_t}
\end{align}
where $A'$ is given in Eq.(\ref{A_prime}) and 
\be
  \dot{A}=\frac{dA}{dt}.
\ee
Comparing the acceleration outside the domain wall for the radial and time dependent metric coefficient, Eq.(\ref{acceleration_t}), to that of the acceleration outside the domain wall for the radially dependent metric coefficient, Eq.(\ref{acceleration}), we see that the acceleration outside the domain in the new case is just the acceleration in the radial case modified by an additional term which depends on $t$-derivatives of the metric coefficient. This is not an unexpected result. 

Substituting Eq.(\ref{acceleration_t}) into Eqs.(\ref{perp1}) and (\ref{perp2}) then yields the equations of motion
\begin{align}
  (\alpha+\beta)R_{\tau\tau}=&-2\frac{\eta}{\sigma}\frac{\alpha\beta(\alpha+\beta)}{R}-\frac{\alpha A'}{2}-2\frac{\alpha\beta}{\sigma}S^{ab}\tilde{\pi}_{ab}\nonumber\\
      &-\frac{\dot{A}\dot{R}\alpha}{2A^3}\left[A^2+2\dot{R}^2(A-\beta)-3A\beta\right]\label{plus2}\\
  (\alpha-\beta)R_{\tau\tau}=&4\pi\alpha\beta G_N(\sigma-2\eta)-\frac{\alpha A'}{2}\nonumber\\
      &-\frac{\dot{R}\dot{A}\alpha}{2A^3}\left[A^2+2\dot{R}^2(A-\beta)-3A\beta\right].\label{minus2}
\end{align}
Taking the ratio of Eqs.(\ref{plus2}) and (\ref{minus2}) allows us to eliminate $R_{\tau\tau}$ from the expression, so we then find
\begin{align}
  0=&\sigma(\sigma-2\eta)+\frac{(1-A)\eta}{2\pi(\alpha+\beta)G_NR}-\frac{A'\sigma}{4\pi(\alpha+\beta)G_N}\nonumber\\
  &+\frac{(1-A)S^{ab}\tilde{\pi}_{ab}}{4\pi(\alpha+\beta)^2G_N}\nonumber\\
  &-\frac{\dot{A}\dot{R}\sigma}{4\pi G_NA^3(\alpha+\beta)}\left[A^2+2\dot{R}^2(A-\beta)-3A\beta\right]
  \label{EOM_t}
\end{align}

Here we make some general comments on Eqs.(\ref{plus2}) and (\ref{minus2}). First, we again see that the first three terms in Eq.(\ref{plus2}) and the first two terms in Eq.(\ref{minus2}) are identical to the radially dependent metric coefficients only, where the last term comes from the time dependence of the metric coefficients. Second, it is not as obvious in this case the behavior of the domain wall. In the case of gravitational collapse, $\dot{R}<0$, making the last term positive.

\section{Examples}

In this section we present some examples using the equation of motion in Eq.(\ref{EOM}). First, we will investigate the case of a massive domain wall. We will show that Eq.(\ref{EOM}) automatically leads to the equation of motion arrived at by Ipser and Sikivie, see Ref.\cite{Ipser}. Second, we will investigate the case of a massive-charged domain wall. We will show that Eq.(\ref{EOM}) automatically leads to the equation of motion arrived at by L\'opez, see Ref.\cite{Lopez}.

Here we note that the usual procedure for determining the metric coefficients is to consider the asymptotic region of space-time (see for example Ref.\cite{Weinberg}).  Here one writes the Ricci tensor, which gives the equations of motion for the the metric coefficients. Then using the asymptotic requirements of the space-time, one integrates the equations of motion for the metric coefficients and fixes the integration constant. As stated above, we will just start with the metric coefficients to find the conserved quantities for the collapsing domain wall.

\subsection{Massive Domain Wall}

It is well known that asymptotic flatness and spherical symmetry require the exterior geometry to be Schwarzschild. Therefore we can write Eq.(\ref{out_metric}), the exterior metric, as
\be
  (ds^2)_+=-\left(1-\frac{2GM}{r}\right)dt^2+\left(1-\frac{2GM}{r}\right)^{-1}dr^2+r^2d\Omega^2.
\ee
Comparing the Schwarzschild metric with Eq.(\ref{out_metric}), one can then identify
\be
  A(r)=1-\frac{2GM}{r}.
\ee
Since the domain wall only contains mass, the stress-energy is only present on the domain wall. Hence $T_{ab}$ vanishes outside of the domain wall. Therefore using Eq.(\ref{EOM}) we can immediately write
\be
  \sigma(\sigma-2\eta)-\frac{2GM}{R^2}\frac{(\sigma-2\eta)}{4\pi(\alpha+\beta)G}=0,
\ee
or rearranging the terms we have
\begin{align}
  M=&\frac{1}{2}(\alpha+\beta)4\pi\sigma R^2\nonumber\\
     =&\frac{1}{2}\left[\sqrt{1+R_{\tau}^2}+\sqrt{1-\frac{2GM}{R}+R_{\tau}^2}\right]4\pi\sigma R^2
\end{align}
where in the second line we use the definition of $\alpha$ and $\beta$, Eqs.(\ref{alpha}) and (\ref{beta2}) respectively. This is identical to Eq.(3.8) in Ref.\cite{Ipser}, for the case of the massive domain wall. 

\subsection{Massive-Charged Domain Wall}

Since the domain wall is charged, and spherically symmetric, the geometry outside the domain wall is given by the Reissner-Nordstr\"om solution to Einstein equations. Therefore we can write Eq.(\ref{out_metric}), the exterior metric, as
\begin{align}
  (ds^2)_+=&-\left(1-\frac{2GM}{r}+\frac{Q^2}{r^2}\right)dt^2+\left(1-\frac{2GM}{r}+\frac{Q^2}{r^2}\right)^{-1}dr^2+r^2d\Omega^2.
\end{align}
Comparing the Reissner-Nordstr\"om metric with Eq.(\ref{out_metric}), one can then identify
\be
  A(r)=1-\frac{2GM}{r}+\frac{Q^2}{r^2}.
\ee
In this case the domain wall contains both mass and charge, thus the stress-energy outside of the shell is taken from Maxwell's tensor, since the inside portion of the spherically symmetric domain wall will not feel the influence of the charge. The only nonvanishing components outside the domain wall are
\be
  T_0{}^0=T_1{}^1=-T_2{}^2=-T_3{}^3=-\frac{Q^2}{8\pi r^4}.
\ee

By taking the difference of Eq.(\ref{Gauss}) on opposite sides of $S$, one finds
\be
  -\frac{2}{\sigma}S^{ab}\tilde{\pi}_{ab}=\frac{Q^2}{4\pi\sigma R^4}.
\ee
Therefore using Eq.(\ref{EOM}) we can write
\be
  \left[(\sigma-2\eta)+\frac{Q^2}{4\pi(\alpha+\beta)R^3}\right]\left[\sigma-\frac{(GM-\frac{Q^2}{2R})}{2\pi(\alpha+\beta)R^2}\right]=0.
\ee
Although this is an algebraic equation of second degree in $\sigma$, only one of the two roots holds
\be
  \sigma=\frac{(GM-\frac{Q^2}{2R})}{2\pi(\alpha+\beta)R^2},
\ee
which, using Eqs.(\ref{alpha}) and (\ref{beta2}) can be put in the form
\be
  \alpha-\beta=4\pi\sigma GR.
\ee
Therefore, solving for the mass yields
\be
  M=\frac{Q^2}{2GR}+4\pi\sigma R^2\left[\sqrt{1+R_{\tau}^2}-2\pi\sigma GR\right]
\ee
which is identical to Eq.(61) in Ref.\cite{Lopez}, for the case of the massive-charged domain wall.

%% file: rho_t.tex
\chapter{$\rho(t)$ Equation}
\label{ch:rho_t}

This work was originally completed in Ref.\cite{Stojkovic}, here we will outline the results.

In the range $t<0$, $\omega$ is a constant and the solution to Eq.(\ref{gen_rho}) is
\be
  \rho(\eta)=\frac{1}{\sqrt{\omega_0}}.
\ee

In the range of interest, during the time of gravitational collapse, we do not have an analytical solution to Eq.(\ref{gen_rho}). However, we can find certain useful properties of $\rho(t)$. First note that in terms of $\eta$
\be
  \omega^2=\frac{\omega_0^2}{1-\eta/R_s}.
\ee
Then after rescaling, Eq.(\ref{gen_rho}) can be written as
\be
  \frac{d^2f}{d\eta'^2}=-(\omega_0R_s)^2\left[\frac{f}{1-\eta'}-\frac{1}{f^3}\right]
  \label{f_t eq}
\ee
where $\eta'=\eta/R_s$, $f=\sqrt{\omega_0}\rho$. The boundary conditions are then
\be
  f(0)=1, \hspace{2mm} \frac{df(0)}{d\eta'}=0.
\ee

The last term in Eq.(\ref{f_t eq}) becomes singular as $f\rightarrow0$. We can then consider a more well behaved function for $1/f^3$. For example
\be
  \frac{d^2g}{d\eta'^2}=-(\omega_0R_s)^2\left[\frac{g}{1-\eta'}-g\right]
  \label{g_t eq}
\ee
with boundary conditions
\be
  g(0)=1, \hspace{2mm} \frac{dg(0)}{d\eta'}=0.
  \label{IC g_t}
\ee
Eq.(\ref{g_t eq}) implies that $g(\eta')$ is a monotonically decreasing function as long as $g(\eta')>0$. Furthermore, it is decreasing faster than the solution for $f$ as long as $f<1$, since the $1/f^3$ in Eq.(\ref{f_t eq}) is a larger ``repulsive" force than the $g$ term in Eq.(\ref{g_t eq}). So
\be
  g(\eta')\leq f(\eta')
\ee
for all $\eta'$ such that $g(\eta')>0$.

Eq.(\ref{g_t eq}) with initial conditions Eq.(\ref{IC g_t}) can be solved in terms of degenerate hypergeometric functions. The important part for us is that $g$ is positive for all $\eta'$ and, in particular, $g(1)>0$ for all the values of $\omega_0R_s$ that we have checked. Therefore $f(\eta')$ is positive, at least for a wide range of $\omega_0R_s$. 

We can find some more properties of $\rho(t)$. Let $f_1=f(1)\not=0$. Then the equation for $f$ can be expanded near $\eta'=1$. 
\be
  \frac{d^2f_1}{d\eta'^2}\sim-(\omega_0R_s)^2\left[\frac{f_1}{1-\eta'}-\frac{1}{f_1^3}\right].
  \label{f1_t eq}
\ee
This shows that 
\be
  \frac{df}{d\eta'}\sim(\omega_0R_s)^2f_1\ln(1-\eta')\rightarrow-\infty
\ee
as $\eta'\rightarrow1$.

Hence $\rho(\eta=R_s)$ is strictly positive and finite while $\rho_{\eta}(\eta=R_s)=-\infty$ for finite and non-zero $\omega_0$. Since $f=\sqrt{\omega_0}\rho$ and $f\rightarrow1$ for $\omega_0\rightarrow0$, we also see that $\rho\rightarrow\infty$ and $\rho_{\eta}\rightarrow0$ as $\omega_0\rightarrow0$. 

In the range $t_f<t$, $\omega$ is a constant. However, the solution for $\rho$ is not constant, unlike in the range $t<0$, since the constant solution $1/\sqrt{\omega(t_f)}$ does not necessarily match up with $\rho(t_f-)$ to ensure a continuous solution. Yet it is easy to check that in this range $\dot{N}=0$ and so there is no change in the occupation numbers. So we need only find $N(t_f-,\bar{\omega})$ to determine $N(t\rightarrow\infty,\bar{\omega})$.

%% file: rho_tau.tex
\chapter{$\rho(\tau)$ Equation}
\label{ch:rho(tau)}

To get an understanding of the number of particles created in the region near the horizon we need to investigate the behavior of the function $\rho(\tau)$ near the Schwarzschild radius.

Near the horizon we can then write the velocity term as
\be
  \left|R_{\tau}\right|\approx\textrm{const}\equiv A.
\ee
In this limit the position of the shell is then, from Eq.(\ref{R_tau_0})
\be
  R(\tau)\approx \tilde{R}_0-A\tau
  \label{A}
\ee
where, as stated in Chapter \ref{ch:Classical}, $\tilde{R}_0$ is the initial position of the shell, we can write
\be
  \frac{\sqrt{1+R_{\tau}^2}}{\left|R_{\tau}\right|}\equiv C.
  \label{C}
\ee
Therefore the frequency becomes
\be
  \omega^2\approx\frac{\omega_0^2}{CB}.
  \label{approx omega}
\ee
Therefore the auxiliary equation becomes
\bd
  \rho_{\eta\eta}+\omega_0^2\frac{R_s}{C((R_0-R_s)-A\tau)}\rho=\frac{1}{\rho^3}
\ed
or using Eq.(\ref{eta_tau}) we can write this as,
\be
  \frac{1}{C^2}\frac{d^2\rho}{d\tau^2}+\omega^2_0\frac{R_s}{C((R_0-R_s)-A\tau)}\rho=\frac{1}{\rho^3}.
\ee
After rescaling we can write this as
\be
  \frac{d^2f}{d\tau'^2}=-\frac{A^2\omega_0^{3/2}R_s^{3/4}C^{5/4}}{(R_0-R_s)^{11/4}}\left[\frac{f}{1-\tau'}-\frac{1}{f^3}\right]
  \label{f_tau eq}
\ee
where $\tau'=A\tau/(R_0-R_s)$, and $f=\sqrt{\omega_0}(R_s/C(R_0-R_s))^{1/4}\rho$. The boundary conditions are then
\be
  f(0)=\left(\frac{R_s}{C(R_0-R_s)}\right)^{1/4}, \hspace{2mm} \frac{df(0)}{d\tau'}=0.
  \label{f_tau IC}
\ee
The last term with the $1/f^3$ becomes singular as $f\rightarrow0$. Let us consider another equation with this term replaced by another more well behaved function. For example consider,
\be
  \frac{d^2g}{d\tau'^2}=-\frac{A^2\omega_0^{3/2}R_s^{3/4}C^{5/4}}{(R_0-R_s)^{11/4}}\left[\frac{f}{1-\tau'}-g\right]
  \label{g_tau eq}
\ee
where the boundary conditions in Eq.(\ref{f_tau IC}) become
\be
  g(0)=\left(\frac{R_s}{C(R_0-R_s)}\right)^{1/4}, \hspace{2mm} \frac{dg(0)}{d\tau'}=0.
  \label{g_tau IC}
\ee
Eq.(\ref{g_tau eq}) implies that $g(\tau')$ is a monotonically decreasing function as long as $g(\tau')>0$. It is decreasing faster than the solution for $f$ as long as $f<1$, since the $1/f^3$ term in Eq.(\ref{f_tau eq}). Therefore we have
\be
  f(\tau')\geq g(\tau')
\ee
for all $\tau'$ such that $g(\tau')>0$. 

The solution for $g$ is positive for all $\tau'$ and, in particular, $g(1)>0$ for all the values $A^2\omega_0^{3/2}C^{5/4}R_s^{3/2}/(R_0-R_s)^{11/4}$ that we have checked. Therefore $f(\tau')$ is positive, at least for a wide range.

Let $f_1=f(1)\not=0$. Then the Eq.(\ref{f_tau eq}) can be expanded near $\tau'=1$,
\be
  \frac{d^2f_1}{d\tau'^2}=-\frac{A^2\omega_0^{3/2}R_s^{3/4}C^{5/4}}{(R_0-R_s)^{11/4}}\left[\frac{f_1}{1-\tau'}-\frac{1}{f_1^3}\right]
  \label{f1_tau eq}
\ee
Integrating Eq.(\ref{f1_tau eq}) we can then write
\be
  \frac{df}{d\tau'}\sim\frac{A^2\omega_0^{3/2}R_s^{3/4}C^{5/4}}{(R_0-R_s)^{11/4}}f_1\ln(1-\tau')\rightarrow-\infty
\ee
as $\tau'\rightarrow1$. Hence $\rho(\tau=(R_0-R_s)/A)$ is strictly positive and finite while $\rho_{\tau}(\tau=(R_0-R_s)/A)=-\infty$ for finite and non-zero $\omega_0$.

We are calculating the occupation number $N$ as a function of frequency $\omega$ at some fixed time. From Eq.(\ref{approx omega}) we see that, in order to keep $\omega$ fixed in time, $\omega_0\rightarrow0$ as $B\rightarrow0$. Thus, $\omega$ varies with $\omega_0$ and not with time. Since $f=(R_s/C(R_0-R_s))^{1/4}$, and  $f=(R_s/C(R_0-R_s))^{1/4}$ for $\omega_0\rightarrow0$, we see that $\rho\rightarrow\infty$ and $\rho_{\tau}=0$ as $\omega_0\rightarrow0$. This implies taht the occupation number in Eq.(\ref{OccNum}) diverges as $\tau\rightarrow\tau_c$ since $B\rightarrow0$ as $\tau\rightarrow\tau_c$.

%% file: PhD_Thesis.bbl
\newpage

\begin{thebibliography}{99}
\addcontentsline{toc}{chapter}{Bibliography}

\bibitem{DeWitt} B. S. DeWitt, Phys. Rev. {\bf 160}, 1113-1148 (1967).

\bibitem{Lewis} H. R. Lewis, Jr. and W. B. Riesenfeld, J. Math. Phys. (N.Y.) {\bf10}, 1458 (1969). 

\bibitem{Pedrosa} C. M. A. Dantas, I. A. Pedrosa and B. Baseia, Phys. Rev. A \textbf{45}, 1320 (1992).

\bibitem{Ipser} J. Ipser and P. Sikivie, Phys. Rev. D {\bf 30}, 712 (1984).

\bibitem{Israel} W. Israel, Nuovo Cimento {\bf 44B}, 1 (1966).

\bibitem{Lopez} C. A. L\'opez, Phys. Rev. D {\bf38} 3662 (1988).

\bibitem{Wheeler} ``Gravitation," C. W. Misner, K. S. Thorne and J. A. Wheeler, Freeman 1973.

\bibitem{FrolovNovikov} ``Black Hole Physics," V. P. Frolov and I. D. Novikov, Kluwer Academic Publishers, Dordrecht 1998.

\bibitem{PenroseHawking} ``The Nature of Space and Time," S. Hawking and R. Penrose, Princeton University Press 1996. 

\bibitem{BogojevicStojkovic} A. Bogojevic and D. Stojkovic, Phys. Rev. D {\bf61}, 084011 (2000) [arXiv:gr-qc/9804070].

\bibitem{Trodden} M. Trodden, V. F. Mukhanov and R. H. Bradenberger, Phys. Lett. B {\bf316}, 483 (1993) [arXiv:hep-th/9305111].

\bibitem{Borstnik} N. Mankoc Borstnik, H. B. Nielsen, C. D. Froggatt and D. Lukman, arXiv:hep-ph/0512061.

\bibitem{Shankaranarayanan} S. Shankaranarayanan and N. Dadhich, Int. J. Mod. Phys. D {\bf13}, 1095 (2004) [arXiv:gr-qc/0306111].

\bibitem{Lowe} D. A. Lowe and L. Thorlacius, Phys. Rev. D {\bf 60}, 104012 (1999) [arXiv:hep-th/9903237]. Phys. Rev. D {\bf73}, 104027 (2006) [arXiv:hep-th/0601059].

\bibitem{Horowitz} G. T. Horowitz and J. M. Maldecena, JHEP {\bf0402}, 008 (2004) [arXiv:hep-th/0310281].

\bibitem{Giddings} S. B. Giddings, Phys. Rev. D {\bf74}, 106005 (2006) [arXiv:hep-th/0605196]. Phys. Rev. D {\bf74}, 106009 (2006) [arXiv:hep-th/0606146].

\bibitem{1975Hawking} S. W. Hawking, Commum. Math. Phys. {\bf 43}, 199 (1975) [Erratum-ibid. {\bf 46}, 206 (1976)].

\bibitem{Goldstein} ``Classical Mechanics," H. Goldstein, Addison-Wesley 1980.

\bibitem{Fulling} P. C. W. Davies, S. A. Fulling and W. G. Unruh, Phys. Rev. D {\bf13} (1976) 2720.

\bibitem{Bekenstein} Lett. Nuovo Cimento {\bf4}, 737 (1972). J. D. Bekenstein. Phys. Rev. D \textbf{7} (8): 2333-2346.

\bibitem{GibbonsHawking} G. W. Gibbons and S. W. Hawking, Phys. Rev. D \textbf{15}, 2752 (1977).

\bibitem{Sakurai} ``Modern Quantum Mechanics," J. J. Sakurai, Addison-Wesley 1994.

\bibitem{Birrell} ``Quantum Fields in Curved Space," N. D. Birrell and P. C. W. Davies, Cambridge Monographs on Mathematical Physics 1999.

\bibitem{Landau} ``Statistical Physics" 3$^{rd}$ Edition Part 1, L. D. Landau and E. M. Lifshitz, Elsevier 1980.

\bibitem{Carroll} ``Spacetime and Geometry," Sean M. Carroll, Addison-Wesley 2004.

\bibitem{Thermal} \textit{Thermal Field Theories and Their Applications}, S. P. Kim, edited by Y. X. Gui, F. C. Khanna, and Z. B. Su (World Scientific, Singapore, 1966).

\bibitem{Kim} S. P. Kim hep-th/9809091. S. P. Kim, and C. H. Lee. hep-ph/0005224. S. P. Kim and D. N. Page. quant-ph/0205006. S. P. Kim. cond-mat/9912472.

\bibitem{Davies} P. C. W. Davies, S. A. Fulling and W. G. Unruh, Phys. Rev. D {\bf13}: 2720-2723 (1976).

\bibitem{Unruh} G. W. Unruh, Phys. Rev. D {\bf14}, 870 (1976).

\bibitem{Stojkovic} T. Vachaspati, D. Stojkovic and L. M. Krauss, Phys. Rev. D \textbf{76}, 024005 (2007). T. Vachaspati and D. Stojkovic, gr-qc/0701096 (2007).

\bibitem{GreenStoj} E. Greenwood and D. Stojkovic, JHEP {\bf 0802}, 042 (2008) [arXiv:0802.4087 [gr-qc]].

\bibitem{Greenwood} E. Greenwood and D. Stojkovic, JHEP {\bf0909}, 058 (2009) [arXiv:0806.0628 [gr-qc]].

\bibitem{Saida} H. Saida, gr-qc/0505089.

\bibitem{Strominger} A. Strominger and C. Vafa, arXiv:hep-th/9601029.

\bibitem{tHooft} G. 't Hooft, Nucl. Phys. {\bf B256}, 727 (1985).

\bibitem{Susskind} L. Susskind, J. Math. Phys. {\bf36}, 6377.

\bibitem{EG_ent} E. Greenwood, JCAP {\bf0906}, 032 (2009) [arXiv:0811.0816 [gr-qc]].

\bibitem{Jacobson} ``Introductory Lectures on Black Hole Thermodynamics," T. Jacobson.

\bibitem{Sorkin} R.D. Sorkin, General Relativity and Gravitation, proceedings of the GR10 Conference, Padova, 1983.

\bibitem{Bombelli} L. Bombelli, R. K. Koul, J. Lee and R. D. Sorkin, Phys. Rev. D {\bf34} (1986) 373.

\bibitem{Frolov} V. Frolov and I. Novikov, Phys. Rev. D {\bf48} (1993) 4545.

\bibitem{Weinberg} ``Gravitation and Cosmology," Steven Weinberg, John Wiley and Sons.

\bibitem{Vach} T. Vachaspati, arXiv:0711.0006 [gr-qc].

\bibitem{Wiki} \textit{http://en.wikipedia.org/wiki/Fractionalcalculus}.

\bibitem{WangGreenStoj} J. E. Wang, E. Greenwood and D. Stojkovic, Phys. Rev. D {\bf} (2009) [arXiv:0906.3250 [hep-th]].

\bibitem{EG_RNrad} E. Greenwood, JCAP {\bf1001}, (2010) [arXiv:0910.0024 [gr-qc]].

\bibitem{GreenHalHao} E. Greenwood, E. Halstead and P. Hao, arXiv: [gr-qc].

\bibitem{Sekiwa} Y. Sekiwa, arXiv:0802.3266 [hep-th].

\bibitem{Volovik} G. E. Volovik, arXiv:0803.3367 [gr-qc].

\bibitem{Coleman_deLuccia} S. R. Coleman and F. De Luccia, Phys. Rev. D {\bf 21}, 3305 (1980).

\bibitem{AbbottColeman} L. F. Abbott and S. R. Coleman, Nucl. Phys. B {\bf259}, 170 (1985).

\bibitem{GreenHalPolStoj} E. Greenwood, E. Halstead, R. Poltis and D. Stojkovic, Phys. Rev. {\bf D} 79:103003 (2009) [arXiv:0810.5343 [hep-ph]].

\end{thebibliography}
